\documentclass[]{spie}  

\usepackage{amsmath,amsfonts,amssymb}
\usepackage{graphicx}
\usepackage[colorlinks=true, allcolors=blue]{hyperref}

\def\muz{\mu_0}

\def\epsr{\epsilon_r}
\def\ohm{\Omega}
\def\ohmtxtnosp{$\ohm$}
\def\ohmtxt{\ohmtxtnosp\ }
\def\mus{\mu\mathrm{s}}
\def\mustxtnosp{$\mus$}

\def\mmsq{(\mathrm{mm})^2}
\def\mmsqtxtnosp{$\mmsq$}
\def\mmsqtxt{\mmsqtxtnosp\ }
\def\mum{\mu\mathrm{m}}
\def\mumtxtnosp{$\mum$}
\def\mumtxt{$\mum$\ }
\def\mumcu{(\mu\mathrm{m})^3}
\def\mumcutxtnosp{$\mumcu$}
\def\mumcutxt{\mumcutxtnosp\ }
\def\muev{\mu\mathrm{eV}}
\def\muevtxtnosp{$\muev$}
\def\muevtxt{\muevtxtnosp\ }
\def\muw{\mu\mathrm{W}}
\def\muwtxtnosp{$\muw$}
\def\muwtxt{\muwtxtnosp\ }
\def\wrthz{\mathrm{W\ Hz}^{-1/2}}
\def\wrthztxtnosp{$\wrthz$}
\def\wrthztxt{\wrthztxtnosp\ }
\def\NEP{\mathrm{NEP}}

\def\deg{^\circ}
\def\degtxtnosp{$\deg$}
\def\degtxt{\degtxtnosp\ }

\def\arcmin{^{\,\prime}}
\def\arcmintxtnosp{$\arcmin$}
\def\arcmintxt{\arcmintxtnosp\ }
\def\arcsec{^{\,\prime\prime}}
\def\arcsectxtnosp{$\arcsec$}

\def\lnu{L_{\nu}}
\def\newathena{\textit{NewAthena}}
\def\planck{{\it Planck}}

\def\sioxnosp{SiO$_x$}
\def\siox{\sioxnosp\ }
\def\sinxnosp{SiN$_x$}

\def\tinxnosp{TiN$_x$}
\def\tinx{\tinxnosp\ }

\def\Ccal{\mathcal{C}}
\def\Lcal{\mathcal{L}}
\def\fr{f_r}
\def\Qr{Q_r}
\def\Qi{Q_i}
\def\Qc{Q_c}
\def\Tb{T_{bath}}
\def\Ts{T_{substrate}}
\def\Tl{T_{load}}
\def\Tx{T_{exc}}

\def\Ta{T_{atm}}
\def\Tt{T_{tel}}
\def\deltls{\delta_{TLS}}
\def\Nqp{N_{qp}}

\def\tqp{\tau_{qp}}
\def\Sparam{S_{21}}
\def\f3db{f_{3dB}}
\def\Sdff{S_{\delta f/f}}
\def\SdQinv{S_{\delta(1/Q)}}
\def\SNqp{S_{\Nqp}}
\def\STLSdff{S^{TLS}_{\delta f/f}}
\def\etaopt{\eta_{opt}}
\def\etapb{\eta_{pb}}
\def\Pqp{P_{qp}}
\def\Popt{P_{opt}}

\title{NEW-MUSIC: The Next-generation Extended-Wavelength
Multiband Sub/millimeter Inductance Camera}

\author[a]{Sunil R. Golwala}
\author[b]{Andrew D. Beyer}
\author[b]{Daniel Cunnane}
\author[b]{Peter K. Day}
\author[b]{Fabien Defrance}
\author[b]{Clifford F. Frez}
\author[c,d]{Xiaolan Huang}
\author[a,f]{Junhan Kim}
\author[a]{Jean-Marc Martin}
\author[a]{Jack Sayers}
\author[c]{Shibo Shu}
\author[a,e,g]{Shiling Yu}

\affil[a]{Division of Physics, Mathematics, and Astronomy, California Institute of Technology, Pasadena, CA, USA, 91125}
\affil[b]{Jet Propulsion Laboratory, California Institute of Technology, 4800 Oak Grove Ave., Pasadena, CA, USA, 91109}
\affil[c]{Institute of High Energy Physics, Chinese Academy of Sciences, Beijing, China, 100049}
\affil[d]{Shanghai Normal University, Shanghai, China, 201418}
\affil[e]{National Astronomical Observatories of China, Beijing, China, 100101}
\affil[f]{Korea Advanced Institute for Science and Technology, Daejeon, 34141, Republic of Korea}
\affil[g]{University of the Chinese Academy of Sciences, Beijing, China, 101408}

\authorinfo{Send correspondence to SRG; email: golwala@caltech.edu}

\begin{document} 
\maketitle

\begin{abstract}
The Next-generation Extended Wavelength-MUltiband Sub/millimeter Inductance Camera (NEW-MUSIC) on the Leighton Chajnantor Telescope (LCT) will be a first-of-its-kind, six-band, trans\-mil\-li\-me\-ter-wave (``trans-mm'') polarimeter covering 2.4 octaves of spectral bandwidth to open a new window on the trans-mm time-domain frontier, in particular new frontiers in energy, density, time, and magnetic field.  NEW-MUSIC's broad spectral coverage will also enable the use of the Sunyaev-Zeldovich effects to study accretion, feedback, and dust content in the hot gaseous haloes of galaxies and galaxy clusters.  Six-band spectral energy distributions, with polarization information, will yield new insights into stellar and planetary nurseries.  NEW-MUSIC will employ hierarchical, phased arrays of polar\-i\-za\-tion-sensitive superconducting slot-dipole antennas, coupled to photolithographic bandpass filters, to nearly optimally populate LCT's 14\arcmintxt field-of-view with six spectral bands over 80--420 GHz (1:5.25 spectral dynamic range; 2.4 octaves).  Light will be routed to Al or AlMn microstripline-coupled, parallel-plate capacitor, lumped-element kinetic inductance detectors (MS-PPC-LEKIDs), an entirely new KID architecture that substantially enhances design flexibility while providing background-limited performance.  Innovative, wide-bandwidth, etched silicon structures will be used to antireflection-treat the back-illuminated focal plane.  NEW-MUSIC will cost-effectively reuse much of the MUSIC instrument, initially deploying a quarter-scale focal plane capable of the bulk of NEW-MUSIC science followed later by a full-FoV focal plane needed for NEW-MUSIC wide-area survey science.  
\end{abstract}

\keywords{kinetic inductance detector, photolithographic antenna, two-level systems, dielectric loss, dielectric noise, millimeter-wave instrumentation}

\section{INTRODUCTION}
\label{sec:intro}  

The time-domain sky at submillimeter and millimeter wavelengths is only just now beginning to be explored thanks to advances in observing facilities at these wavelengths.  Sources include various explosions associated with stellar death, outbursts from and accretion onto the remnants, young stars growing their mass, flaring stars, and active galactic nuclei.  This spectral range traces some of the most energetic phenomena, penetrates deep into some of the densest environments, accesses the earliest times and shortest-timescale variability, and probes the highest magnetic field environments.  Critical to understanding these sources are multi-band spectral energy distribution (SED) data over a large spectral range.  For synchrotron sources, the SED spectral slopes and breaks help to constrain the energies of the emitting electrons and thus the shocks that accelerate them, and the SED time evolution constrains dynamics of the explosion and the outflow.  For local dusty sources, the spectral slope and curvature constrain the temperature and grain emissivity, and temporal information tells us about the episodic nature of evolution.  

Accretion and feedback play central roles in the evolution of galaxies and galaxy clusters, impacting the hot circumgalactic medium (CGM) and intracluster medium (ICM).   Via Sunyaev-Zeldovich effect observations yielding total thermal content, pressure, and density, it is possible to probe deviations from equilibrium due to accretion and feedback processes: non-thermal pressure and bulk motions in galaxy clusters, and deviations of the CGM from self-similar scaling.  The complex nature of the spectral signature of the SZ effects, combined with the presence of contaminating foreground and background sources, necessitates multi-band data through the trans-mm (0.7--3.8~mm) regime.

Six-band spectral energy distributions, with polarization information, will yield new insights into stellar and planetary nurseries.

The very broadband trans-mm SED information needed for these various applications will be provided by the Next generation Extended Wavelength MUltiband Sub/millimeter Inductance Camera (NEW-MUSIC) on the Leighton Chajnantor Telescope.  NEW-MUSIC will provide 2.4 octaves of spectropolarimetric coverage in six spectral bands from 80 to 420~GHz (0.7--3.8~mm).  It will be deployed on the Leighton Chajnantor Telescope (LCT), the re-siting of the 11~\mumtxt rms, 10.4~m Leighton Telescope of the Caltech Submillimeter Observatory to Cerro Toco in the Atacama Desert in Chile.  

A number of key elements enable this broad spectral coverage with fundamental-noise-limited performance.  Hierarchical phased arrays of slot-dipole antennas using low-loss, hydrogenated amorphous silicon (a-Si:H) dielectric make it possible to couple incoming light to detectors across the 2.4-octave bandwidth while also matching pixel size to the diffraction spot size so the detector count and sensitivity requirements are not unnecessarily demanding.  The antennas are also inherently polarization-selective.  We couple light into the antennas through the silicon substrate using metamaterial, silicon, antireflective structures.  We sense the light from the antennas with Al or AlMn microstrip-coupled, parallel-plate capacitor, lumped-element KIDs (MS-PPC-LEKIDs), an innovative new KID design that combines the low two-level-system noise of a-Si:H with a flexible KID design that is also inherentely shielded against direct absorption.  These KIDs provide sensitivity limited only by photon statistics and generation-recombination noise (the sum being what we term ``fundamental noise'').  This revolutionary focal plane technology will be integrated into the existing MUSIC cryostat and relay optics and make use of existing KID readout systems to enable quick deployment.

\section{SCIENTIFIC MOTIVATION}

\subsection{The Trans-mm Time-Domain Frontier}
\label{sec:tda}


NEW-MUSIC on LCT will have transformative impact via simultaneous observations in six spectral bands from 80 to 420~GHz, covering the critical spectral range where transient and time-domain synchrotron emission shows peaks and spectral breaks and where dust thermal emission is accessible in most weather conditions at an excellent site.  NEW-MUSIC/LCT will build on enormous investments in O/IR time-domain surveys and the growing transient alert capacity of mm-wave CMB surveys. 

\subsubsection{Explosive Stellar Death}

\begin{table}[t]
\begin{center}
\includegraphics*[width=3in,viewport=72 504 340 666]{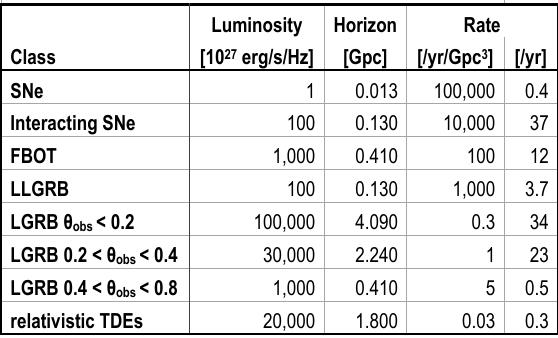} \\
\end{center}
\caption{\textbf{Estimated luminosities, detection horizons, and volumetric and total detection rates for various classes of exotic explosions.}  The rates assume a $5\sigma$ flux limit of 5~mJy at 90~GHz (feasible in a 3-minute LCT integration: see Table~\ref{tbl:sensitivity}).  Simultaneous multi-band observations would yield $5\sigma$ flux limits at 150, 230, 290, 345, and 405 GHz of 10, 9, 13, 25, and 46 mJy (Table~\ref{tbl:sensitivity}).  The rate of SNe is given as a reference point for luminosities and rates, not because they are a viable science target.  SNe, interacting SNe rates from~\cite{perley_rates_2020}, and we assume LLGRBs occur at 1\% of the SNe rate.  FBOT, LGRB, and TDE rates and luminosities from~\cite{eftekhari2022}.  For jetted LGRBs and TDEs, the luminosity is not  intrinsic but  observed.}
\label{tbl:transient_rates}
\end{table}

\paragraph*{The Nearby Universe} \hfill \\

\textbf{Death omens} 
Core-collapse supernovae (CCSNe) explode in a circumstellar medium (CSM) sculpted by mass loss during the star’s life. There is increasing evidence that many, or perhaps most, massive stars undergo intense eruptive mass loss within days to years of core collapse, producing a dense CSM. However, 
emission from supernova-driven 
shocks in dense CSM (``Interacting SNe'') is absorbed at early times (when the explosion is still close to the source; days to weeks) \textit{except in the trans-mm}. NEW-MUSIC/LCT's simultaneous six-band observations with mJy sensitivity (Table~\ref{tbl:transient_rates}) will track the explosion size and the structure of the surrounding medium (CSM density and magnetic pressure) via the time evolution of the synchrotron self-absorption (or free-free absorption) peak.  A program to study nearby ($\lesssim$100 Mpc) events would comprehensively assess pre-supernova mass-loss rates across the zoo of CCSNe.  Insight into the CSM structure may also shed light on the nature of the progenitors.

\textbf{Relativistic outflows into the CSM?} Fast blue optical transients (FBOTs) are a recently discovered class of transients, remarkable because of their blue colors and short ($\approx 10$ day) durations.  Many models have been put forth (see, e.g.,~\cite{fbots2023} for a review); one is that they are a particularly rare (0.1\%) class of interacting SNe, interesting because the newly formed compact object launches a mildly relativistic outflow into the CSM that enhances the luminosity by 100 times.  
Seven have been discovered to-date, informing the rate estimate.  NEW-MUSIC/LCT should detect of order ten FBOTs per year (Table~\ref{tbl:transient_rates}) to substantially enhance the sample with trans-mm SEDs well-sampled in time, which will reveal the dynamics of the outflow and the explosion environment and directly probe the particle acceleration process in relativistic shocks.

\textbf{Low-luminosity gamma-ray bursts} LLGRBs are thought to be cases of the fairly common broad-line SNe Ic that emit gamma-rays like gamma-ray bursts (discussed below) but with lower luminosity, perhaps due to smothering by a dense CSM --- effectively, another case of CCSNe exploding into a dense CSM, but now adding gamma-rays.  NEW-MUSIC/LCT should detect a handful a year, adding to our understanding of the LLGRB emission mechanism and CSM.

\paragraph*{The Distant Universe} \hfill \\

\begin{figure}[t]
\begin{center}
\begin{tabular}{cc}
\includegraphics*[width=2.75in]{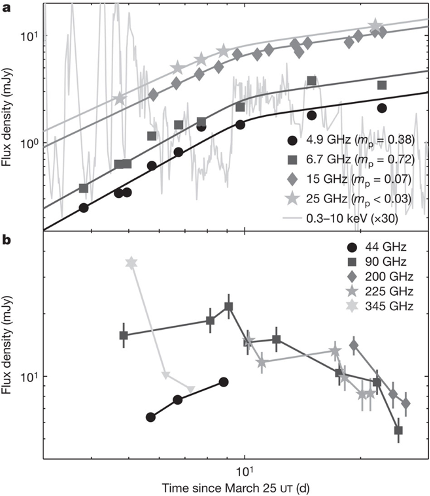} & 
\parbox[b]{3.5in}{
\includegraphics*[width=3.4in]{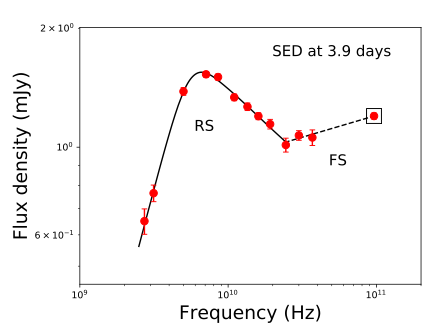} 
\vspace{0.3in}}
\end{tabular}
\end{center}
\caption{(Left)~\textbf{The tidal disruption event (TDE) \textit{Swift} J1644.493+57.3451.}  Radio (top) and trans-mm (bottom) lightcurves~\cite{zauderer2011}. While the smooth radio light-curves are affected by self-absorption within the source, the structure in the trans-mm lightcurves probe the rapidly flickering relativistic jet.  (Right)~\textbf{GRB afterglow SED~\cite{laskar2019}.}  At earlier times ($<$~1 day), the reverse shock will peak in the trans-mm. }
\label{fig:tde}
\label{fig:grb}
\end{figure}

\textbf{Gamma-ray bursts} Long-duration gamma-ray bursts (LGRBs) are thought to arise from relativistic jets launched in the collapse of massive stars. Thousands of GRBs have been discovered, but only a handful have the trans-mm data that provide unique information about the jet.  In particular, the reverse shock (RS) dominates the trans-mm at early times ($\lesssim$1~day~\cite{laskar2019}) and provides information about the baryon content and magnetization of the ejecta, while the forward shock (FS) dominates from $\mathcal{O}(1)$~day onward and provides information about the CSM, including its magnetic field geometry via polarization~\cite{laskar2019}.  Figure~\ref{fig:grb} shows a typical spectrum at 3.9~days and 
Table~\ref{tbl:transient_rates} shows luminosities and rates for different viewing angles.  CSM information may be critical to testing whether LGRBs are a special case of 
LLGRBs, one in which the CSM is tenuous and the full gamma-ray emission can escape~\cite{nakar2015}.

\textbf{Jets from shredded stars} Relativistic tidal disruption events (TDEs) are a rare case of TDEs in which a synchrotron-emitting jet is launched~\cite{at2022cmc_ami_outflow} when the supermassive black hole shreds a passing star. Only 1--10\% of all TDEs are relativistic, so only four have been observed, with \textit{Swift}~J1644+57 being an exemplar (Figure~\ref{fig:tde}).  NEW-MUSIC/LCT could detect a relativistic TDE once every year or two (Table~\ref{tbl:transient_rates}), yielding trans-mm SEDs well-sampled in time that, like for FBOTs, will reveal the environment and outflow dynamics  and directly probe relativistic shock particle acceleration.  Non-relativistic TDEs are 100 times less luminous~\cite{eftekhari2022}, so their 10--100$\times$ higher rate does not compensate the smaller detection volume.

\begin{figure}[t]
\begin{center}
\includegraphics*[width=4.5in,viewport=0 0 512 188]{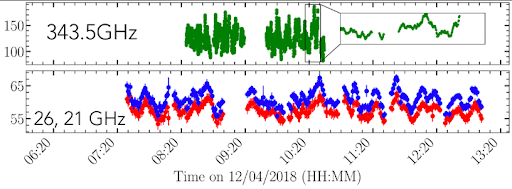}
\end{center}
\caption{\textbf{ALMA and VLA lightcurves of the black hole X-ray binary MAXI~J1820+070~\cite{tetarenko2021}.}  (Top)~ALMA; (bottom)~VLA.  The flux density units are mJy.  The high-frequency lightcurve is much richer and exhibits faster-timescale variation than those at lower frequencies.}
\label{fig:tetarenko}
\end{figure}

\textbf{Black Hole Accretion} Black holes accreting from companion stars power transient jets when in outburst.  Trans-mm monitoring near the spectral power-law break probes jet power variability that is more difficult to observe at lower frequencies due to synchrotron self-absorption~\cite{tetarenko2021}. As these sources evolve over the weeks following the outbursts, mm flares are still observed, likely tracing discrete jet ejections.  
For the handful of such systems studied, this field has been plagued by sparse data relative to the minimum variability timescales of $\ll$~hours, with recent dedicated observations revealing just how significant the activity is on short timescales and at high frequencies (see Figure~\ref{fig:tetarenko}).   NEW-MUSIC/LCT minute-by-minute measurements, with $\approx$10~mJy rms at 345~GHz, would be sufficient to track the variability detected by ALMA, to do so in many more sources, and to provide substantial SED information to characterize the time evolution of the spectral break.

\textbf{Novae} These thermonuclear explosions on accreting white dwarfs yield shocks that accelerate particles to radiate in the radio, X-rays, and gamma-rays.  Internal shocks determine the morphology of nova ejecta and eventually lead to dust and molecule formation in the interstellar medium.  Nova synchrotron emission should be brighter in the trans-mm early on, but data are lacking.  V1324 Sco, one of the most gamma-ray-luminous novae, showed mJy fluxes at 33~GHz with a spectral index $F_\nu \sim \nu^2$ at tens of days~\cite{finzell_v1324sco_nova_2018}.  NEW-MUSIC/LCT would have been able to provide a full SED with mJy noise in 3 minutes, and these SEDs and their time evolution would be much more diagnostic of the electron spectrum and the gamma-ray source than radio data~\cite{chomiuk_araa2021}.

\subsubsection{Outbursts and Pulses from Stellar Remnants}

\begin{figure}[t!]
\begin{center}
\includegraphics*[height=2in,viewport=0 0 426 263]{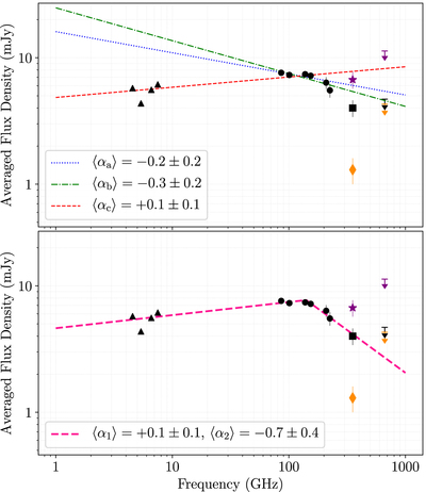}
\includegraphics*[height=2.25in,viewport=75 500 400 750]{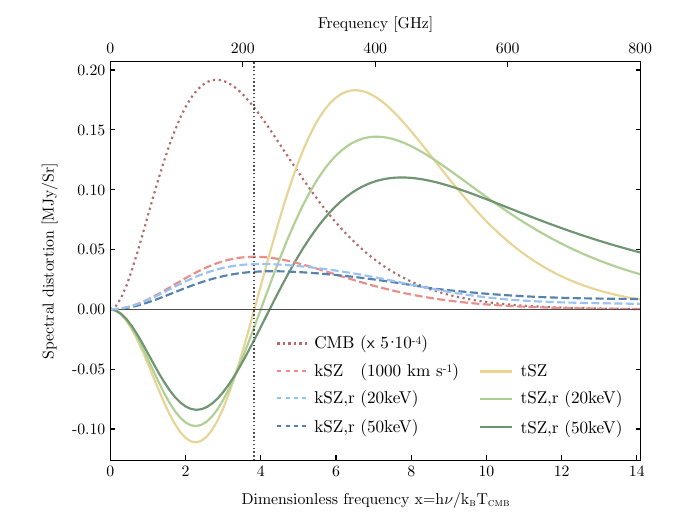}
\end{center}
\caption{(Left)~\textbf{Mean spectrum of the magnetar XTE~J1810-197~\cite{torne2022}.}  Different colored points indicate measurements at different epochs; the power-law fit is done for one epoch.  The ``averaged flux density'' is the integral of the flux density over the pulse divided by the period; the peak pulse flux is of order 20 times larger, easily detectable with NEW-MUSIC/LCT, while no quiescent flux has been detected.  NEW-MUSIC/LCT would cover the spectral peak. (Right)~\textbf{Sunyaev-Zeldovich effect spectra\cite{mrockzowski_review_2019}.} Solid: 
thermal (tSZ); dashed: kinematic (kSZ).  Relativistic corrections are included, differentiating the spectral shapes for different electron temperatures (in keV).  The assumed optical depth is $\tau = 10^{-2}$ and the Comptonization parameter is $y = 10^{-4}$. The dotted curve illustrates the shape of the unscattered CMB spectrum, which has been scaled by a factor of $5\,\times\,10^{-4}$.  The importance of multiple trans-mm spectral bands for measuring and separating the CMB, tSZ, and kSZ is clear.}
\label{fig:torne}
\end{figure}

\textbf{Magnetars} Magnetars, incredibly magnetized neutron stars with $10^{13-15}$~G fields, can have pulsed trans-mm emission synchronized to their radio pulsar activity, but the trans-mm emission mechanism is not known.  It is even conceivable the pulsed flux, and the SED shape and position of the break, can vary with time~\cite{torne2022}.  Of the thirty known magnetars, six are radio-loud, but only two have been detected in the trans-mm: XTE~J1810-197 (Figure~\ref{fig:torne}),  which exhibits a spectral break in the trans-mm, and the galactic center magnetar SGR~J1745-2900,  2--290~GHz with a $\nu^{0.4}$ spectrum.   NEW/MUSIC can both monitor the spectral shape of these known sources and search for trans-mm emission from others, even those with past non-detections given potential variability.

\subsubsection{Mass Buildup in Infant Stars}

Young stellar objects (YSOs) are generally underluminous compared to expected mass accretion rates.  The wide range of YSO variability (e.g., Figure~3 of~\cite{fischer_review_2023}), with some outbursts showing 100$\times$ increase in luminosity lasting for decades, support the idea of episodic accretion.   Trans-mm variability can be cleanly used to measure the change in $T_{dust}$ driven by stellar luminosity variations~\cite{johnstone2013, baek2020, fischer_review_2023}.  The JCMT Transient Survey~\cite{johnstone2018b, lee_jcmt_transient_survey_2021} has monitored eight nearby star-forming regions monthly since 2015/2016, observing 10--30\% variations in $\gtrsim 20\%$ of sources at a rms sensitivity per half-hour monthly observation of 14~mJy/beam at 345~GHz.  NEW-MUSIC/LCT will complement and expand on this work by accessing the southern sky and achieving a factor of 2 better sensitivity (see Table~\ref{tbl:transient_rates} caption).  Six-band SEDs will enable a search for time-variable free-free emission.  

\subsubsection{Active Stars and Exoplanet Habitability}
\label{sec:stars}

The number of millimeter flares from across the stellar landscape is growing, mainly via serendipitous searches in wide-area CMB surveys~\cite{guns2021, li_act_transients_2023}.  The flares range over six orders of magnitude in luminosity, entirely unexpected from prior radio data.   Several flares are in binary systems (including YSOs) and may be triggered by reconnection events in the interacting magnetospheres.  Only one flare has spectral information above 300~GHz.  The short flare timescales of several minutes to several hours, and extremely low duty cycles, motivate rapid ($\lesssim$1 hr) follow-up with NEW-MUSIC/LCT.  The flares seen to date, with flux changes of tens to hundreds of mJy at 90--220~GHz, are eminently detectable with NEW-MUSIC/LCT's 1--10 mJy rms across this specral band in 3 minutes (Table~\ref{tbl:transient_rates} caption), and NEW-MUSIC/LCT will add three more bands of data, especially in the poorly sampled region above 300~GHz.  Multi-band lightcurves will reveal the flare energetics, critical for understanding the emission mechanism.  Cyclotron emission is likely and will be circularly polarized.  Such flares have deep implications for exoplanet habitability: life on Earth relies on the relatively modest solar flaring activity and protection by Earth's magnetic field.

\subsubsection{Active Galactic Nuclei: A View Deep into the Jet}

Most radio galaxies have compact cores with flat spectra ($\lnu \sim \nu^0$). 
The flux at a given frequency is dominated by a radius $r  \propto \nu^{-1}$ for a conical jet. In low-luminosity sources such as M87, the innermost parts of the jet (at about 5--10 Schwarzschild radii) emit at $\sim$200 GHz, while for higher luminosity sources, and more strongly relativistically beamed ones, the innermost parts emit at 400~GHz or above. 
As shocks propagate down the jets and compress the magnetic field, there are flux outbursts and swings of polarization correlated with inverse-Compton-scattered high-energy gamma-ray emission.  
Trans-mm monitoring of $\sim$100 ``interesting'' sources on a few-day cadence would complement the long-running 15~GHz monitoring program on the Caltech OVRO 40~m telescope~\cite{ovro40m2012}, more completely characterizing variability over a range of length scales deep into the jet. Especially exciting would be correlations with gamma-rays (CTA) and neutrinos (IceCube, Baikal, KM3NeT).  Many of these sources are bright ($> 30$~mJy, so SNR~$> 30$ in three minutes (Table~\ref{tbl:transient_rates} caption)) and used as pointing calibrators, ensuring a large database of observations.


\subsection{Using Hot Gas Haloes to Study Accretion and Feedback in Galaxy and Galaxy Cluster Evolution}
\label{sec:sz}

Astro2020~\cite{astro2020} highlighted the emerging study of accretion and feedback via the hot, ionized CGM's thermal energy density (=~pressure) and electron density distributions as measured by, respectively, the thermal and kinetic Sunyaev-Zeldovich (tSZ, kSZ) effects\footnote{CMB distortions due to scattering with free electrons~\cite{sz1972a}}.   kSZ can also reveal coherent flows of ionized gas (non-thermal gas motions).

\subsubsection{The Dynamics of Accretion from the IGM onto the ICM in Galaxy Clusters}

\begin{figure}[t!]
\begin{tabular}{ll}
\hspace{-0.25in}
\parbox[b]{2.5in}{\includegraphics*[height=1.85in,viewport=300 540 540 705]{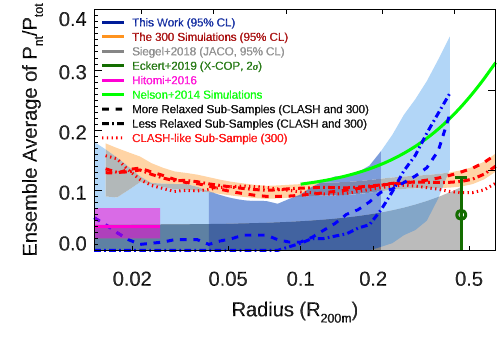} \vspace{-0.2in}}
&
\hspace{0.02in}
\begin{tabular}[b]{lll}
\includegraphics*[width=1.82in,viewport=50 630 158 730]{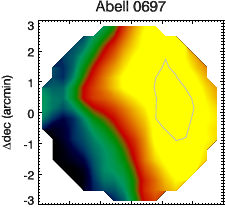} & \hspace{-0.2in}
\includegraphics*[width=1.54in,viewport=67 527 158 627]{figures/sayers2019fig7.pdf} & \hspace{-0.2in}
\includegraphics*[width=0.39in,viewport=515 525 565 725]{figures/sayers2019fig7.pdf} \\
\includegraphics*[width=1.85in,viewport=50 505 160 524]{figures/sayers2019fig7.pdf}  & \hspace{-0.47in}
\includegraphics*[width=1.85in,viewport=50 505 160 524]{figures/sayers2019fig7.pdf} & \\
\end{tabular}
\end{tabular}
\caption{(Left)~\textbf{Non-thermal pressure fraction as a function of radius from an analysis of 16 galaxy clusters~\cite{sayers_clump3d_2021},} compared to prior results~\cite{Siegel2018, Eckert2019} and simulations.  
(Right)~\textbf{Illustration of the use of kSZ for non-thermal motions~\cite{sayers_ksz2019}.}  The contours are $2\sigma$ and $3\sigma$.  A moving subclump in MACSJ0717.5 is clearly detected, and both mergers show prominent cluster-wide trends, but current sensitivity only permits the detection of extreme bulk motions in mergers.  
}
\label{fig:tsz_galaxies}
\label{fig:tsz}
\label{fig:ksz}
\end{figure}

Galaxy clusters serve as high-mass, high-SNR  galaxy analogues for studying accretion from the intergalactic medium (IGM).  The accretion shock heats infalling gas to near-virial temperatures, but significant support against gravity is also provided by residual coherent motions of order 500~km/s~\cite{Battaglia2012, Nelson2014, Biffi2016}.
Such support should coincide with a tSZ thermal pressure deficit relative to hydrostatic equilibrium~\cite{Siegel2018, Sereno2017, morandi_bolocam_2012}, for which there is modest evidence~\cite{morandi_bolocam_2012,Sereno2017, Siegel2018, Eckert2019, sayers_clump3d_2021, kim_chexmate_2023} (Figure~\ref{fig:ksz}).
kSZ measurements of such motions have been sensitivity-limited to extreme mergers (relative velocity $\sim$3000~km/s~\cite{sayers_macsj0717_ksz2013, sayers_ksz2019}; Figure~\ref{fig:ksz}), though the recently fielded TolTEC/LMT~\cite{toltec_spie2020} will have 5--10$\times$ better sensitivity and 5$\times$ finer angular resolution.  

With six bands and better sensitivity than prior instruments on CSO, NEW-MUSIC/LCT will be able to spectrally separate tSZ, kSZ, and dust/synchrotron contaminants in a large program on 20 well-studied galaxy clusters.  The $\gtrsim$20$\times$ improvement in tSZ SNR will enable \%-level constraints on non-thermal pressure, yielding a high-significance detection and distinguishing among simulations (Figure~\ref{fig:tsz}).  The $10\times$ kSZ uncertainty improvement to 100~km/s will enable mapping of bulk motion  in typical clusters, complementing TolTEC/LMT in field size.  Future X-ray spectroscopic imaging (XRISM, \newathena, X-ray probe mission) of such motions will also be complementary, mainly probing core regions at lower $z$.  The high-frequency bands will constrain the ICM dust content 
and deliver SZ constraints on the mass-weighted temperature via relativistic corrections.

\subsubsection{Feedback and the Relation between Galaxies and their Circumgalactic Medium}

Simulations show that, on galaxy scales, deviations from self-similar, gravity-only predictions for the radial tSZ profile of stacks of galaxies reflect feedback mechanisms including supernova winds~\cite{LeBrun2015, vandevoort_cgm2016, Crain2015}, AGN-driven outflows~\cite{su_hopkins_2021, kim_cgm_2021, simba_agn_tsz_cgm_2023}, and cosmic-ray pressure~\cite{ji_cr_2020, kim_cgm_2021}.  Both deficits and excesses have been observed and there are varying degrees of consistency with simulations that do or do not incorporate AGN feedback~\cite{planck_cgm2013, Greco2015, Gralla2014, Crichton2016, Verdier2016, amodeo2021, meinke_tsz_cgm_2021, meinke_tsz_cgm_2023}.  Current work focuses only on quiescent, high-mass galaxies rather than comparing star-forming and quiescent samples, and thus conclusions rely on comparison to imperfect, incompletely calibrated simulations.  Empirical comparison of different samples may be more conclusive, but deeper maps are needed.  Dust contamination is also a serious problem~\cite{planck_cgm2013, meinke_tsz_cgm_2021}.

Relative to prior work with \planck, SPT, and ACT, NEW-MUSIC/LCT's angular resolution, broader spectral coverage, and narrower, deeper survey will reach to 3--10$\times$ lower mass.  Its high-frequency bands will better constrain dust than the 220~GHz data used to date.  These improvements will enable differential measurements between galaxy stacks with different star-formation rates and potentially a detection of a deviation of the tSZ-mass relation from self-similarity.

\subsection{Dust in Stellar and Planetary Nurseries}
\label{sec:other_science} 

NEW-MUSIC/LCT will provide six-band SEDs and polarimetry of dust thermal emission in star-forming regions, protostellar cores, and protoplanetary disks, with many potential applications.  Magnetic field orientation measurements on scales between ALMA (sub-arcminute) and \planck\ (degree) can help quantify the role of magnetic fields in regulating star formation.  The frequency dependence of dust polarization on small scales can reveal whether different dust temperatures and populations are needed~\cite{vaillancourt_araa_2015}, impacting shielding in many environments.  
SEDs of protoplanetary disks can yield sizes and environments of large dust grains, testing if protoplanetary dust forms \textit{in situ} and how grain size affects shielding and clumping, impacting the speed of protoplanetary disk evolution.

\section{INSTRUMENT PARAMETERS}
\label{sec:instr_params}

Table~\ref{tbl:parameters} summarizes the spectral bands we are targeting for NEW-MUSIC, motivated by a combination of the science goals outlined above, atmospheric transmission windows, and appropriateness of the technologies being developed.

The prime driver for this specific frequency range is science using the Sunyaev-Zeldovich effects, as this spectral range overlaps the atmospheric windows where the effect is bright and spectrally distinguishable from contaminating sources such as the primary anisotropy of the cosmic microwave background, radio galaxies, and, most importantly, dusty, star-forming galaxies.  As outlined in the prior section, this frequency range is also an excellent match to unfulfilled needs for the study of time-domain sources in this spectral band.  In particular, the large proposed spectral range (1:5.25) will provide a large lever arm for: measuring spectral indices and looking for spectral breaks in non-thermal sources that constrain their engines, emission mechanisms, and explosion and outflow dynamics; for constraining $T/(1+z)$ for extragalactic dusty sources; and, for separating synchrotron, free-free, and thermal dust emission for local sources.  

The specific choice of bands is driven by the available atmospheric windows (see Figure~\ref{fig:bandpasses}), with the additional requirement that we split the very wide 190--310~GHz window into two bands to obtain spectral information across that window.  Practical considerations limit the spectral bands at high and low frequencies.  Above 420~GHz, there are no atmospheric windows with useful fractional bandwidths until the 650 and 850~GHz windows, which approach the Nb gap and thus require a fundamentally different optical coupling technology.  The rarity of good observing conditions at these higher frequencies also argue for different instrumentation.  At low frequencies, there is again a large gap down to the next atmospheric window below 45~GHz.  The combination of degraded angular resolution and the need for a much lower $T_c$ material render other technical approaches more appropriate.

We note that we evaluate the atmospheric optical load at the approximate 25th percentile for the site, 0.55~mm PWV, because we expect that, under \textit{better} conditions, LCT will be in use for 650 and 850~GHz observations.  (Comprehensive weather statistics are actually not available for Cerro Toco.  The one existing study of Cerro Toco~\cite{cortes2016pwv} suggests that the PWV there is, on average, 90\% of that on the plateau.  Thus, we use the percentile for 0.6~mm PWV on the plateau~\cite{cortes2020pwv, radfordpvt}.)   We use the \textit{best} planned observing conditions for our calculations to obtain the most stringent requirements on instrument sensitivity so that, under all conditions, the instrument is background limited.

Table~\ref{tbl:parameters} also summarizes the focal plane parameters: pixel size in mm and $(f/\#)\lambda$, beam FWHM, and number of pixels.  The $f/\#$ of the optics is chosen so that the pixel size is between $1(f/\#)\lambda$ and $2(f/\#)\lambda$ at all frequencies.  The band centers work out such that the pixel size is closer to $1(f/\#)\lambda$ at high frequency, where angular resolution is important for dusty, star-forming galaxies, while the pixel size is more conservative (closer to $2(f/\#)\lambda$) at low frequencies where sidelobes and stray light are more of a concern.

The total number of pixels in each band is chosen to approximately fill the LCT 14\arcmintxt field of view.  (Gaps between pixels for KIDs and the readout feedline make the focal plane larger than $\ell_{pix} \sqrt{N_{pix}}$.)  Initial deployment will use a quarter-scale focal plane, and the final instrument will consist of four copies of the quarter-scale focal plane.

\begin{table}
\begin{center} 
 \begin{tabular}{r|cccccc} 
\hline\hline 
Quantity & B1 & B2 & B3 & B4 & B5 & B6 \\ \hline \hline
$\nu$ [GHz] &  90 & 150 & 230 & 275 & 350 & 400 \\ 
$\lambda$ [mm] & 3.3 & 2.0 & 1.3 & 1.1 & 0.87 & 0.74 \\
$\Delta\nu$ [GHz] &  35 &  47 &  45 &  40 &  34 &  30 \\ \hline
$\Ta$ [K] &   5 &   6 &  10 &  14 &  30 &  57 \\ 
$\Tt$ [K] &  13 &  13 &  13 &  13 &  27 &  27 \\ 
$\Tx$ [K] &  14 &  14 &  17 &  20 &  41 &  61 \\ 
$\Tl$ [K] &  32 &  33 &  40 &  47 &  98 & 145 \\ \hline
$\ell_{pix}$ [mm] & 6.66 & 6.66 & 3.33 & 3.33 & 1.66 & 1.66 \\
$(F/\#) \lambda$ & 1.17 & 1.94 & 1.46 & 1.84 & 1.11 & 1.30 \\
FWHM [\arcsectxtnosp] & 76 & 53 & 32 & 29 & 20 & 17 \\
$N_{pix}$ & 64 & 64 & 256 & 256 & 1024 & 1024 \\ 
\hline \hline 
\end{tabular} 
\end{center} 
\caption{\textbf{NEW-MUSIC spectral bands, expected optical loads, pixel sizes, beam FWHMs, and pixel counts.} $\nu$, $\lambda$ = spectral band center; $\Delta \nu$ = spectral bandwidth; $\Ta$, $\Tt$, $\Tx$, $\Tl$ = Rayleigh-Jeans optical loading from atmosphere, telescope, instrument, and total.  $\Ta$ assumes $\approx$25th percentile precipitable water vapor conditions (PWV = 0.55~mm) for Cerro Toco; see text for motivation for this choice.  $\Tt$ assumes 1\% or 2\% emissivity per aluminum optical reflecting surface (see \S\ref{sec:optics} for optical configuration) but neglects panels gaps and feedleg scattering.  $\Tx$ is assumed to be such that the total loading is increased by an ad hoc factor of $\sqrt{3}$ from telescope and sky loading alone.  The bandwidths and optical loadings have intentionally been optimistically chosen to place the most stringent demands on instrument sensitivity.  The choice of pixel sizes and numbers of pixels is discussed in the text.  The beam FWHMs on the sky derive from the pixel size and the optical configuration.}
\label{tbl:parameters}
\end{table}

\section{TECHNICAL APPROACH}
\label{sec:tech_approach}

\subsection{Focal Plane Architecture -- Design}

\subsubsection{Polarized, Hierarchical Antennas using Low-Loss Hydrogenated Amorphous Silicon (a-Si:H)} 
\label{sec:design:antenna}

\paragraph{Phased Arrays of Slot-Dipole Antennas}

\begin{figure}[t!]
\begin{center}
\begin{tabular}{cc}
\multicolumn{2}{c}{\hspace{0.25in}
\includegraphics*[width=1in,viewport=0 -10 90 90]{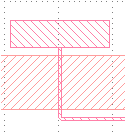}
\hspace{0.25in}
\includegraphics*[width=5in]{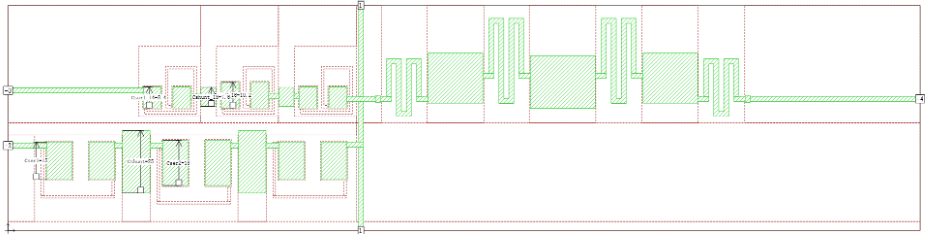}\vspace{0.2in}} \\ 
\multicolumn{2}{c}{\includegraphics*[width=6.5in]{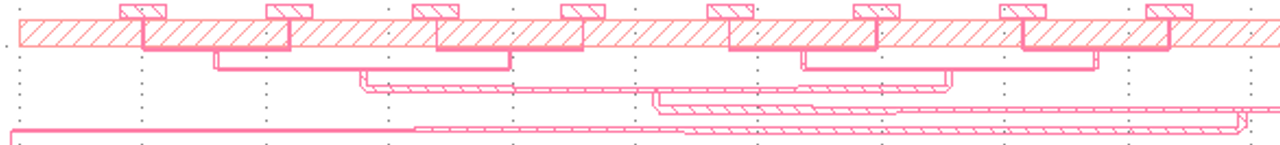}} \\ 
\includegraphics*[width=2.75in]{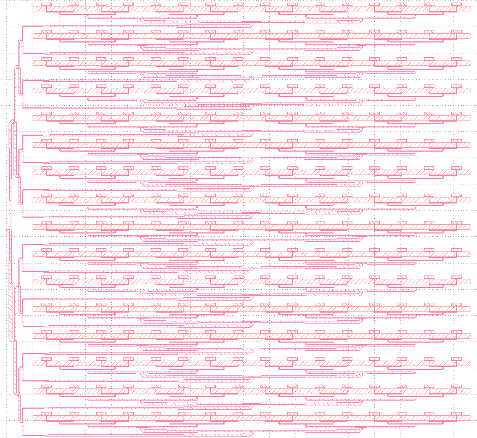} &
\includegraphics*[width=3.5in,viewport=0 0 650 500]{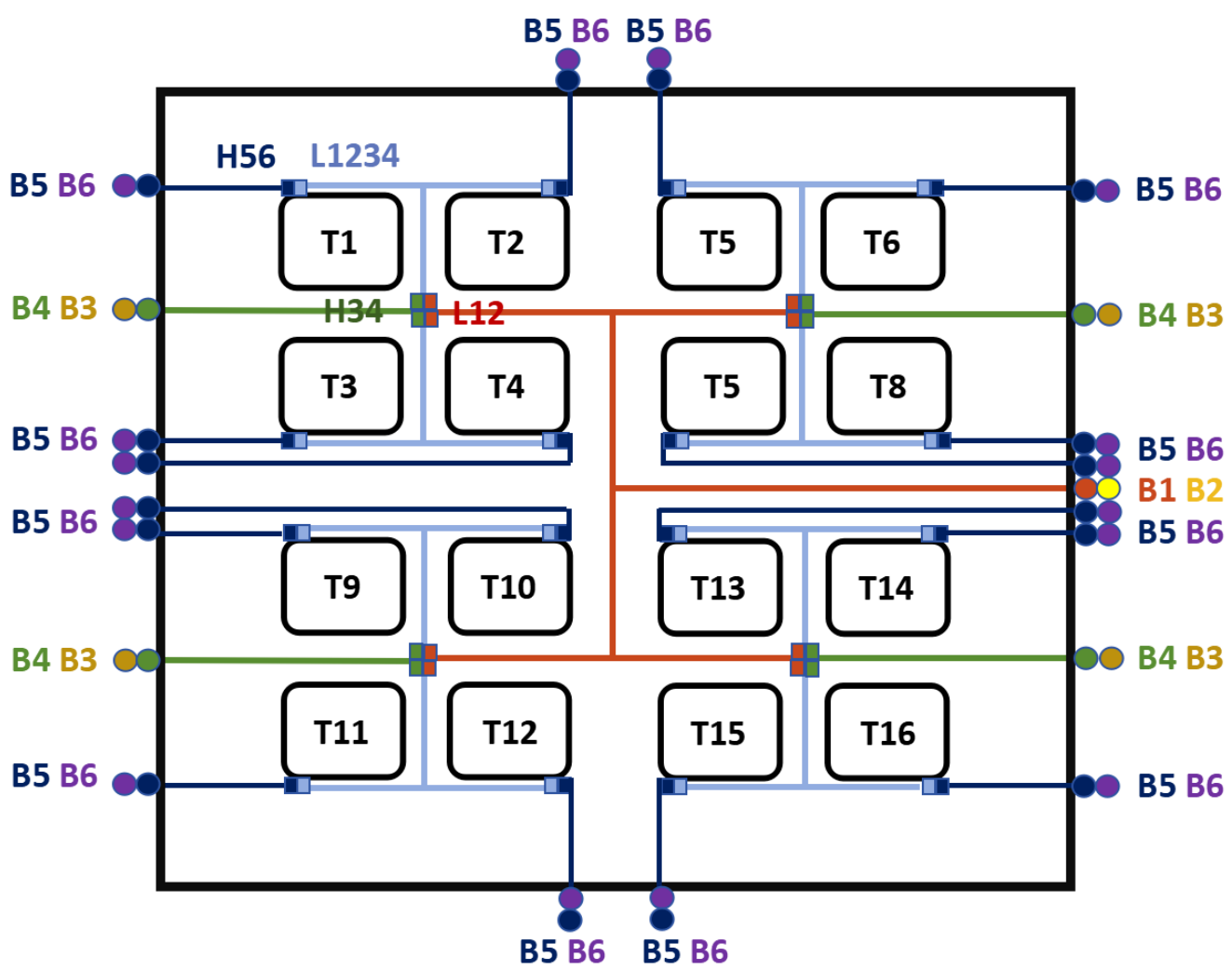}
\\ 
\end{tabular}
\end{center}
\vspace{-6pt}
\caption{\textbf{Hierarchical phased-array antenna and filterbank example.} (Top left)~Close-up view of one 1~\mumtxt wide feed crossing the 18~\mumtxt wide slot and shunted to the ground plane by a $37\,\mum \times 10\,\mum$ capacitor at the top. (Top right)~Summing junction/BPF-LPF filterbank example.  The microstriplines entering at top and bottom carry the in-phase signals after summing of two $16\times 16$ fundamental elements.  The rectangular shapes are capacitors, either shunt capacitors to the ground plane (single elements) or series capacitors (paired elements) where one plate is on the microstripline layer and one plate is in a window in the ground plane.  The meanders connecting them are inductors, sitting in the same windows in the ground plane.  To provide a scale, the capacitor at lower left has 15~\mumtxt vertical dimension.  The filters at the left are B3 and B4 BPFs.  The filter at right is a LPF that passes B1 and B2.  A similar BPF/LPF combination must be placed at the output of each $16\times 16$ fundamental element, but it only requires a single input since it does not perform summing, just filtering.  After the final summing junction at the center of the entire hierarchical array, there will be a B1-B2 BPF filterbank with no LPF.  (Middle)~Close-up view of half of one slot with its summing tree.  The slot is left-right mirror symmetric except for the output leg of the summing tree.  (Bottom left)~$16\times 16$ fundamental antenna element. (Bottom right)~Schematic showing how antenna elements' outputs are routed, selectively filtered, and summed for the three-scale antenna.  Each square ``TX'' is a 16~$\times$~16 fundamental element (level 0).  Of the four elements of a level 1 group (e.g., T1, T2, T3, T4), the left ones match the schematic at bottom left while the right ones have been mirrored through a vertical line so the summing tree exits to the right, not the left.  (There is no mirroring through a horizontal line because that would cause a 180\degtxt phase shift of the feed excitation.) The elements of each level 1 group are separated by 208~\mumtxt while the elements of the level 2 group are separated by 312~\mumtxt to accommodate the summing trees.  The gaps provide sufficient space for the required summing junctions and filterbanks.}
\label{fig:antenna}
\end{figure}

Light is received at the focal plane by a superconducting phased-array antenna~\cite{goldin_ltd9, goldin_spie2002}, back-illuminated through the silicon substrate, as shown in Figure~\ref{fig:antenna}.  The fundamental element is a 1.664~mm long, 18~\mumtxt wide slot in a niobium ground plane.   An incoming EM wave polarized normal to the slot excites a voltage across it, which excites waves on capacitively shunted microstripline (``microstrip'')~\cite{microstripline_wikipedia} feeds crossing the slot.  The feeds have a 54~\ohmtxt impedance given the microstripline geometry.  The capacitors are 37~\mumtxtnosp$\times$~10~\mumtxt and have 40~\ohmtxt reactance at 100~GHz.  The 1~\mumtxt wide microstrip comprises the ground plane (190~nm thick) and a Nb wiring layer (160~nm thick) sandwiching a 1070~nm thick hydrogenated amorphous silicon (a-Si:H) dielectric layer. The ground plane prevents direct excitation of the microstrip by incoming light, both by geometrical blocking and by imposing a zero electric field boundary condition $\ll \lambda$ away from the top conductor.  A binary summing tree combines the trans-mm wave from 16 feeds along a slot and from 16 such slots with equal path lengths.  The feeds and the slots are all spaced by 104~\mumtxt center-to-center.  After each summing junction, the now-widened microstripline is tapered back down in width so the summing tree occupies minimal space between the slots, but then the microstripline exiting the summing tree is allowed stay at the final summing junction output width, 5~\mumtxtnosp.  This wider microstripline is both more ideal (width large compared to the 1.07~\mumtxt dielectric thickness) and more robust against fabrication defects as it travels to the filters and KIDs.  A backshort 150~\mumtxt from the vacuum side improves the antenna forward efficiency averaged over its entire band, and it does this for a very wide bandwidth because of the high permittivity substrate (whose thickness is also optimized in the calculation).  The inductance it adds is tuned out by the capacitive shunts at the feeds.  The intrinsic bandwidth of the antenna is calculated to be very wide, 1:7.5 above 90\% efficiency.

To the extent that the slot impedance is the same for all 16 feeds along a slot, the illumination of the antenna is uniform and the far-field beam will be sinc-like.  Variations of the slot impedance with position may cause the illumination to be tapered along the slots.  At low frequency, the impedance oscillates significantly along the slot (varying by $\pm$50\% at 100~GHz) but the variation is over the entire slot, while, as the frequency increases, the slot impedance is stable over the bulk of the slot and then shows similar $\pm$50\% variations only near the ends.  The resulting coupling variation may result in a small $E$-$H$ plane beam asymmetry.  Fortunately, even these significant impedance mismatches should cause $\lesssim$10\% variations in power coupling from the slot to the feed.  It remains to be determined whether any such effects are observed (see \S\ref{sec:exp:beams}).  If so, they may be corrected by adjusting the dimensions of the phased array.

\paragraph{Hierchical Summing}

The main innovation in the phased-array antennas for NEW-MUSIC is to hierarchically sum the slot antennas in a frequency-selective way so that the pixel size grows with wavelength to roughly track the diffraction $\text{FWHM} = (F/\#)\lambda$.  Figure~\ref{fig:antenna} shows how photolithographic low-pass and band-pass filters (LPFs and BPFs) will be used to do this, starting with 16$\times$16, 1.664~mm antennas.  The main advantage is the nearly optimal trade-off between optical efficiency and angular resolution.
For $\lambda$ for which a fixed antenna size is smaller than $(F/\#)\lambda$, its efficiency is $<50$\%, enhancing the requirement on detector noise.  For $\lambda$ for which the fixed antenna size is larger than $2(F/\#)\lambda$, angular resolution is degraded ($\text{FWHM} > \lambda/D$), which degrades point source sensitivity and confusion limit.

In implementing hierarchical summing, we allow gaps between the fundamental elements, which could have been avoided by use of microstripline crossovers.  While such crossovers have been demonstrated at trans-mm wavelengths\cite{spt3g_posada_sust_2016}, we deemed it simpler, more robust, and easier to model to instead array the $16\times 16$ slot antennas with gaps between them, using the space between them for filtering and microstripline routing without crossovers.  The gaps are 208~\mumtxt between the level 0 (fundamental) antennas and 312~\mumtxt between the level 1 (one level of summing) antennas; the gap grows because the level 1 gap must permit space for the summing trees.  Both are multiples of the feed/slot spacing of 104~\mumtxtnosp.

\paragraph{Low-Loss a-Si:H for Trans-mm Microstripline}
\label{sec:design:asihloss}

An enabling technology for hierarchical phased-array antennas is low-loss dielectric for the trans-mm-wave microstripline, which permits the detectors for even the highest-frequency bands of interest to be at the outer edge of the low-frequency antenna.  We use hydrogenated amorphous silicon (a-Si:H), for which we have demonstrated recipes with RF loss tangent $\delta$ as low as $7\,\times\,10^{-6}$\cite{defrance_asihloss}.  For fabrication convenience here, we used somewhat lossier recipes, with $\delta \approx 3\,\times\,10^{-5}$ (see \S\ref{sec:fab} for details).  The dielectric is 1070~nm thick to provide a microstripline impedance comparable to that obtained with more typical dielectrics ($\epsr \approx 12$ for a-Si:H  vs.\ $\epsr \approx 4$ for \siox and $\epsr \approx 7$ for \sinxnosp, $\mu = \muz$ for all).  For the material used here and assuming a trans-mm loss tangent equal to the RF loss tangent, the microstripline dielectric loss for the 420~GHz band would be 0.6\%.  It is expected the loss tangent will increase between RF and trans-mm frequencies\cite{music_spie2012_golwala, sron_asich_2024}, but even a factor 10 degradation would yield an acceptable 94\% microstripline transmission at 420~GHz.

\paragraph{Polarization Coverage}

While the antennas are inherently polarization-selective, it is well understood that polarization sensitivity must be modulated quickly to mitigate systematic uncertainties in differencing complementary polarizations on the sky.  The first level of such differencing will be provided by rotating the antennas 90\degtxt between adjacent level 2 (B1/B2) pixels, separated by approximately 1.5\arcmintxt on the sky.  At scan speeds of 0.5--1\degtxtnosp/s, there will be only a 25--50~ms time separation, fast enough to freeze atmospheric emission variations (``sky noise'')~\cite{bolocam_skynoise_apj, act_hincks2010}.  The sky noise will thus provide a relative calibration and will difference away well.  Parallactic angle rotation will yield full coverage in $Q$ and $U$ Stokes parameters.  Another, potentially more effective level of differencing, pending evaluation of its consistency with the final optical design, would be provided by placing a polarizing grid between the final lens and the focal plane, feeding complementary polarizations to two focal planes situated at right angles with respect to each other.  Additional modulation could be provided by a rotating or stepped broadband, multi-layer, sapphire half-wave plate~\cite{hanany_waveplate2005, savini_waveplate2006a, savini_waveplate2006b}, cryogenically situated at an internal Lyot stop, though obtaining the necessary 2.4-octave bandwidth may be challenging~\cite{takaku2023}.  A more practical option may be a rotating or stepped broadband, reflective metal-mesh half-wave plate~\cite{pisano_broadband_hwp_2016} placed at one flat relay mirror outside the cryostat.  If circular polarization sensitivity is desired (interesting for stellar flaring cyclotron emission, \S\ref{sec:stars}), a variable-delay polarization modulator~\cite{vpm_chuss_2006, krejny_vpm2008, vpm_chuss_2012} could be used at the second flat relay mirror, employing a $\lambda/8$ air gap rather than the conventional $\lambda/4$ to make it a quarter-wave rather than half-wave plate.  This approach is narrow-band, but observations could be taken sequentially with multiple spacings.

\subsubsection{Microstrip-Coupled, Parallel-Plate Capacitor, Lumped-Element KIDs (MS-PPC-LEKIDs) using Low-Noise a-Si:H} 
\label{sec:design:kid}

\paragraph{Parallel-Plate Capacitor, Lumped-Element KIDs (PPC-LEKIDs)}

\begin{figure}
\begin{center}
\includegraphics*[width=2in,viewport=720 40 1190 350]{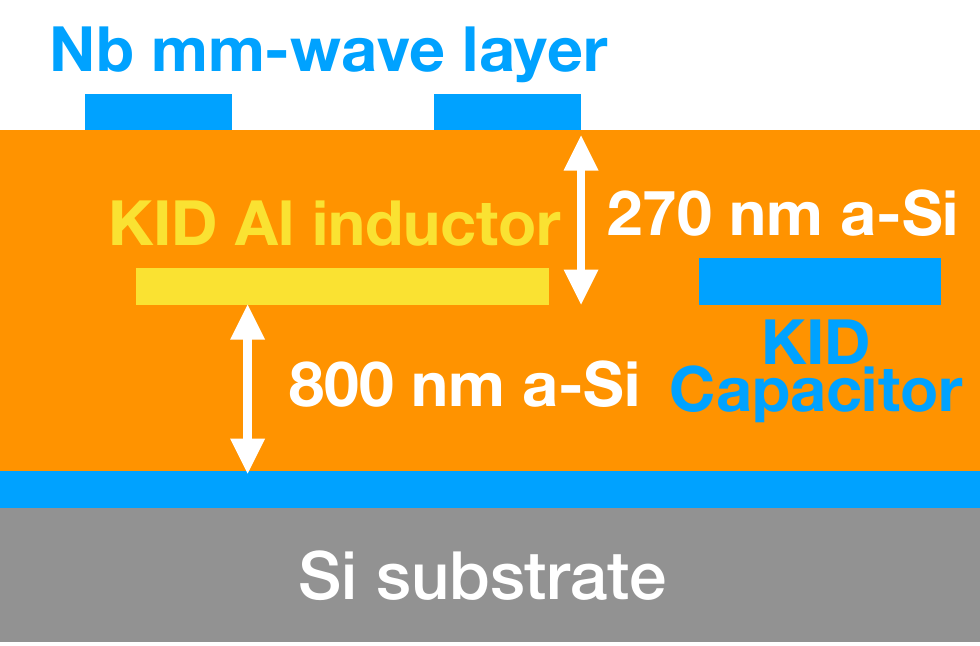} \\
\smallskip
\includegraphics*[width=6.5in]{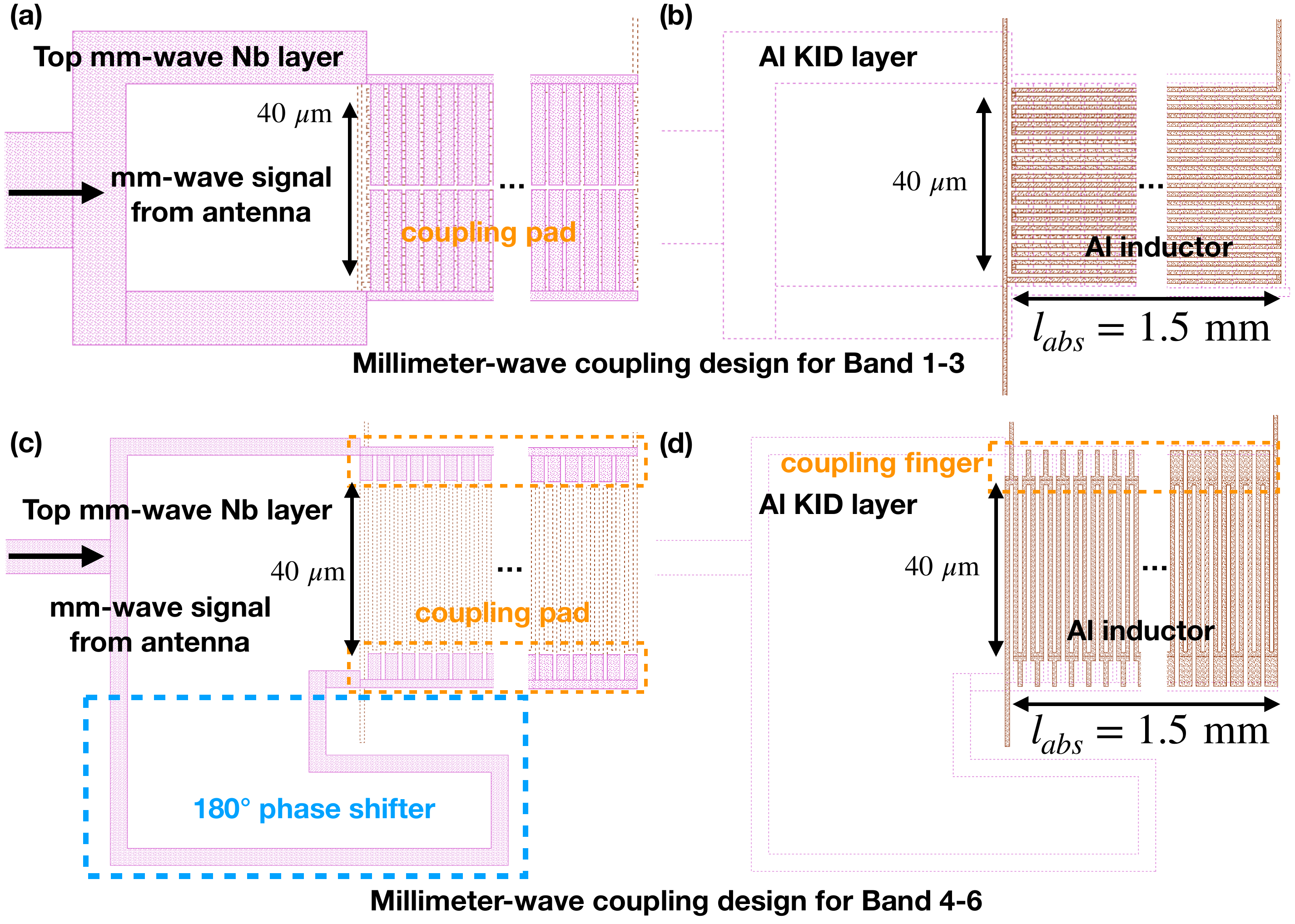} \\
\end{center}
\vspace{-9pt}
\caption{\textbf{Coupling of trans-mm microstripline to KID inductor} (Top)~Layer structure.  (Bottom)~These structures couple trans-mm power from the microstripline entering at the left into the KID inductor.  Different structures are used at lower frequencies (B1--B3) and higher frequencies (B4--B6).  The Nb microstripline top layer is shown in magenta.  The Al KID inductor layer is shown in brown.  The underlying Nb ground plane and the intervening dielectric layers are not shown.  The operation of each of the structures is described in the text.  Figures reproduced from\cite{shu_multiscale_al_ltd19}.}
\label{fig:ms_coupling}
\end{figure}

The trans-mm micro\-strip\-line exiting the BPFs couple to a novel KID design illustrated in Figure~\ref{fig:ms_coupling}~\cite{multiscale_tin_spie2014, shu_multiscale_al_ltd19}.  The two ends of a meandered inductor connect to two plates, all 100~nm thick Al or AlMn.  The structure sits on top of the ground plane and a first 800~nm thick a-Si:H layer.  The top plates form two parallel-plate capacitors (PPCs) with the ground plane, connected in series.  The symmetric KID design makes the shared PPC electrode a virtual ground, obviating isolating it from the ground plane.  While the two plates could in principle couple to an incoming electric field normal to the slot between them, the ground plane shields the KID inductor and capacitor in the same way as it does the trans-mm microstrip.  To prevent out-diffusion of the quasiparticles into the inductor, we deposit Nb (via liftoff to prevent etching damage to the KID material) on the PPC top plates to raise the pair-breaking energy to 3~meV.  This energy corresponds to $\nu = 2\,\Delta_{Nb}/h \approx 740$~GHz, so the Nb is also a poor direct absorber for in-band light (as well as being highly reflective even above 740~GHz).  We couple the KID to a 2~\mumtxt wide microstripline feedline (impedance 32.5~\ohmtxtnosp) via a PPC coupling capacitor attached to one end of the inductor.  Though it is difficult to impedance-match this type of feedline to 50~\ohmtxtnosp, it ensures the ground plane is uninterrupted except at the antenna slots.  A traditional coplanar-waveguide (CPW) feedline inherently breaks the ground plane into two halves, which can at best be coupled intermittently by ground bridges across the CPW.

\paragraph{Microstrip Coupling}
\label{sec:ms_coupling}

The trans-mm microstripline deposits power in the inductor via a unique capacitive coupling.  For B4/B5/B6, the design is similar to the one we detailed earlier for a \tinx inductor~\cite{multiscale_tin_spie2014}, illustrated in Figure~\ref{fig:ms_coupling}.  The microstripline from the BPF encounters a 50-50 splitter.  One of the outputs is delayed by a half-wavelength so that the two complementary microstriplines, now with nearly equal magnitude but opposite sign trans-mm voltages on their top electrodes, present a trans-mm voltage difference.  The KID inductor meanders back and forth in the space between the two microstriplines, with small pads extending from the end of each meander under pads that extend outward from the microstripline top electrode, separated by 270~nm of a-Si:H (the total thickness of 1070~nm less the 800~nm thickness between the KID inductor and the ground plane).  The overlapping pads and each meander of the inductor thus form a trans-mm $C$-$R$-$C$ network, through which the trans-mm voltage on the two microstriplines drives a trans-mm current.  This current dissipates the trans-mm power in the meanders of the KID inductor.  Any single $C$-$R$-$C$ element has a lumped element impedance much larger than that of the microstripline, and the microstripline continues onward, so each meander can be consider a high parasitic impedance between the two microstriplines rather than a terminating impedance.  With many such meanders, the trans-mm power is adiabatically absorbed as the wave propagates down the microstripline pair.  The coupling $C$ can be increased along the microstripline so that equal power, rather than equal fractional power, is absorbed per unit length, ensuring the trans-mm power is absorbed uniformly over the entire inductor rather than with an exponential profile.  The coupling $C$ and the meander width, thickness, and length are all free parameters, providing a great deal of design space to simultaneously obtain a high enough KID responsivity (set by the KID inductance, capacitance, and inductor volume) that generation-recombination (and thus photon noise) dominates over amplifier and TLS noise while also obtaining high optical absorptance.  The details of the optimization have been previously provided for a \tinx inductor\cite{multiscale_tin_spie2014} and for an Al inductor\cite{shu_multiscale_al_ltd19}.   We term this design the ``adiabatic lumped-element'' coupler.

For the lower frequency bands B1/B2/B3, it proved difficult for the above design to provide high optical absorptance with a meander of a reasonable length and width given the low frequency combined with the low resistivity of Al (or AlMn) in comparison to the \tinx for which the design was originally intended\cite{multiscale_tin_spie2014}.  (A narrower Al linewidth could have addressed this challenge, but we worried about fabrication yield given that the Al sits on somewhat rough a-Si:H rather than optically polished crystalline silicon.)  We therefore developed an alternative ``traveling wave'' coupler design, illustrated in Figure~\ref{fig:ms_coupling}.  First, we taper the microstripline to a very large width (low impedance) to match the low impedance of the coupler structure.  Again, there is a 50-50 splitter, but this time with no delay added.  Instead, where the two microstriplines run in parallel, we extend the capacitive pads from each microstripline to become fingers, almost touching between the two microstriplines.  The KID inductor is again sandwiched between the ground plane and the microstripline top electrode, with the same 800~nm/270~nm split of the a-Si:H thicknesses.  In this case, however, the inductor meanders along the same direction as the microstripline, and the microstripline's capactive tabs extend almost completely over the inductor's meanders.  We believe this geometry effectively makes a capacitive voltage divider between the microstripline top electrode, the inductor, and the ground plane, imposing a spatially varying voltage along the KID meanders.  This coupling excites a microstripline mode between the meanders and the ground plane, and the mode's energy is dissipated in the trans-mm-lossy KID inductor material.  Each long KID meander is connected to one of its neighbors at either end in order to form one continuous inductor.  The reflections imposed by these shorts effectively make the excitation in the KID inductive meander microstripline a standing wave that is excited by the incoming wave on the Nb microstripline.  Like the adiabatic lumped-element coupler, the traveling wave coupler is adiabatic in the sense that power is deposited gradually along the coupler's microstripline rather than by terminating the microstripline in a matched impedance.  Again, the details of the optimization have been previously been provided\cite{shu_multiscale_al_ltd19}.

Since there is no voltage difference between the 50-50 split Nb microstriplines in the traveling-wave design, the split may be an unnecessary residual feature of its origin in the adiabatic lumped element coupler.  The use of capacitive fingers is, however, important: it provides the necessary voltage divider coupling to the KID meander while maintaining a higher microstripline impedance than would be obtained by simply widening the microstripline to cover the KID inductor.  The capacitance per unit length $\Ccal$ is increased by about half as much as would have been obtained by widening the microstripline, while the inductance per unit length $\Lcal$ is largely unchanged (very little current flows along the fingers).  Therefore, rather than the impedance decreasing by a factor $\approx w_b/w_c$, where $w_b$ and $w_c$ are the microstripline width exiting the BPF and in the coupler, it decreases by a more modest factor, $\approx \sqrt{2\,w_b/w_c}$.  The increase in microstripline width between the BPF and the coupler can be smaller by the same factor.  

\paragraph{Contrasts with Prior KID Designs}

Our design contrasts with the other standard approaches for coupling mm/submm power to KIDs: direct absorption in the inductor (unmediated by microstripline) and termination of microstripline directly in the KID inductor.  In both cases, the KID inductor must match the impedance of the incoming wave, which may be in vacuum (direct illumination), a vacuum cavity (horn-coupled direct illumination), dielectric (silicon or alumina lens coupling), or microstripline (horn, lens, or antenna coupling), while simultaneously providing high enough responsivity (deriving from the volume and resonant frequency) for photon-background-limited sensitivity.  Impedance depends on film thickness and line width, so yield and uniformity can be fabrication challenges.  The designs used here are quite robust against film thickness and linewidth variations because the adiabatic coupling to the KID avoids dimension-sensitive impedance matching.  

\paragraph{Low-Noise a-Si:H for PPC-LEKIDs}

This design is only feasible because of our development of low-loss a-Si:H, based on~\cite{mazin_microstrip} but with 10$\times$ lower loss tangent~\cite{defrance_asihloss} and at least 2$\times$ lower noise~\cite{defrance_asihnoise}.  Our first work\cite{defrance_asihloss} presented two a-Si:H deposition recipes for two different machines at two sites and found that the loss tangent is stable, both across films fabricated months or years apart~\cite{defrance_asihloss} and in a given film over time ($d\deltls/dt \approx 0.35\,\times\,10^{-6}$/mo in a typical lab environment).  Our KID design optimization\cite{shu_multiscale_al_ltd19} used measurements of TLS noise for recipe A from \cite{defrance_asihloss} taken at $\Tb = 100$~mK.  In this work, we make our measurements at $\Tb \approx 250$~mK.  Assuming the measured $\Tb^{-1.7}$ scaling of TLS noise power spectral density with temperature~\cite{kumar_noise} and a naive (and experimentally unconfirmed) linear scaling with $\deltls$, we should observe TLS noise about 1.5$\times$ lower than the design.  

\paragraph{AlMn KIDs}

Al, especially in thin films, generally has $T_c$ too high to be suitable for B1, whose lower edge is $\approx$75~GHz: even bulk $T_c \approx 1.2$~K and thus $2\,\Delta_{Al}/h \approx 88$~GHz would substantially decrease the B1 bandwidth (Table~\ref{tbl:parameters}), and thin films generally have higher $T_c$\cite{al_thin_film_tc}.  We will therefore use AlMn, an alloy in which Mn suppresses Al $T_c$~\cite{youngimplant, youngimplant2004, deiker_almn, ruggiero_almn} and that has previously yielded $Q > 2\,\times\,10^5$ resonators down to $T_c = 0.69$~K~\cite{jones_almnkid2017}.  

\paragraph{Control and Trimming of $\fr$}

There is reason to believe PPC-LEKID resonant frequencies will be more well controlled than for IDC-based KID designs: because the electric field is so well confined in the PPC, there should be negligible frequency scatter due to parasitic capacitances and inter-resonator couplings.  Fabrication non-uniformities such as variations of inductor linewidth and thickness (both affecting geometrical as well as kinetic inductance), dielectric thickness, or PPC plate dimensions should be smooth functions of position.  Regardless, it is sensible to have a trimming mechanism~\cite{liu_ctrimming2017, shu_ctrimming2018}.  We will etch the edge of the PPC furthest away from the inductor (etching through a-Si:H, Nb, and Al) to avoid collateral damage to the inductor by etch chemicals permeating between layers transversely to the inductor, which was a problem in the past for some of the buffered oxide etches used to clean various layers before or after the second a-Si:H deposition (\S\ref{sec:fab}).

\subsection{Focal Plane Architecture -- Fabrication} \hfill 
\label{sec:fab}

The devices are fabricated on double-side polished, high-resistivity, float-zone silicon wafers, 100~mm in diameter and 375~\mumtxt thick.  We do fine-scale photolithography using a Canon EX3 stepper mask aligner and some coarser features (e.g., a-Si:H trench etch) using a Heidelberg MLA 150 maskless aligner.  Steps:
\begin{enumerate}

\item \textbf{Nb ground plane:} After a buffered oxide etch (BOE) dip of the wafer to remove native oxide, we ion mill and then deposit a 190~nm thick ground plane via RF magnetron sputtering using a 6-inch target at 900~W RF power.  It is patterned with the antenna slots and the BPF windows using a fluorine etch in an ICP RIE machine. 

\item \textbf{First a-Si:H layer (microstripline, KID capacitor):} Another BOE dip is done (to remove oxides on the Nb) and then a 1070~nm a-Si:H layer is deposited by CVD. The machines and recipes available are described in detail elsewhere\cite{defrance_asihloss}.  The lowest loss recipes require deposition in a machine at JPL with 350\degtxtnosp~C substrate temperature, which can cause flaking if the deposition chamber is not properly conditioned.  For the first device studied here, we deposited at 350\degtxtnosp~C but with a machine at the Caltech Kavli Nanoscience Institute that has previously yielded $\delta \approx 1.2\,\times\,10^{-5}$\cite{defrance_asihloss}, and for the second device, we used the JPL machine but at 150\degtxtnosp~C, which previous work had shown yielded $\delta \approx 3\,\times\,10^{-5}$.  We then etch away 270~nm of a-Si:H in windows where the resonators, coupling capacitors, and microwave feedline will reside.

\item \textbf{KID inductor and capacitor top plates:} After a BOE dip to remove \siox that forms on the first a-Si:H layer, we ion mill the a-Si:H and then sputter-deposit the 100~nm Al layer using a 6-inch target and 750~W RF power.  If doing AlMn, we instead use a dedicated machine that co-sputters AlMn (2500 ppm Mn doping) and pure Al, aiming for 750~ppm.  We use chlorine etch in an ICP RIE machine to pattern the Al (AlMn) to obtain the KID inductor and capacitor, coupling capacitor outer (bottom) plates, and readout feedline.  We then ion mill the Al (AlMn) and deposit/pattern, using liftoff, 50~nm Nb and 30~nm Al over the patterned Al (AlMn) \textit{except over the KID inductor} (same 6-inch target sputter tool and powers).  The Nb is intended to prevent outdiffusion of quasiparticles from the Al (AlMn) inductor and to increase the threshold frequency for photon absorption in the capacitor (already mitigated by its high reflectivity and its placement within 1.1~\mumtxt of the Nb ground plane).  It also ensures the feedline has transmission at 4~K, which is useful for device screening.  The Al top layer protects the thin Nb layer from the later ``Nb wiring layer'' etch (step 5 below). 

\item \textbf{Second a-Si:H layer (trans-mm coupler):} A second layer of a-Si:H is then deposited, 270~nm thick, now always at 150~C to prevent damage to the underlying Al film (formation of AlSi at $\approx$166~C) and always using the JPL ICP PECVD (both for convenience and because the KNI PECVD does not make good a-Si:H at 150~C).  No BOE dip is done before this deposition to avoid damage to the now-patterned KID Al (AlMn); any oxide at the interface will only be present in the trans-mm coupler, not in the KID or the trans-mm microstripline.  This a-Si:H layer is etched away almost everywhere except in the window where the prior a-Si:H layer was etched.  The result is an approximately constant thickness of a-Si:H over the entire wafer, but with the KID inductor and capacitor films (and KID coupling capacitor outer plates and readout feedline) residing between the 800~nm and 270~nm a-Si:H layers.  The one exception is that the 2nd a-Si:H layer is etched away over the microwave feedline bondpads so they are accessible.

\item \textbf{Nb wiring layer:} We complete the KID, antenna, and bandpass filters, and also the KID coupling capacitor by ion milling the a-Si:H and depositing a last Nb layer, 160~nm thick, which is now patterned using fluorine-chlorine ICP RIE.  This etch is not highly selective against Si, hence the Al protect layer for the microwave feedline in step 3.

Ideally, because it is more effective at removing \sioxnosp, we would do a BOE dip prior to the Nb layer, but there have been cases in which BOE at this step reduced KID yield, presumably because BOE can permeate through microfissures or pinholes in the a-Si:H to the Al (AlMn) KID layer.  To achieve the best trans-mm loss, we may try to resolve this problem so we may implement a BOE dip at this step.  

\item \textbf{Borders for good electrical and thermal connection:} To ensure good RF and thermal coupling to the device box, we etch a border through all the layers (fluorine-SF$_6$ ICP RIE) to expose the Nb ground plane.  On three sides of the device, we additionally etch through part of the exposed Nb (same fluorine-chlorine ICP RIE as Nb wiring layer) and deposit 10~nm of Ti (sticking layer) and 350~nm of Au by electron-beam evaporation through a liftoff mask that leaves openings over both Si and Nb.  When the device is mounted, Au wirebonds connect the Au pads to the copper box to provide good thermal connection to the silicon substrate (see discussion of substrate heating vs.\ direct absorption in the KIDs in \S\ref{sec:nblc_direct}) as well as electrical connection to Nb on three sides, and Al wirebonds connect the Nb ground plane on the fourth side to the box for RF grounding. 

\end{enumerate}

\subsection{Focal Plane Architecture -- Experimental Validation} \hfill 
\label{sec:exp}

We undertook extensive dark and optical tests of two prototype devices.  Both devices incorporate a two-scale analogue of the three-scale antenna planned for NEW-MUSIC.  For B3--B5, the two-scale prototype uses a fundamental 3.328~mm wide, $32\times 32$ slot array antenna, analogous to the $16\times 16$ slot array antenna presented in \S\ref{sec:design:antenna} and Figure~\ref{fig:antenna}, and it sums four such antennas for B2, so it tests the critical feature: summing of fundamental elements with gaps between.  Both devices also incorporate a limited set of bands as a first step in the complexity of the filter banks.\footnote{In fact, each fundamental element has its own four-band BPF at its output, following by summing of the B2 outputs.  The long-term scheme would not use BPFs for the bands to be summed but only LPFs, with a single BPF to follow after summing.  This latter approach prevents BPF variation among the summed elements from causing beam asymmetries.}  Each antenna feeds four KIDs in each of B3--B5 and one KID in B2.  We additionally have four dark KIDs (no connection to an antenna).  We thus expect to see 56 KID resonances on each die.

The second device incorporates some pixels where an output feeds not one but rather two KIDs for loss and impedance/wave-speed testing.  For the former, a 50-50 splitter is followed immediately by a KID on one leg and by a long length of microstripline terminated in a KID on the other leg.  The relative optical efficiency of the two KIDs measures the transmittance, and thus the loss, of the length of microstripline.  For the latter, the second arm of the splitter is instead followed by a Fabry-Perot cavity consisting of a widened (and thus impedance-mismatched) length of microstripline.  The resulting standing wave pattern's frequency measures the microstripline wave-speed and its amplitude contrast measures the ratio of the widened to standard microstripline impedance.  The total number of KIDs is unchanged, but 14 more are ``dark'' due to the microstrip routing required for the test structures.  Measurements of these test structures are not yet available, so we report only on the non-test-structure KIDs.

\begin{figure}[t!]
\begin{center}
\includegraphics*[width=2.5in,angle=90,viewport=0 90 430 690]{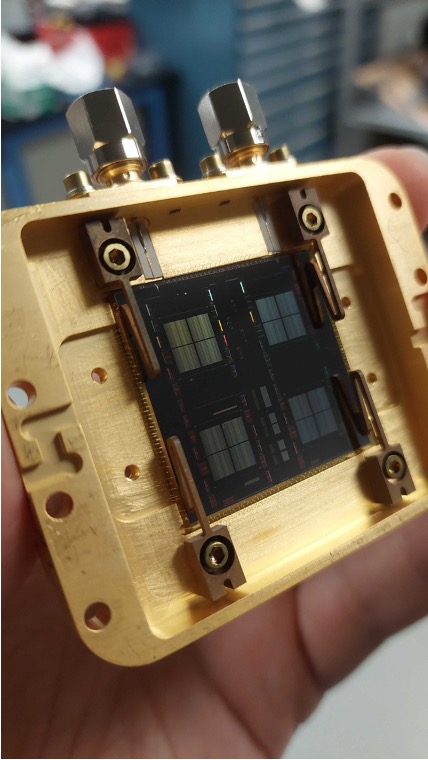}
\end{center}
\caption{\textbf{Fabricated device.} The 32~mm~$\times$~32~mm die is mounted in a gold-plated copper box.  (No Ni sticking layer is used because it can be magnetic.)  Below the wafer (not visible) is a window in the box and a metamaterial silicon antireflection wafer.  Phosphor bronze spring clips hold the device in place.  Numerous wirebonds provide thermal and RF connection to the box (see text).  On the chip, there are four copies of the hierarchical antenna.  Along the midline of the wafer, one can see the Nb LC resonators (the large rectangles) used to measure a-Si:H RF loss as well as four dark KIDs.  Before installation in the cryostat, a Nb backshort is placed above the device, located and fixed by the four screw holes in the wells.  The relative height of the wells and the shoulders for the die are precision-machined so the backshort surface is at the desired 150~\mumtxt distance from the device, with an uncertainty of 10--20~\mumtxtnosp.}
\label{fig:device_picture}
\end{figure}

For optical testing, the devices incorporated a 2-layer metamaterial-structured silicon antireflection wafer (1:1.6 bandwidth, 190--310~GHz)~\cite{cit_2layer_2018} and a niobium backshort (\S\ref{sec:design:antenna}).  They were tested in a pulse-tube-cooled 4~K dewar with a 240~mK Chase $^3$He/$^3$He/$^4$He sorption refrigerator\cite{bhatia_he10fridge_2002}.  The dewar has a UHMWPE window (with single-layer Porex\footnote{\url{https://www.porex.com/products/porous-sheets/}} PM23DR 0.25~mm thick AR coating) with $\approx$30~cm clear aperture, permitting very wide angle beam measurements (up to 40\degtxt off axis).  For blackbody radiation filtering, the dewar has five 3~mm thick Zotefoam\footnote{\url{https://www.zotefoams.com}} HD-30 sheets behind the vacuum window, two PTFE filters (25 and 10~mm thick, with single-layer Porex PM23DR 0.25~mm thick AR coatings) at 50~K, and, at 4~K, a nylon filter (10~mm thick, with single-layer Porex PMV30 0.25~mm thick AR coating) and a 420~GHz low-pass cutoff metal mesh filter~\cite{ade_spie}.  To ensure good heat sinking of the devices, we bond the Au border on three sides of the device to the copper device box every 0.5--1~mm using Au wirebonds.  The remaining side has Al wirebonds to the Nb ground plane and microwave feedline of similar density. Figure~\ref{fig:device_picture} shows the backside of a device mounted for testing.

A magnetic shield, residing at 4~K and consisting of two layers of Amuneal A4K material, enclosed the devices to limit the impact of Earth's magnetic field.  The shield incorporates an aperture at the top to permit optical access for one device.  The device under optical test incorporates an additional single-layer A4K shield, also with an aperture, to improve its shielding as it sits near the aperture in the larger shield.  A combination of stainless steel and NbTi semi-rigid coaxial cables carried the readout signal to the devices, with 30~dB and 10~dB in-line attenuators at 4~K and 0.35~K, respectively, to block 300~K thermal noise.  Similar NbTi coax carried the signal exiting each device to a cryogenic SiGe low-noise amplifier (LNA) at 4~K (ASU 10~MHz--2~GHz or Cosmic Microwave Technologies CITLF2), with a noise temperature of approximately 5~K, followed by stainless steel coax back to 300~K.  Additional LNAs at 300~K ensured the cryogenic LNA dominated the system noise.  We monitored the device temperature using a Stanford Research System (SRS) SIM921 reading a Lakeshore Germanium Resistance Thermometer (GRT) located next to the devices.  The temperature was controlled using a SRS SIM960 analog PID controller supplying a current to a 10~k$\Omega$ heater on the mechanical stage holding the devices. 

Measurements relying on $S_{21}(f)$ scans used a Copper Mountain Technologies SC5065 Vector Network Analyzer.  We used the Python module SCRAPS~\cite{Carter:2017} to fit the $S_{21}(f)$ data to standard forms (e.g., \cite{jonas_arcmp}) to extract the resonance frequency $\fr$ and quality factors $\Qr$, $\Qi$, and $\Qc$.  Measurements of small signal response such as beam maps and Fourier Transform Spectroscopy used an Ettus X310 USRP with a UBX~160 daughter card.  Measurements of noise relied on a standard homodyne mixing setup.  The resonator drive signal and local oscillator for mixing was provided by an Anritsu MG3694A synthesizer, and the mixer was followed by Stanford Research Systems SR560 voltage preamplifiers and a National Instruments NI-9775 ADC.

\subsubsection{KID Parameters and Yield} 
\label{sec:kid_params}

\begin{figure}[t!]
\begin{center}
\begin{tabular}{cc}
\includegraphics*[height=1.55in,viewport=0 0 1320 755]{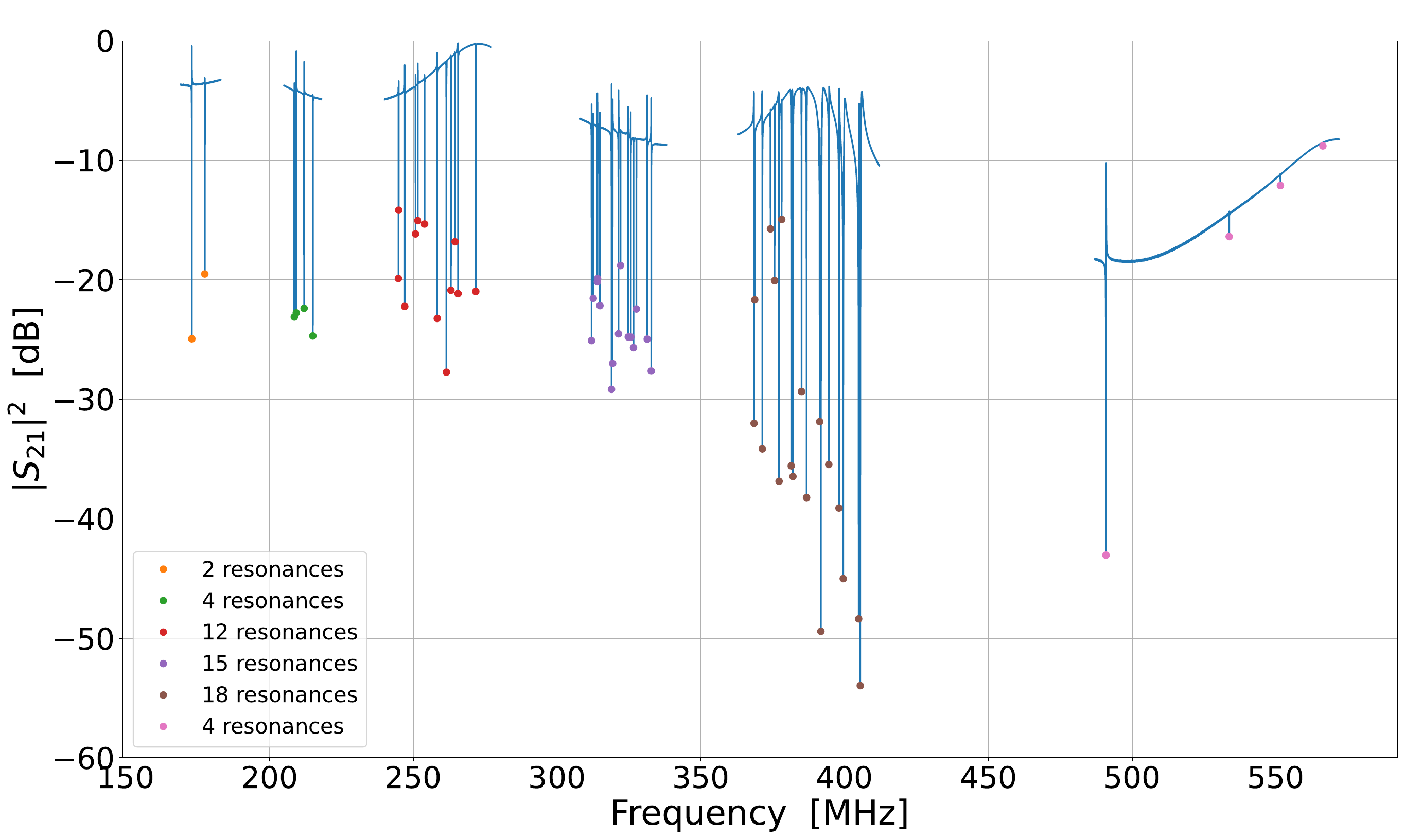}
& \includegraphics*[height=1.55in,viewport=0 0 50 755]{figures/Be210504p2bl_VNA_AnalysisLog353.pdf}
\includegraphics*[height=1.55in,viewport=32 5 330 180]{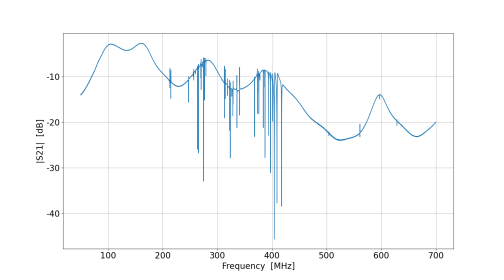} \\
\includegraphics[height=1.55in]{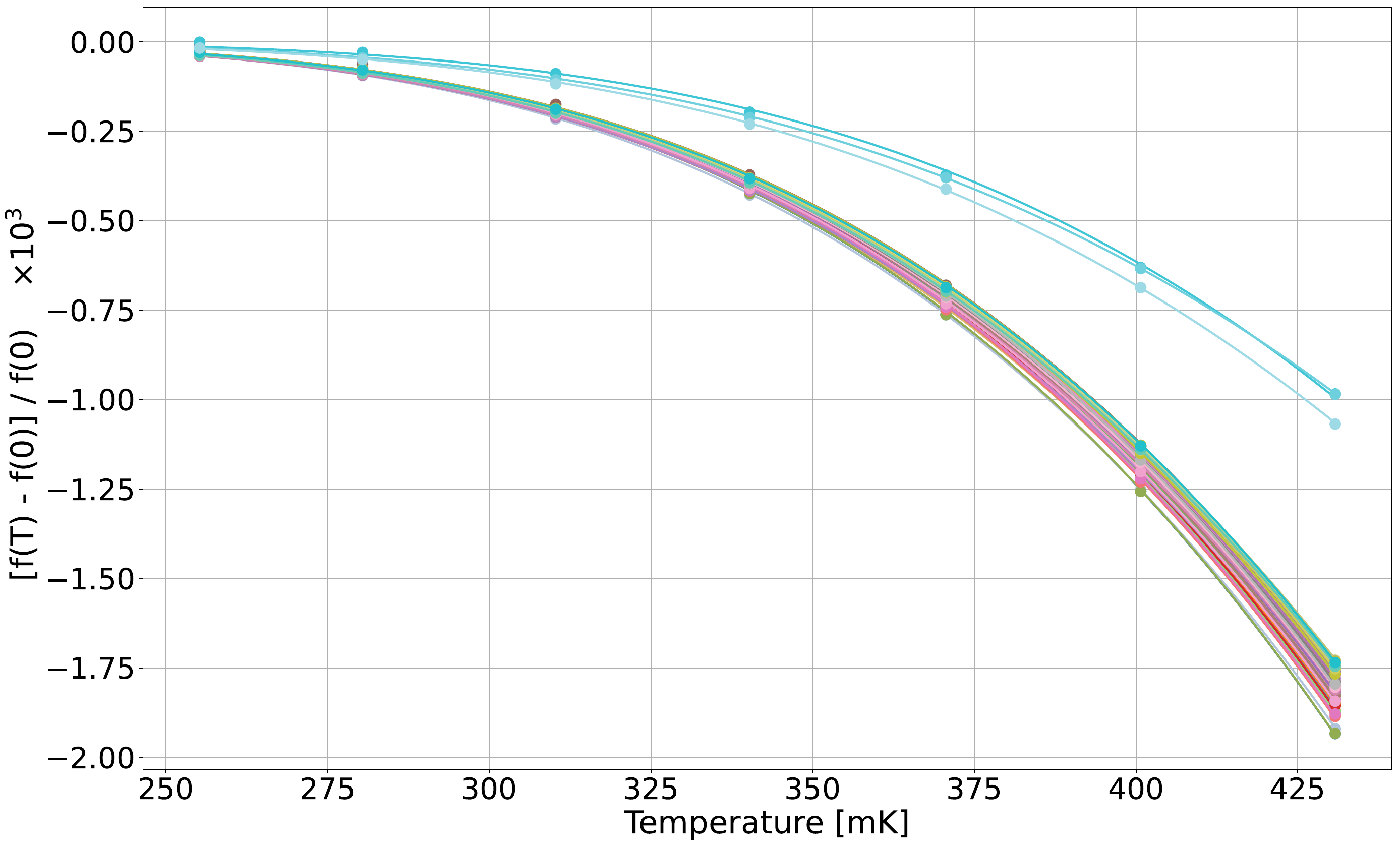}
& \includegraphics[height=1.55in]{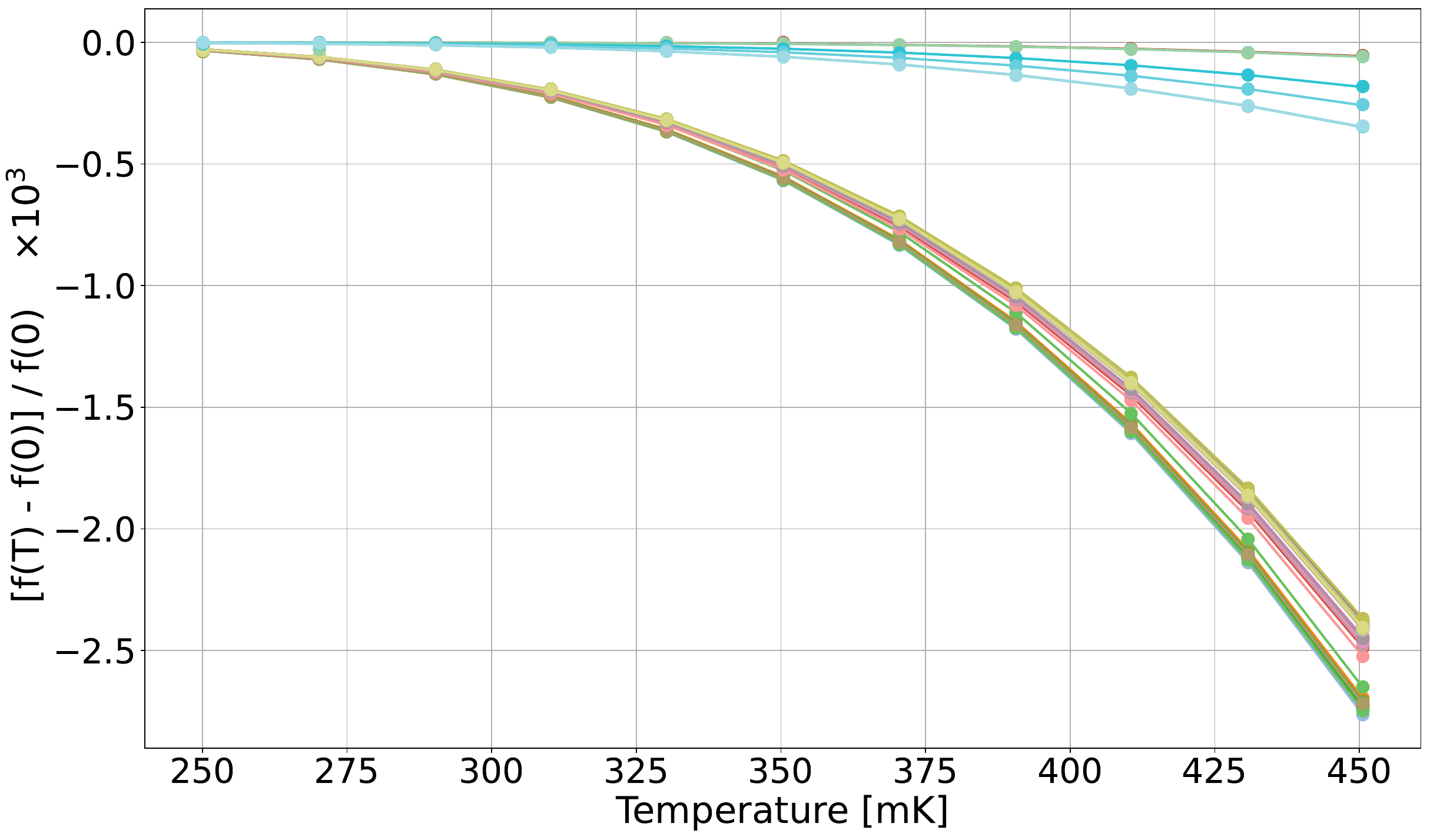} \\
\includegraphics[height=1.55in]{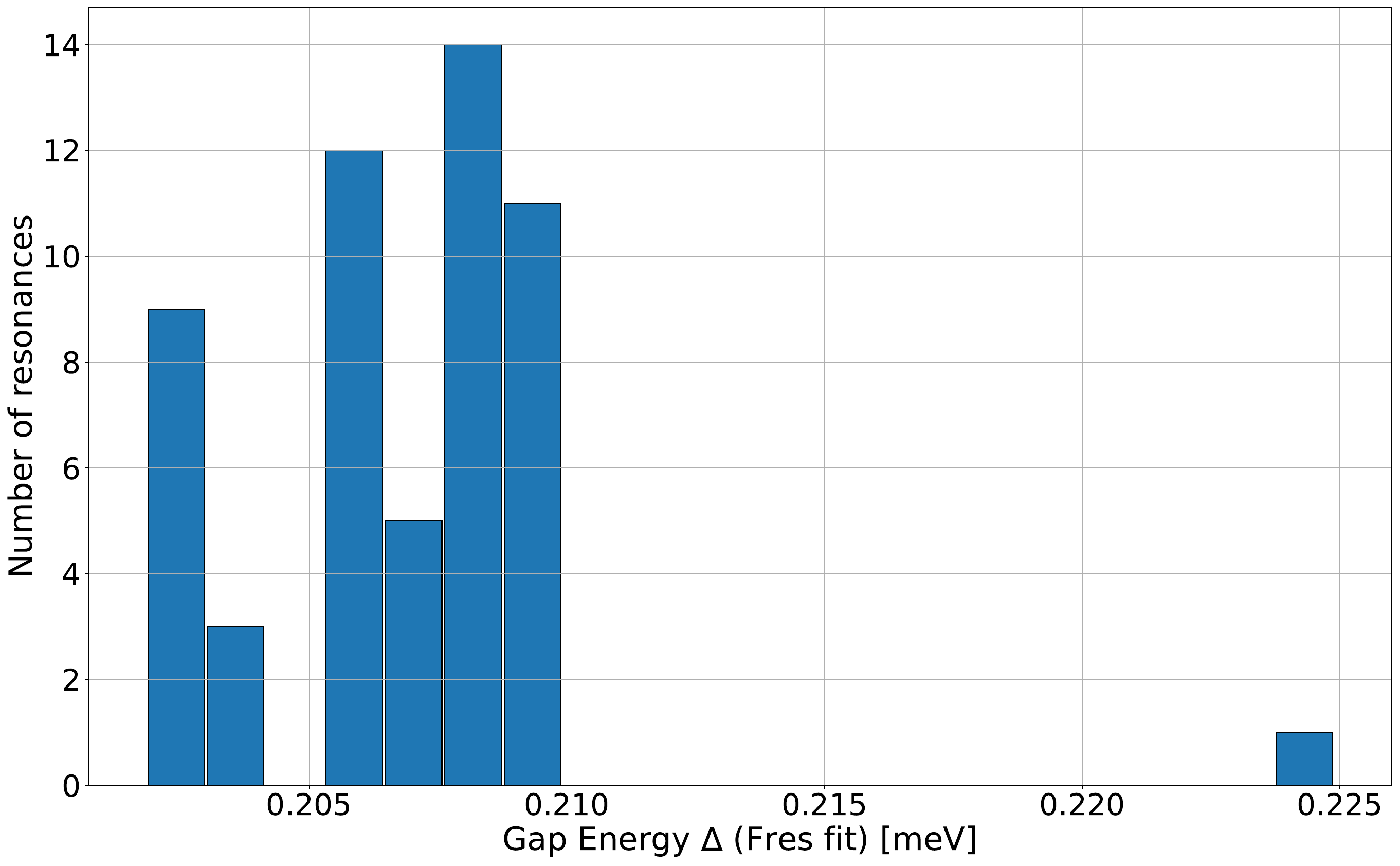}
& \includegraphics[height=1.55in]{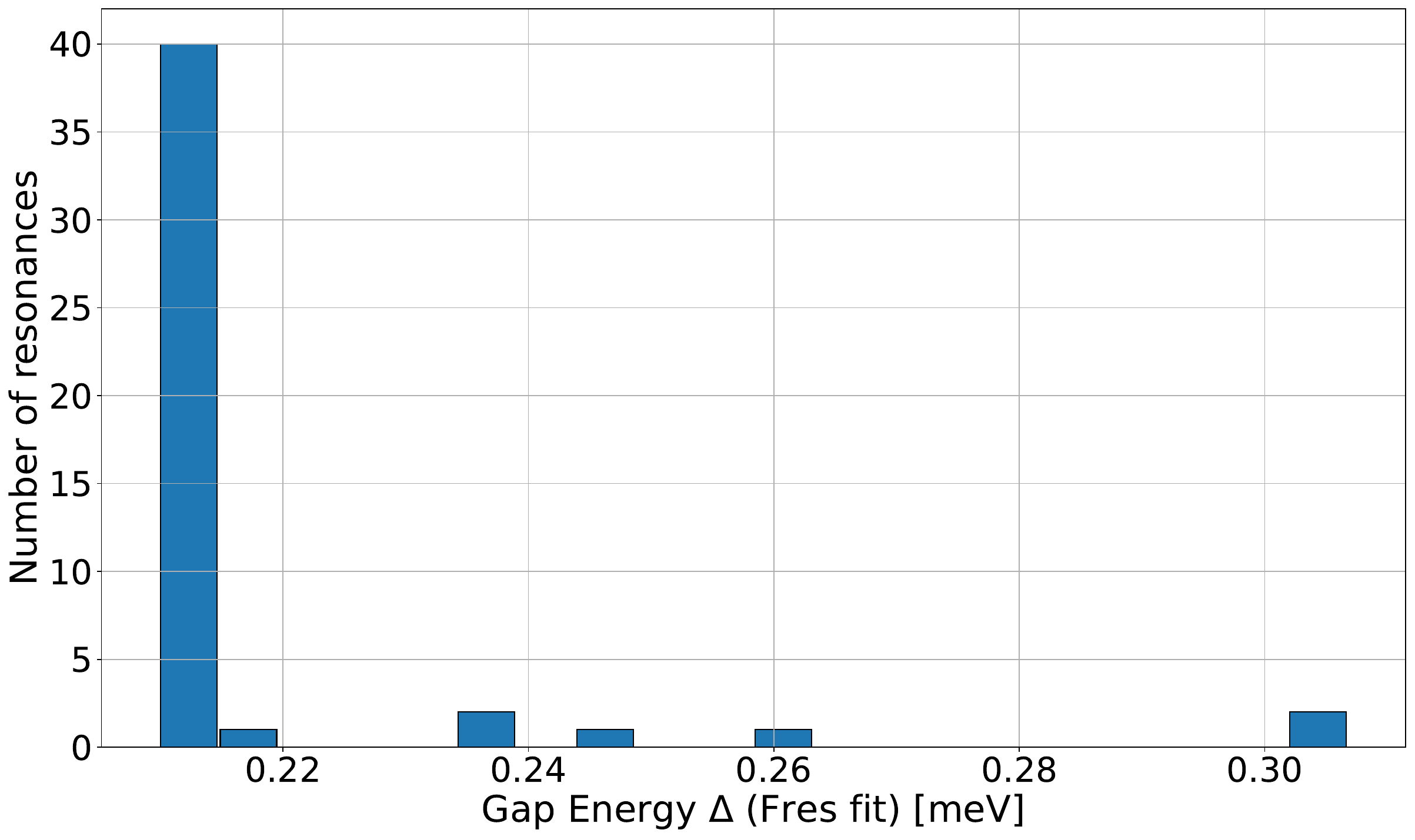} \\
\includegraphics[height=1.55in]{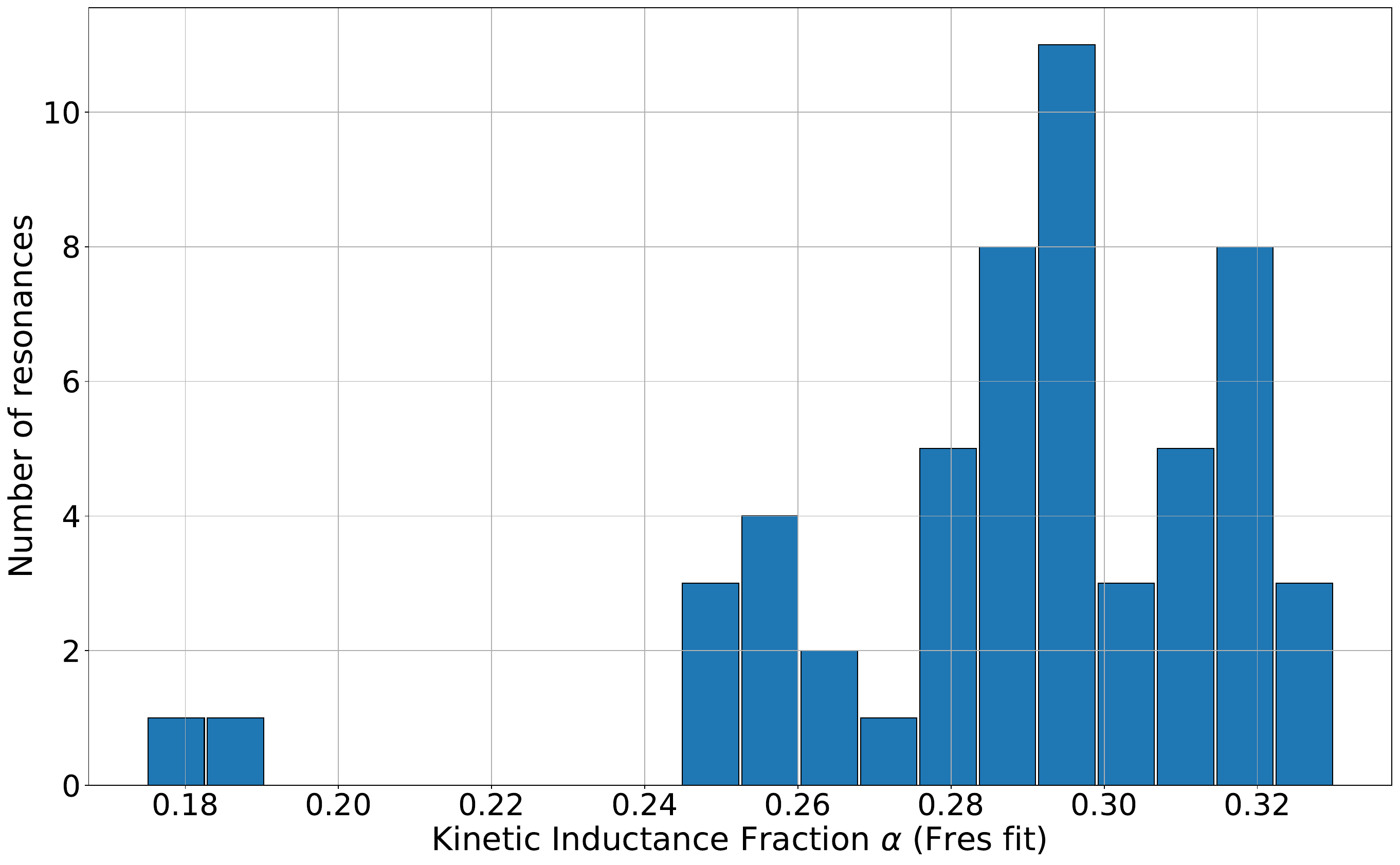}
& \includegraphics[height=1.55in]{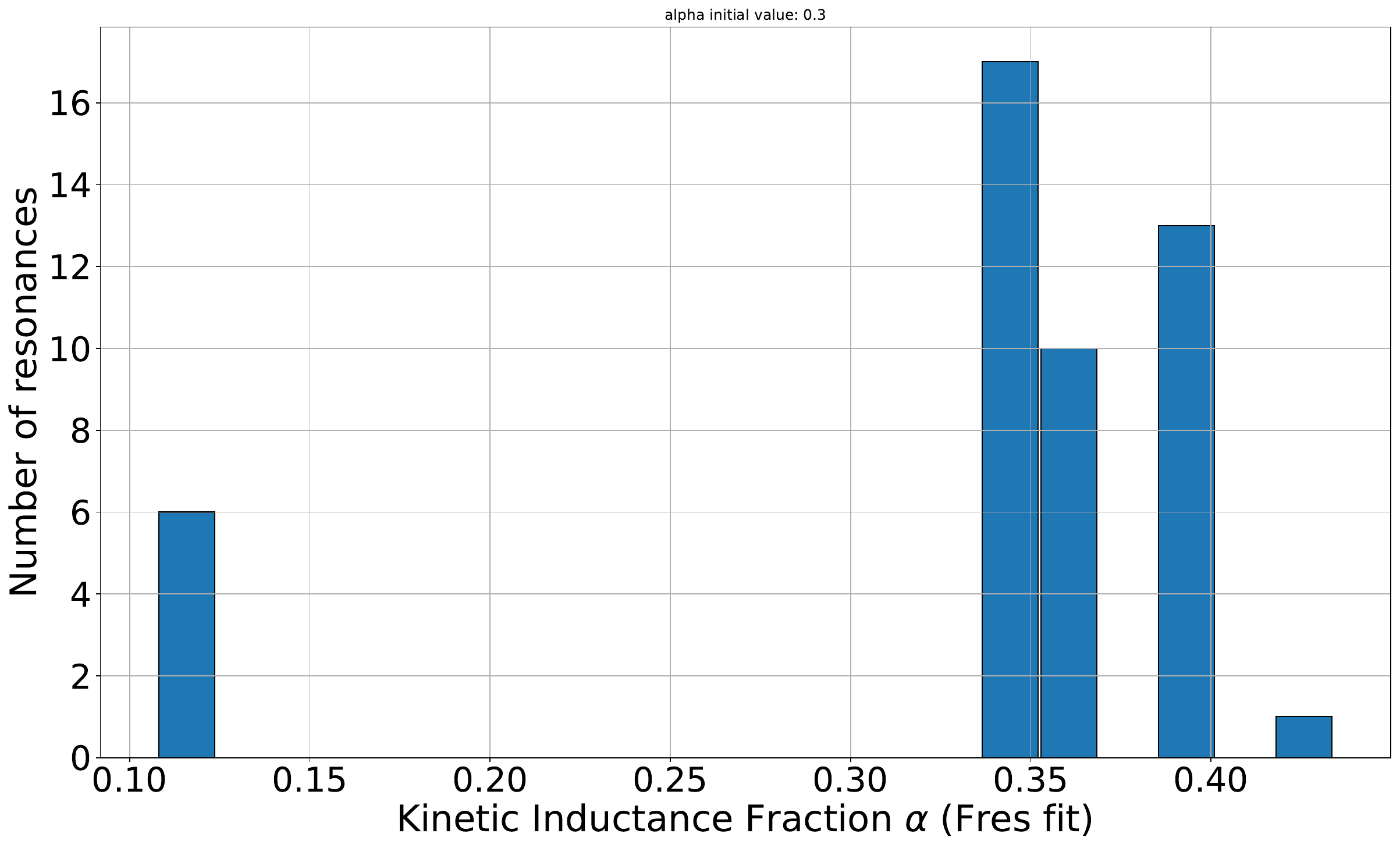} \\
\includegraphics[height=1.55in]{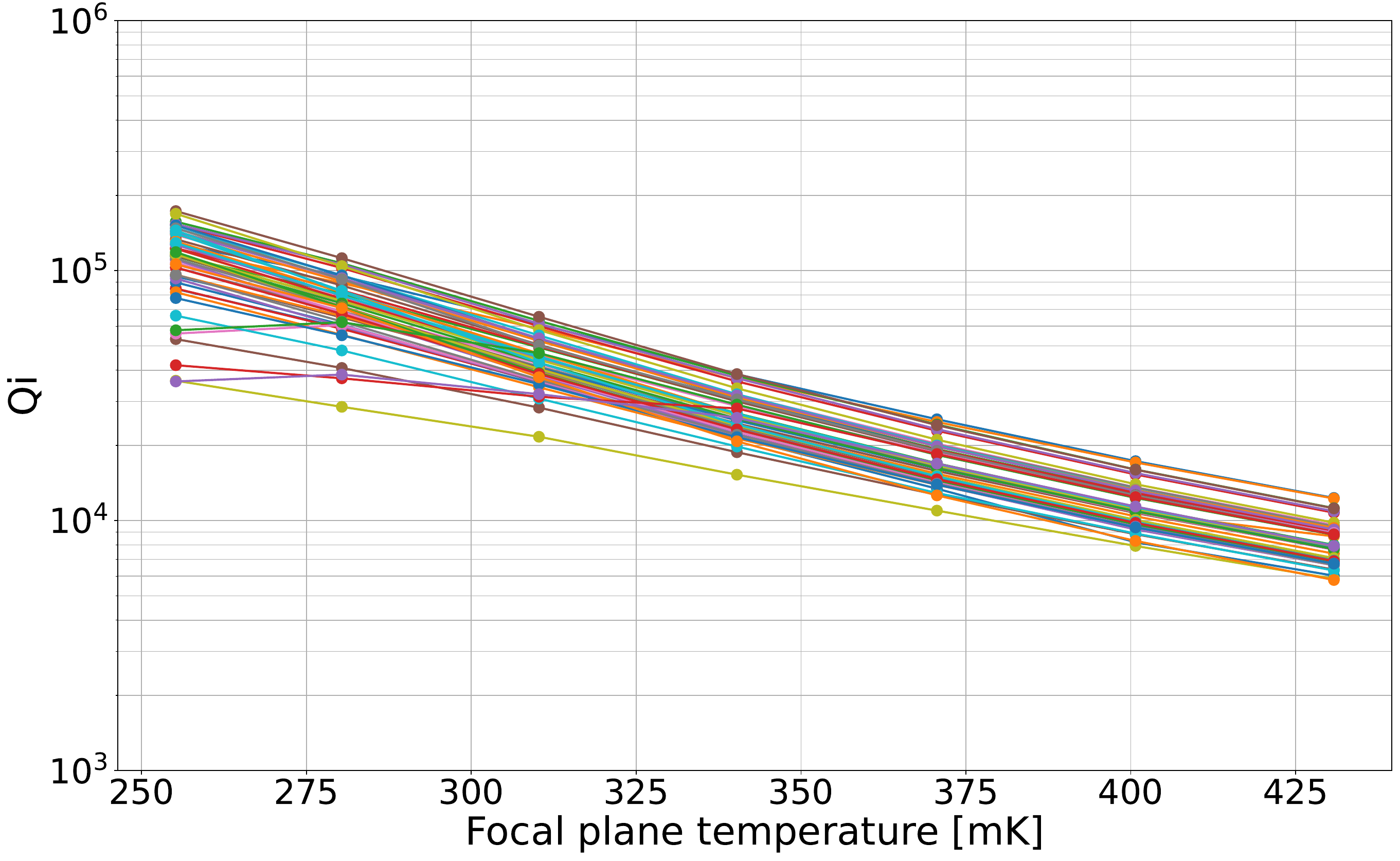}
& \includegraphics[height=1.55in]{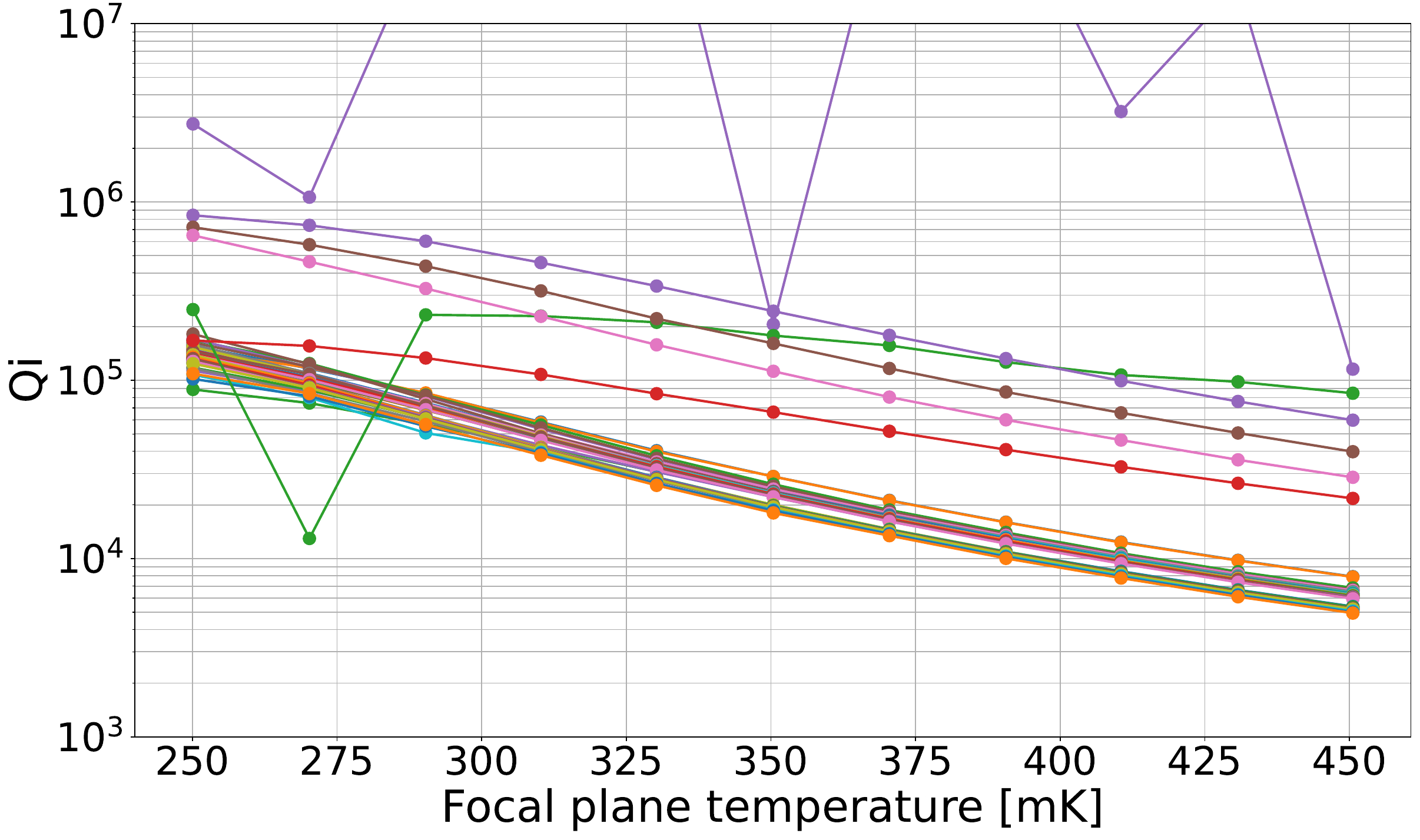} \\
\end{tabular}
\end{center}
\caption{\textbf{Measured KID parameters for two devices.}  See text for discussion.}
\label{fig:kid_params}
\end{figure}

We use a vector network analyzer (VNA) to measure $\Sparam(f)$ over the frequency range containing the resonances at a range of $\Tb$ values under dark conditions.  We fit these resonance scans using standard techniques (e.g., \cite{gao_thesis}).  Figure~\ref{fig:kid_params} shows frequency scans, $\delta\fr/\fr$ and $\Qi$ vs.\ $\Tb$, and kinetic inductance fraction $\alpha$ and gap parameter $\Delta$ inferred from fits of the $\delta\fr/\fr$ vs.\ $\Tb$ data to  Mattis-Bardeen theory~\cite{mattisbardeen}.  
The different trans-mm spectral bands also have different $\fr$ design ranges to enable disambiguation of bands without FTS data.  Yield and uniformity is fairly high already.  The two devices for which data are shown have yields of 55/56 and 47/56 resonators.  Two additional devices show yields of $>$50/56 and 40/56 resonators.  A detailed inspection has not yet been done, but, given the uniformity of the KIDs that do appear, we suspect catastrophic fabrication defects caused the failed resonators.  Better process control will likely enable regular achievement of $>$95\% yield.

From the Mattis-Bardeen fits, we infer $\Delta \approx 0.20$--0.22~meV, which would imply $2\,\Delta_{Al}/h \approx 100$~GHz.  We find $\alpha \approx 0.25$--0.40, fairly consistent with the design value $\alpha = 0.26$, which assumed 100~nm Al with a normal state sheet resistance of 0.069~\ohmtxtnosp/$\Box$\cite{shu_multiscale_al_ltd19}.   We speculate that the systematic differences in $\Delta$ and $\alpha$ between wafers are due to slight differences in Al film thickness: the penetration depth in particular, which determines $\alpha$, is a strong function of thickness for $d \lesssim 100$~nm\cite{gao_thesis}.  Fortunately, these variations in $\alpha$ only affect $\fr$ as $1/\sqrt{1 + \alpha/(1-\alpha)}$, so the $\fr$ bands are very similar, 209--406~GHz and 213--417~MHz.  (In the first device, two B3 resonators moved below 200~GHz, but the remainder did not move, so these shifts are likely due to a defect specific to those two resonators.)  The design frequency range was 180--336~MHz, so the shift relative to design varies from 16\% to 24\% across the octave band.  Given the good match of $\alpha$ to expectations, this shift seems likely to be due to unmodeled parasitic reactances.

Each device has a handful (3 and 6) of resonators that have anomalous $\delta\fr/\fr$ vs.\ $\Tb$ curves.  These all happen to be dark and also seem to have moved from their design $\fr$ values (374--388 MHz for nominal darks, mixed with the other bands for the darks due to the test structures) by a larger factor than the optically sensitive KIDs, up to $\sim$500--600~MHz.  They all seem to have low $\alpha$, too, which causes the anomalous $\delta\fr/\fr$ vs.\ $\Tb$.  Lower $\alpha$ can explain part of the upward shift in $\fr$, but the films are the same thickness as those of the other KIDs, so the kinetic inductance $L_k$ should not have changed.  One potential explanation for all of this behavior is that the removal of the microstripline in the trans-mm coupler (unnecessary since these resonators are intended to be dark) increases the geometrical inductance $L_g$ (decreases $\alpha$) while also reducing $C$ by a larger factor, making higher $\fr \propto 1/\sqrt{LC}$ possible in spite of the larger $L_g$.  We will re-implement the trans-mm coupler for these dark KIDs (with no incoming microstripline) to eliminate this systematic difference.

For the majority of resonators, we find $\Qi \gtrsim 10^5$ and still rising at the lowest temperatures for which we have data, $\Tb \approx 250$~mK.  There is good uniformity in the behavior.  Thus, in spite of the unconventional device structure, with the Al inductor material on a thick a-Si:H layer and with many parasitic capacitive and inductive couplings, the loss is still dominated by quasiparticles at $\Tb/T_c  \approx 0.15$--0.2.  

One challenge to be addressed is large variation of $\Qc$, over an order of magnitude (not shown).  We suspect impedance mismatches arising from the use of the 32.5~\ohmtxt microstripline feedline, causing reflections at the interfaces to the 50~\ohmtxt readout wiring and standing waves on the feedline.  Our initial choice of feedline width was conservative due to the catastrophic impact of feedline failure, but we have not had a single device with a failed feedline, we will try narrowing the feedline in future devices to increase the impedance.

\subsubsection{Hierarchical Antenna Beams} 
\label{sec:exp:beams}

\begin{figure}[t!]
\begin{center}
\begin{tabular}{cc}
\hspace{0.12in}\includegraphics[width=3.05in]{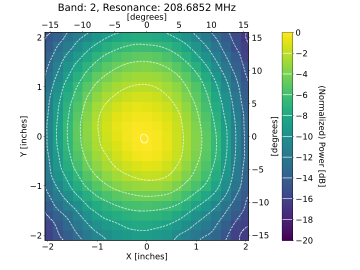} & 
\includegraphics[width=2.7in]{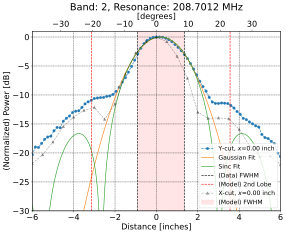}
\\
\includegraphics[width=2.65in]{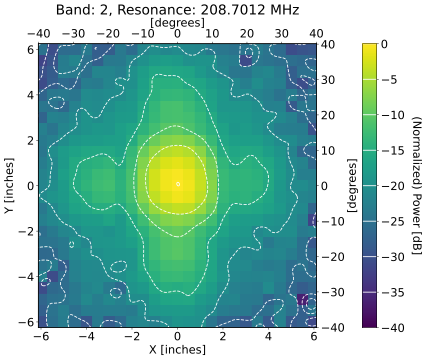} & 
\hspace{0.18in}\includegraphics[width=3in]{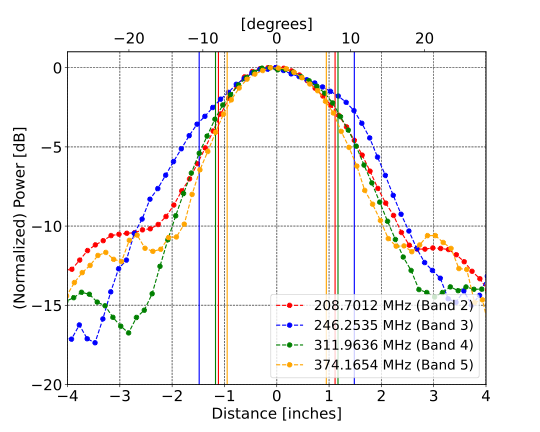}
\\
\end{tabular}
\end{center}
\vspace{-9pt}
\caption{\textbf{Hierarchical phased-array antenna beam measurements.}  (Left)~B2 beam maps, main lobe (top), and wider map including sidelobes (bottom). (Top right)~B2 beam cross sections and comparisons to expected FWHM and fits to gaussian and sinc functions.  The slight difference in beam width in the two directions is being confirmed before corrective action is taken. (Right)~Cross sections of beams for all four bands B2--B5.  The FWHMs are within 10\% of expectations~\cite{martin_ltd20_2024}.  The near equality of the B2 and B4 FWHMs indicates the B2 summing works.  The somewhat elevated sidelobes for B2 and B5 are being investigated.}
\label{fig:beams}
\end{figure}

Figure~\ref{fig:beams} shows experimental validation of beams.  The measurements were done using a chopped hot blackbody source\footnote{A commercial ceramic heater source (e.g., \url{https://www.amazon.com/Infrared-Ceramic-Heater-Forming-Element/dp/B0C394KWJB}), coated in Bock black\cite{bock_thesis}.} behind a 18.35~mm aperture at a distance of 189~mm.  The main expected features are visible.  Most importantly, the beam FWHM scales with frequency as expected, accounting for summing for B2, rendering the B2 FWHM similar to the B4 FWHM.  The B3 and B4 beams show sidelobes similar in shape to the sinc expectation, though the nulls are insufficiently deep and the level is 2--5~dB too high.   The B2 and B5 beams show shoulder-like features, as if the sinc function null were filled in.  The B2 beam may show some asymmetry between the $E$ and $H$ planes, which we are confirming before taking corrective action.   While we are still trying to understand some details, it is clear that hierarchical summing works well in spite of the gaps between the fundamental elements.  Moreover, these sidelobes will be terminated on a cold Lyot stop in the NEW-MUSIC optical configuration (\S\ref{sec:optics}).

\subsubsection{Spectral Bandpasses} 
\label{sec:bandpasses}

\begin{figure}[t!]
\begin{center}
\includegraphics[width=6.5in]{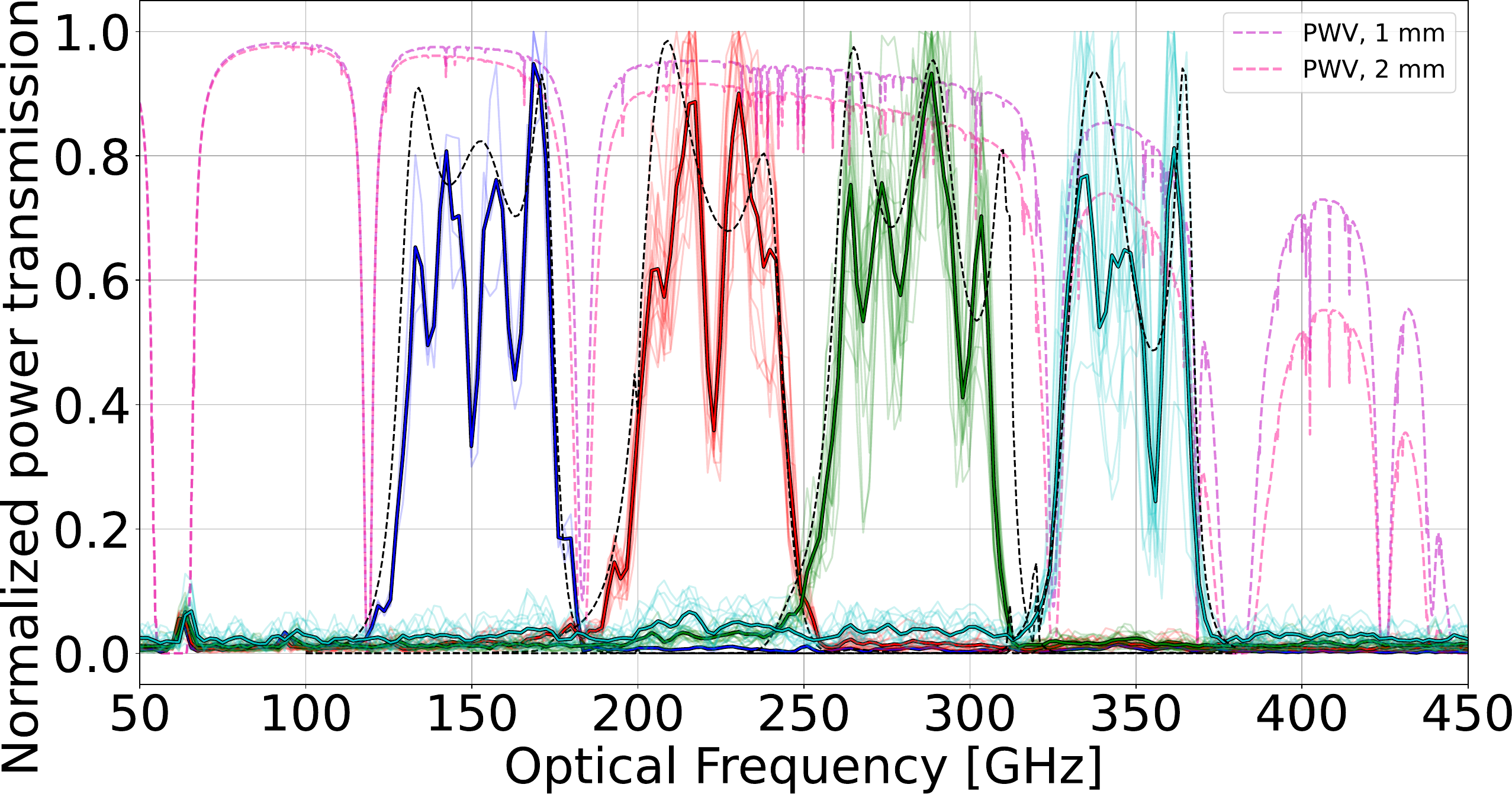}
\end{center}
\caption{\textbf{Spectral bandpass measurements.} Bandpasses for B2--B5 measured using Fourier Transform Spectroscopy.  Dashed black: design.  Thin solid: individual KIDs.  Thick solid: average by band.  Dashed magenta: atmospheric transmission for LCT site with 1 and 2~mm~precipitable water vapor (PWV; 50th and 75th percentile conditions), showing the as-yet unpopulated B1 and B6 windows.}
\label{fig:bandpasses}
\end{figure}

Figure~\ref{fig:bandpasses} shows bandpass measurements using a Martin-Puplett Fourier Transform Spectrometer fed by a chopped $\approx$1100\degtxtnosp~C cavity blackbody\footnote{CI-Systems SR-200N}.  The measurements are overlaid on expectations from Sonnet for the BPF banks alone and expected atmospheric transmission for 1~mm and 2~mm PWV at the LCT site.  There is good qualitative agreement of the measurements with expectations in terms of band centers and edges.  The measured spectra show higher contrast ripples, and the B5 upper edge approaches the upper edge of its atmospheric window too closely (a design error).  As noted above, the optical train includes a 190--310~GHz silicon antireflection wafer and UHWMPE, PTFE, and nylon plastic windows/filters coated with single-layer AR coatings, and an uncoated metal-mesh filter, so some reflections are to be expected.  More detailed modeling is in process.  Nevertheless, the basic elements of the design appear work, with refinement of the design and the measurement setup necessary to reduce non-idealities.

\subsubsection{Optical Efficiency} 
\label{sec:opt_eff}

\begin{figure}[t!]
\begin{center}
\begin{tabular}{cc}
\includegraphics*[height=3.5in,viewport=10 257 490 807]
{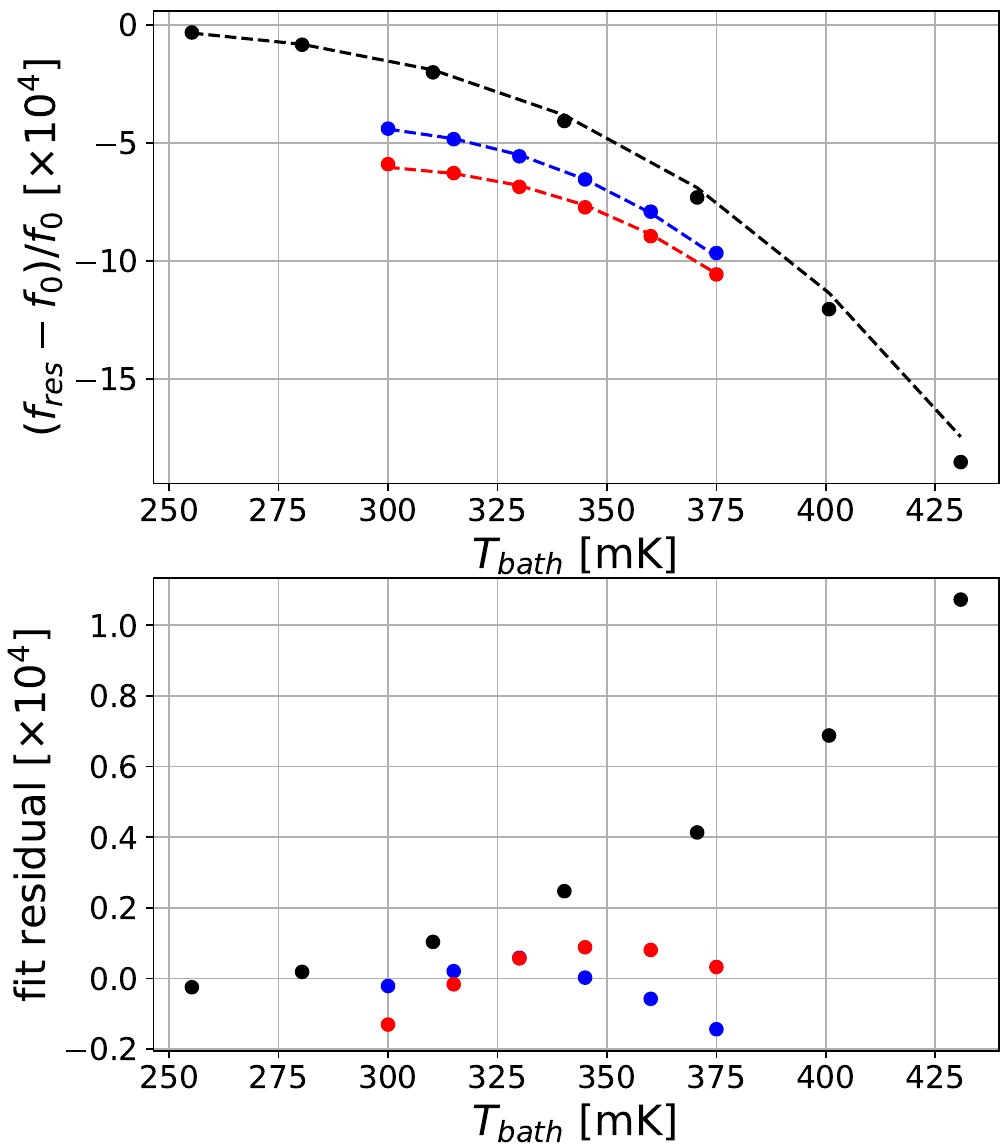} & 
\includegraphics*[height=3.5in,viewport=0 278 500 828]
{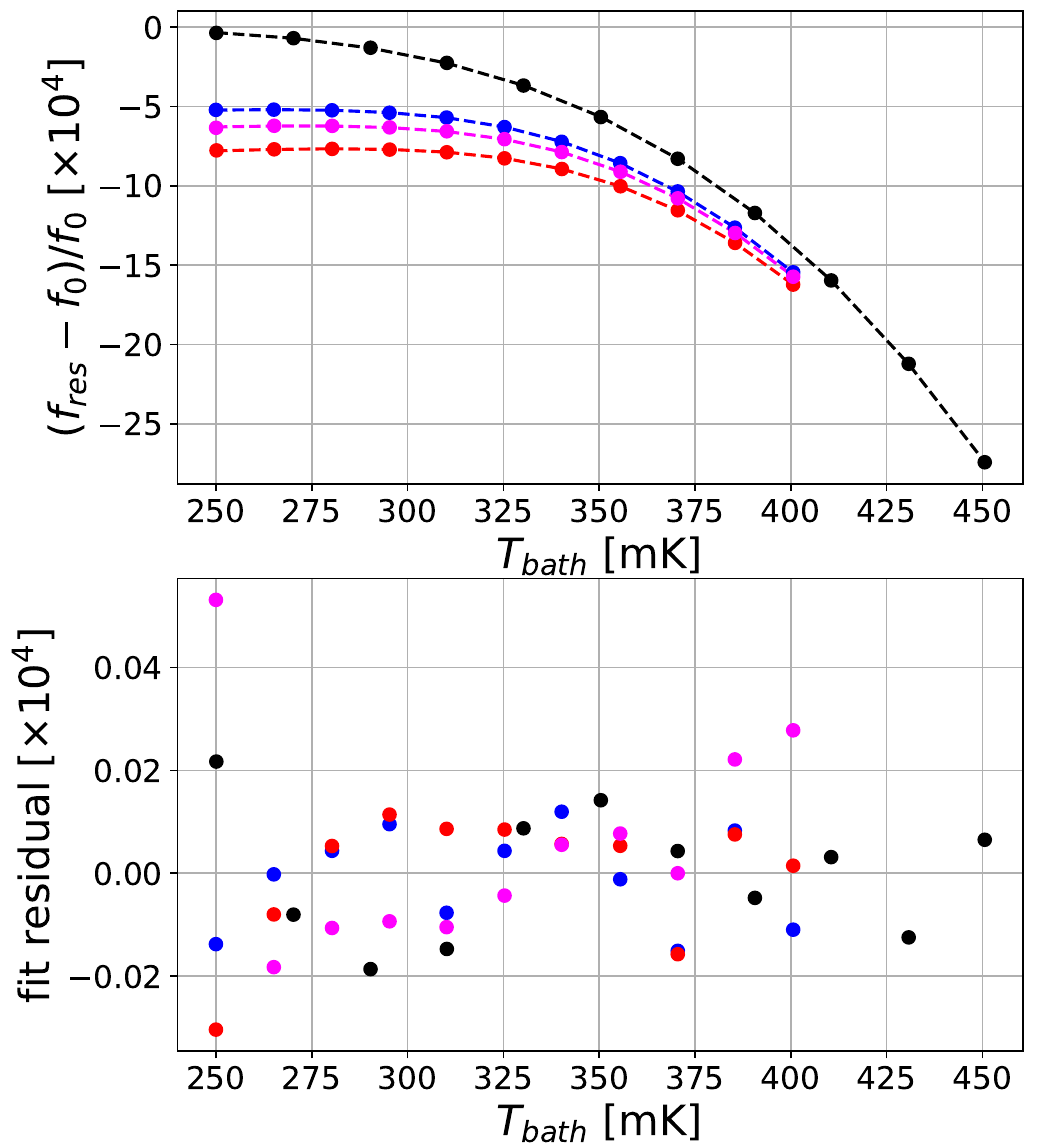}\\
\hspace{0.2in}\includegraphics*[width=3in,viewport=10 10 1490 820]{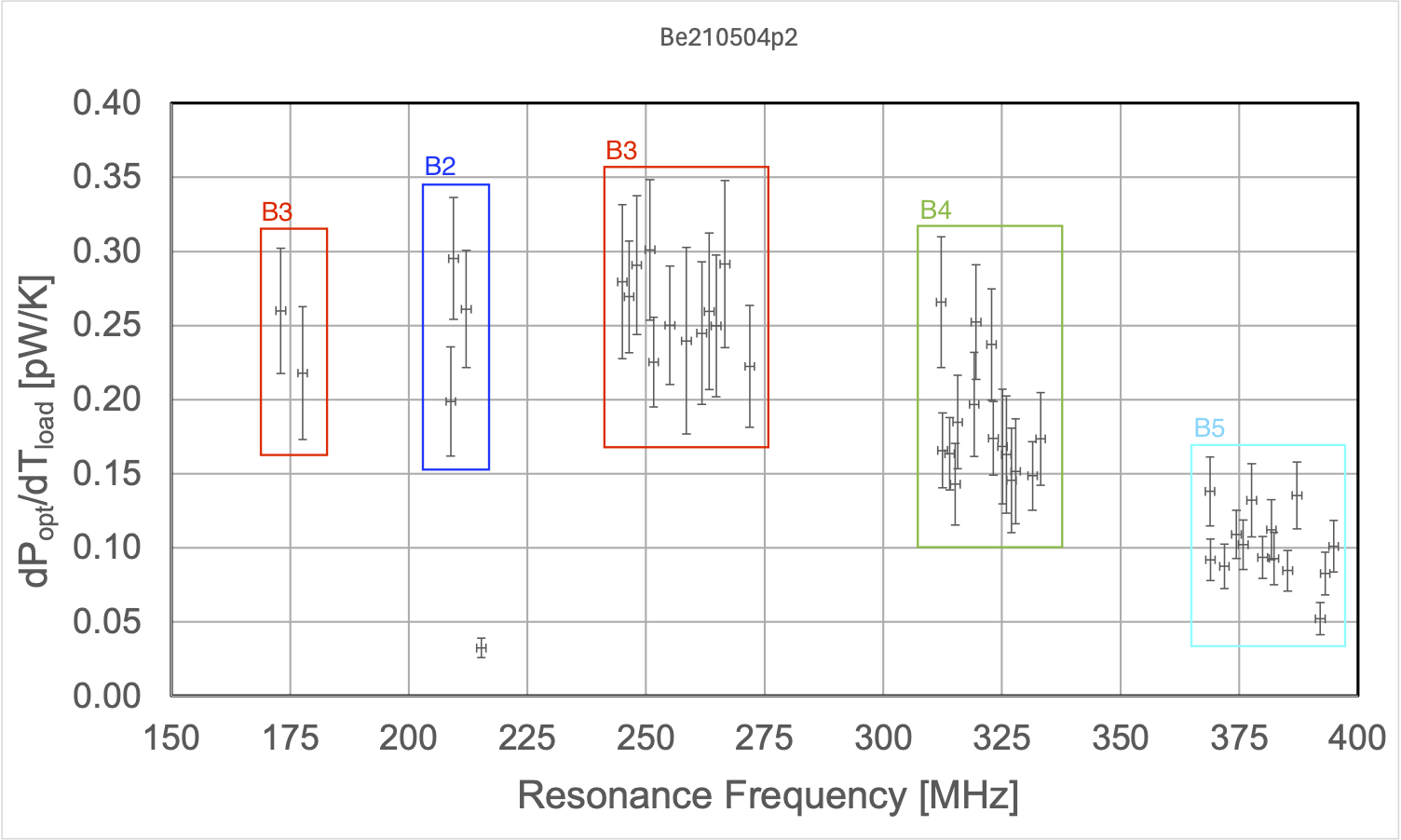} &
\hspace{0.2in}\includegraphics*[width=3in,viewport=10 10 1490 820]{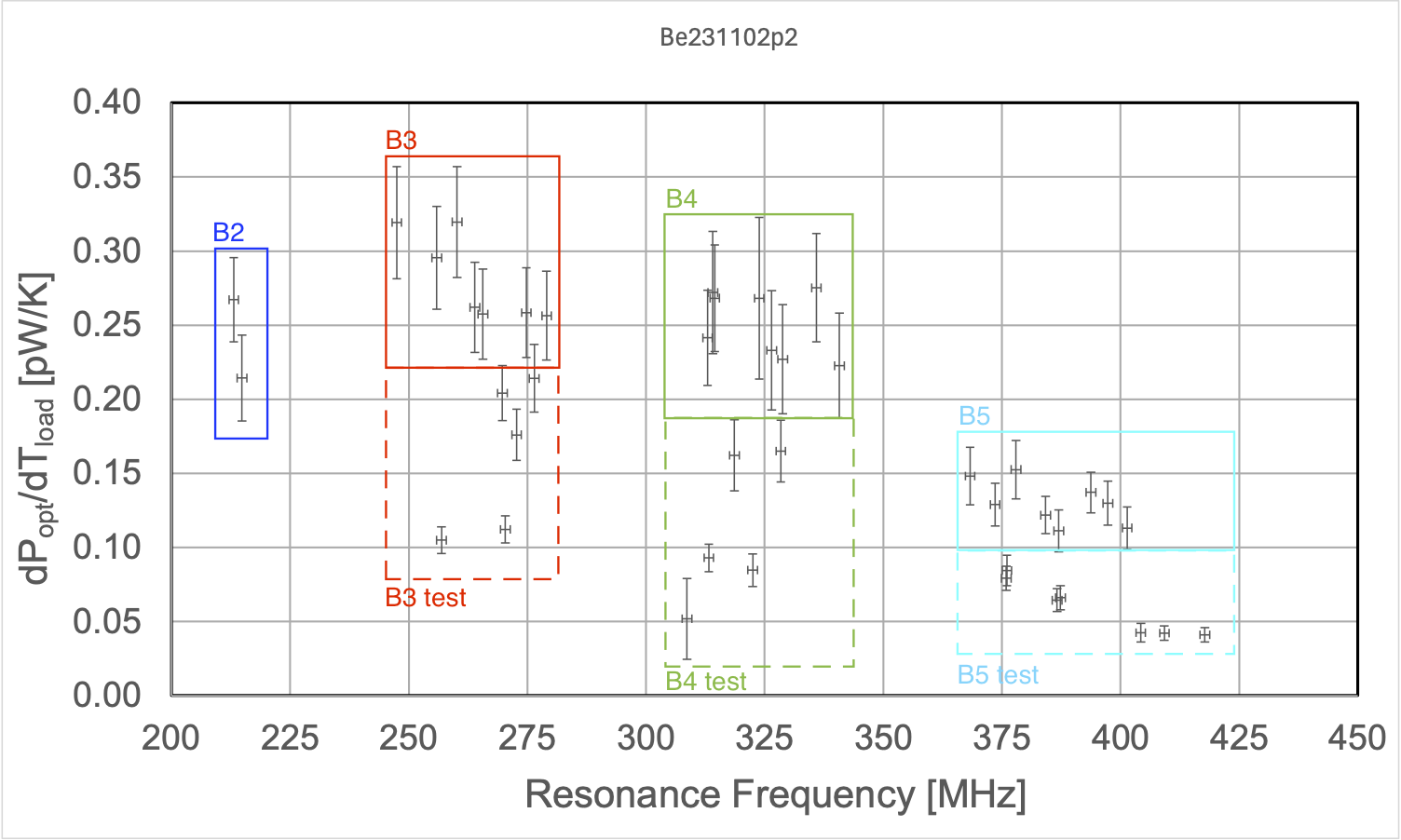}
\end{tabular}
\end{center}
\vspace{-9pt}
\caption{\textbf{Data used to measure trans-mm power absorbed in quasiparticle system for hot and cold blackbody loads.} (Top)~Typical $\delta\fr/\fr$ data sets used to fit for $dP_{qp}/d\Tl$ and $\Tx$ for the two devices discussed here.  Black, blue, magenta, and red data points and lines correspond to dark, cold, mirror, and hot data and fits.  (The ``mirror'' data uses a reflective cover over the window and is not used in this analysis.  There is no ``mirror'' data for the first device.)  Data under optical load for the first device are over a restricted range of $\Tb$ due to sub-Kelvin cryogenics problems during that cooldown. (Middle)~Fit residuals.  There is a greater systematic difference between dark data and model for the first device, but it does not appear to have significant impact on the results.  Both fits allow for heating of the device wafer, $\Ts \ne \Tb$.  For the second device, the full power-law heating model provided in the text is used.  For the first device, due to the limited $\Tb$ range under optical load, a simple fixed temperature offset $\Ts - \Tb$ was allowed.  In both cases, however, the heating model shows results inconsistent between different resonators, suggesting it is simply accommodates a systematic error in the modeling of $\delta\fr/\fr$ vs.\ $\Tb$. (Bottom)~$d\Popt/d\Tl$ inferred from the above fits as explained in the text.  The spectral bands are indicated, and ``test'' indicates resonators attached to test structures that should have lower response.}
\label{fig:opt_eff}
\end{figure}

We measure optical efficiency using beam-filling cold (liquid nitrogen) and hot (room temperature) blackbody loads\footnote{61~cm~$\times$~61~cm pieces of WAVASORB\textsuperscript{\textregistered} VHP (\url{https://www.ecanechoicchambers.com/pdf/WAVASORB\%20-\%20VHP.pdf}).  For the cold load, we immerse the blackbody in liquid nitrogen contained in a closed-cell ethylene vinyl acetate (EVA) foam container that we assembled from individual layers cut by Rapid Die Cut (\url{https://rapiddiecut.com}).  The layers were provided with adhesive on one side to aid assembly.}.  We use an air knife\footnote{Exair Super Air Knife, \url{https://www.exair.com/products/air-knives/super-air-knives.html}} to prevent condensation on the large clear-aperture window during measurements.  We measured $\fr$ for the two values of $\Tl$ and a range of $\Tb$.  We then fit the $\delta \fr/\fr$ data to a model that incorporates resonator-specific measurements of $(\alpha, \Delta)$ from dark data (\S\ref{sec:kid_params}) and determines $\Pqp(\Tl)$ and $\Tx$ for each resonator, where $\Pqp$ is the optical power absorbed in the KID quasiparticle system (i.e., that can affect $\fr$ and $\Qi$) and $\Tx$ is the ``excess load'' due to emission from the dewar (especially the dewar windows) converted to a Rayleigh-Jeans load temperature outside the dewar.  If $\Popt$ is the power incident on the KID from the microstripline, then $\Pqp = \etapb\,\Popt$ where $\etapb$ is the efficiency with which incoming trans-mm photons break Cooper pairs (with the remainder of the energy lost to sub-$2\Delta$ phonons).   Figure~\ref{fig:opt_eff} shows an example of these data and fits.  To infer $\Popt(\Tl)$, we use $\etapb$ as given in Table~\ref{tbl:sensitivity}.

We obtain from $\Popt(\Tl)$ two quantities: $d\Popt/d\Tl$ and the optical efficiency, $\etaopt$.  To do so, we make a typical set of assumptions: 1) the antennas are single-moded and thus have throughput $A\,\Omega = \lambda^2$ at 100\% efficiency; 2) the blackbody loads are in the Rayleigh-Jeans limit at the frequencies of interest; and, 3) the antenna is sensitive to a single polarization.  These assumptions imply $\Popt = \etaopt\,k_B\,\Tl\,\Delta \nu$ where $\Delta \nu$ is the spectral bandpass width and $\etaopt$ is the end-to-end optical efficiency between the blackbody source and the KID, excluding $\etapb$.  We may thus calculate $d\Popt/d\Tl = \Popt/\Tl$ trivially.  As always, there is a degeneracy between $\Delta \nu$ and $\etaopt$.  Peak normalization of the spectral bandpasses is one choice that is frequently made.  Another choice would be to take $\Delta \nu$ to be equal to the design values from Table~\ref{tbl:parameters} so that $\etaopt$ both incorporates non-idealities in the spectral bandpasses and absorption.  Figure~\ref{fig:opt_eff} provides plots of the unambigous $d\Popt/d\Tl$ quantity, and Table~\ref{tbl:opt_eff} provide minimum, maximum, and mean values for each band and device for $d\Popt/d\Tl$ and for $\etaopt$ under the two different assumptions for $\Delta \nu$.  We find the two choices for $\etaopt$ normalization differ fractionally at the $\lesssim 10$--15\% level, resulting in no sigificant difference in interpretation of the results.

In B2, the $d\Popt/d\Tl$ values are comparable to the best typically achieved for sub-Kelvin detectors given the significant blackbody filtering required, 0.24~pW/K~\cite{bicep2_antennas2015}. B3 has comparable performance to B2 in both $d\Popt/d\Tl$ and $\etaopt$.  There is a significant difference between \textit{mean} B4 performance for the two devices.  The second device again yields $d\Popt/d\Tl$ and $\etaopt$ performance comparable to B2 (and B3).  The first device's best detectors also match this performance, but it has a number of detectors with response lower by about 1/3 fractional.   The B5 detectors show appreciably lower $\etaopt$ than all the other bands, also down by 40\% relative to B2 and B3 but now uniformly.  (Comparison of B5's $d\Popt/d\Tl$ other bands is not useful because of its lower design bandwidth.)

While we have not modeled the optical efficiency in detail, we note at least two potential causes for variations in optical efficiency among bands (and especially in B5): loss in the plastic windows/filters (expected transmittances of 0.76, 0.71, 0.66, and 0.62 in the four bands) and reflection loss due to insufficiently wide-band AR coatings on optical elements (for the silicon AR wafer, $<$0.5\% reflectance in B3 and B4 but $\approx 10$\% and $\approx 5$\% reflectance in B2 and B5, respectively~\cite{cit_2layer_2018}).  Together, these effects result in an expected transmittance relative to B3 of 0.96, unity, 0.93, and 0.83.  Motivated by the dispersion among detectors noted above, which is not due to measurement uncertainty but rather to individual detectors suffering performance non-idealities, Table~\ref{tbl:opt_eff} compares these expected relative transmittances to B3-normalized \textit{maximum} efficiencies.  The B2--B4 relative transmissions are generally in line with expectations while the B5 relative transmission is not.  A more extensive analysis is underway (\S\ref{sec:bandpasses}) and may explain this discrepancy.  We also plan measurements with a blackbody at the 4~K stage of the cryostat to eliminate filter transmission uncertainty.

We have developed a broader-bandwidth three-layer AR structure~\cite{defrance_ltd18_2020, defrance_grin_2024} and are finalizing the development of an even broader-bandwidth four-layer AR design, and we anticipate implementing lower-loss windows and filters with better AR coatings, so we expect significant improvement in B5's performance as well as modest gains in the other bands.

\begin{table}[t!]
\begin{center}
{\small
    \begin{tabular}{cc|ccc|ccc|cccc|cc} \hline \hline
     & &  \multicolumn{3}{c|}{$d\Popt/d\Tl$} & \multicolumn{3}{c|}{$\etaopt^{peak-norm}$} 
     &  \multicolumn{4}{c|}{$\etaopt^{design}$} & \multicolumn{2}{c}{expected}   \\
     & &  \multicolumn{3}{c|}{}              & \multicolumn{3}{c|}{}
     & & & & max/ &  \multicolumn{2}{c}{$\eta_{filt} \eta_{AR}$} \\
     & &  \multicolumn{3}{c|}{}              & \multicolumn{3}{c|}{}
     & & & & B3 &  & value/  \\
Band & Device & min  & max  & mean & min & max & mean & min & max & mean & max & value & B3
\\ \hline \hline
2    & a & 0.20 & 0.30 & 0.25 & 0.33 & 0.49 & 0.42 & 0.31 & 0.46 & 0.39 & 0.95 & 0.68 & 0.96 \\
2    & b & 0.21 & 0.27 & 0.24 &      &      &      & 0.33 & 0.41 & 0.37 & 0.81 & 0.68 & 0.96 \\ \hline
3    & a & 0.22 & 0.30 & 0.26 & 0.39 & 0.53 & 0.46 & 0.35 & 0.48 & 0.41 & 1    & 0.71 & 1    \\
3    & b & 0.26 & 0.32 & 0.28 &      &      &      & 0.41 & 0.51 & 0.45 & 1    & 0.71 & 1    \\ \hline
4    & a & 0.14 & 0.27 & 0.18 & 0.23 & 0.42 & 0.29 & 0.26 & 0.48 & 0.33 & 1 & 0.66 & 0.93 \\
4    & b & 0.22 & 0.28 & 0.25 &      &      &      & 0.40 & 0.50 & 0.45 & 0.98 & 0.66 & 0.93 \\ \hline
5    & a & 0.08 & 0.14 & 0.10 & 0.16 & 0.27 & 0.21 & 0.18 & 0.29 & 0.22 & 0.60 & 0.59 & 0.83 \\
5    & b & 0.11 & 0.15 & 0.13 &      &      &      & 0.24 & 0.33 & 0.28 & 0.65 & 0.59 & 0.83 \\ \hline \hline
    \end{tabular}
    }
    \end{center}
\caption{\textbf{Optical efficiency estimates.} We convert the minimum, maximum, and mean values of $d\Popt/d\Tl$ for each spectral band and each device to optical efficiency under the two different assumptions noted in the text: peak-normalized bandpasses and design bandpasses.  The differences between the two approaches are at the 10\% fractional level.  We have not measured bandpasses for the second device, so we do not provide $\etaopt^{peak}$ values for it.  The penultimate column shows the expected transmittance of the plastic windows and the silicon metamaterial antireflection wafer.  For $\etaopt^{design}$ and the expected $\eta_{filt} \eta_{AR}$, we compute the values relative to the B3 values to assess whether the relative transmissions match expectations.  (For the measurements, we use the maximum values as explained in the text.)  The B2--B4 relative transmissions are generally in line with expectations while the B5 relative transmission is not.}
\label{tbl:opt_eff}
\end{table}

\subsubsection{a-Si:H Loss Tangent} 
\label{sec:nblc_loss}

\begin{figure}[t!]
\begin{center}
\begin{tabular}{cc}
\includegraphics*[height=1.75in,viewport=0 0 1367 750]
{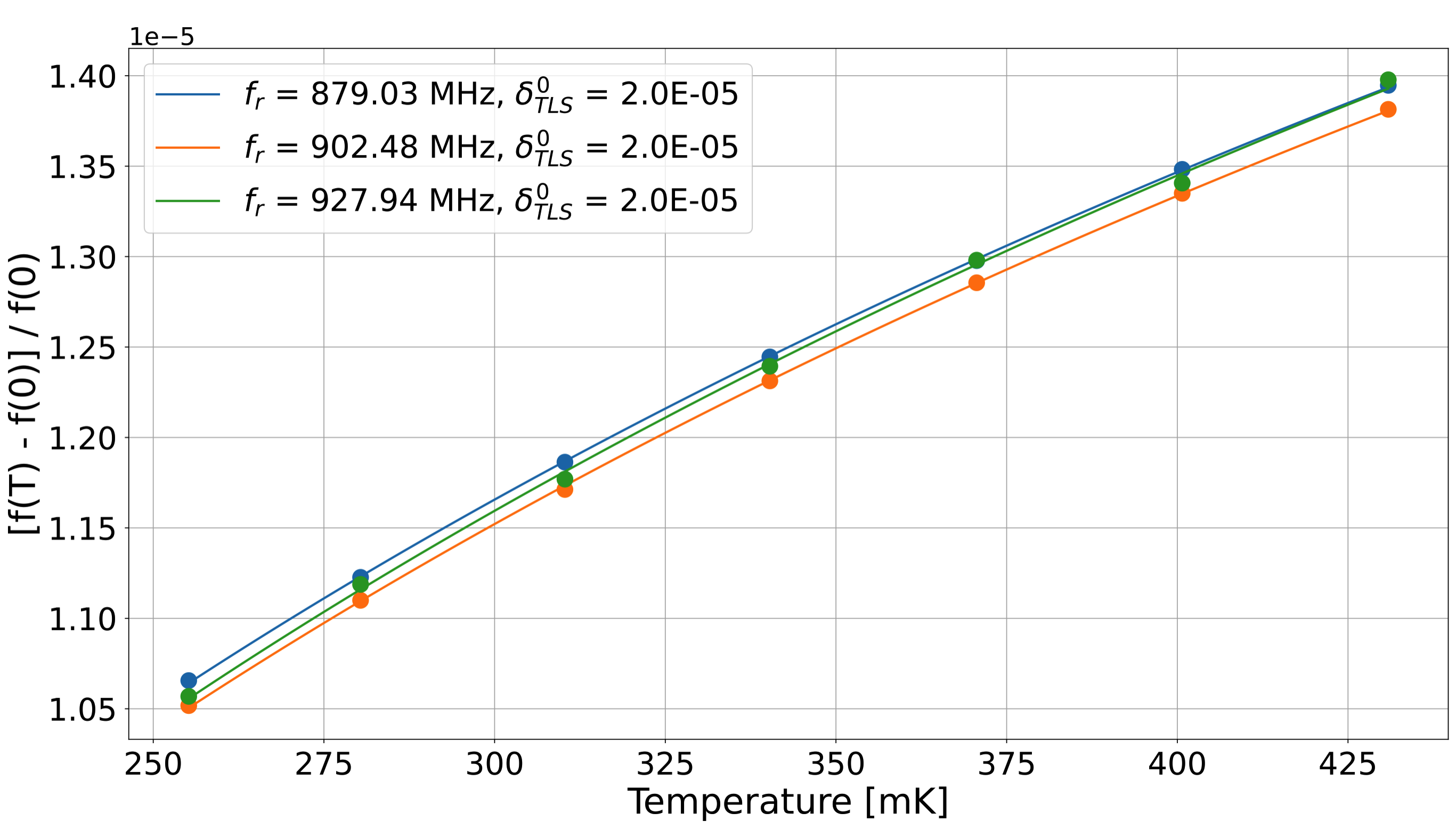} & 
\includegraphics*[height=1.75in,viewport=0 0 1139 700]
{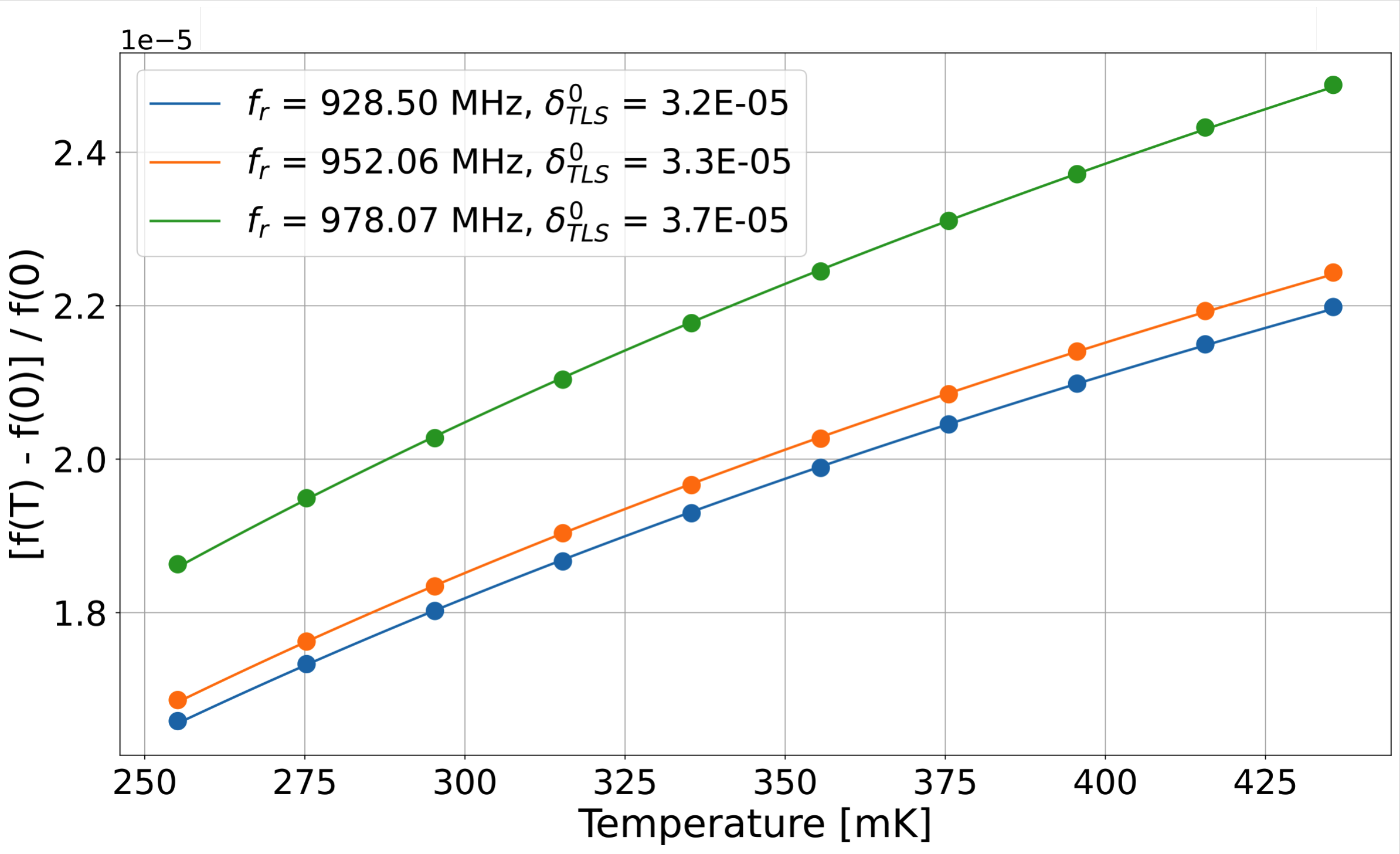} \\
\includegraphics*[width=3.25in,viewport=0 0 1600 688]
{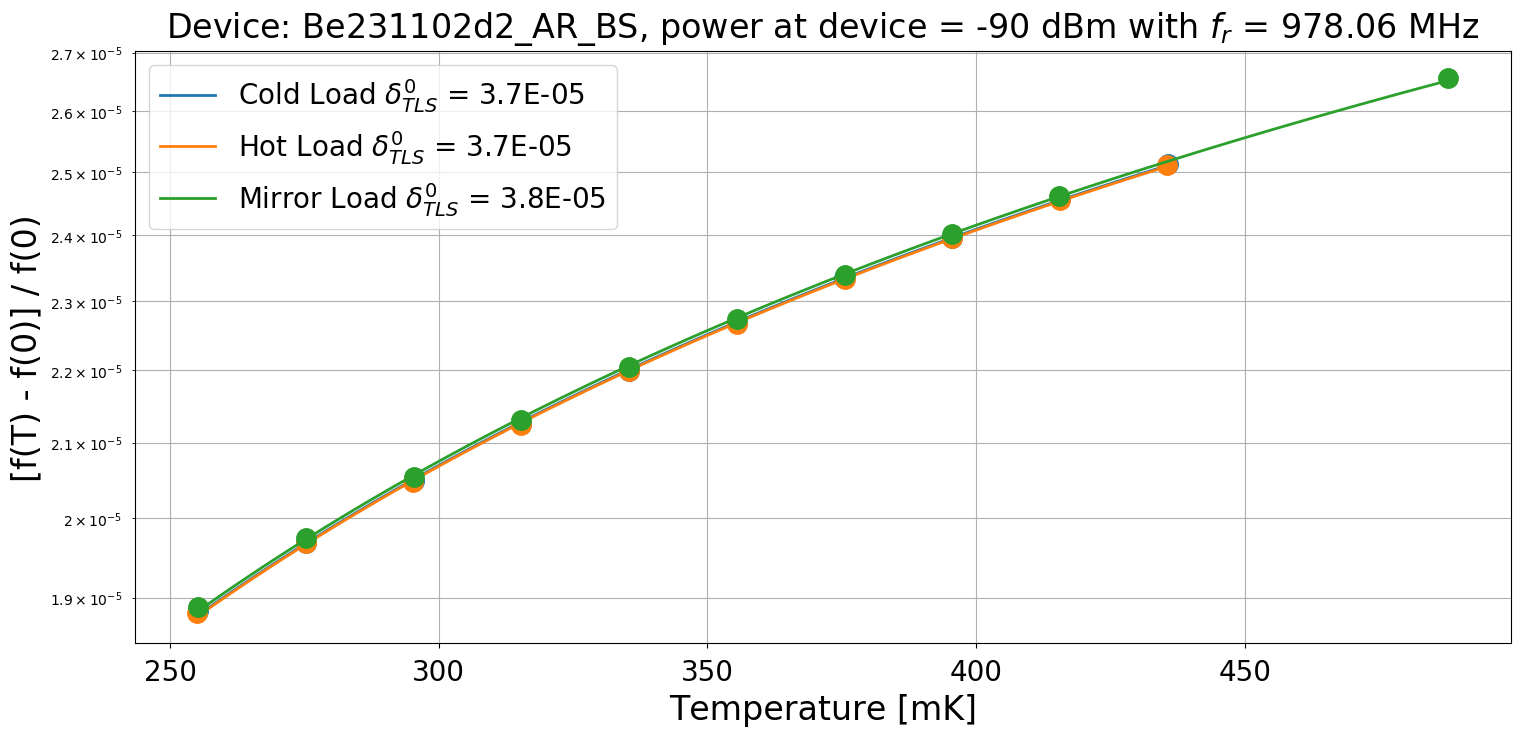} & 
\includegraphics*[width=3.25in,viewport=0 0 1600 688]
{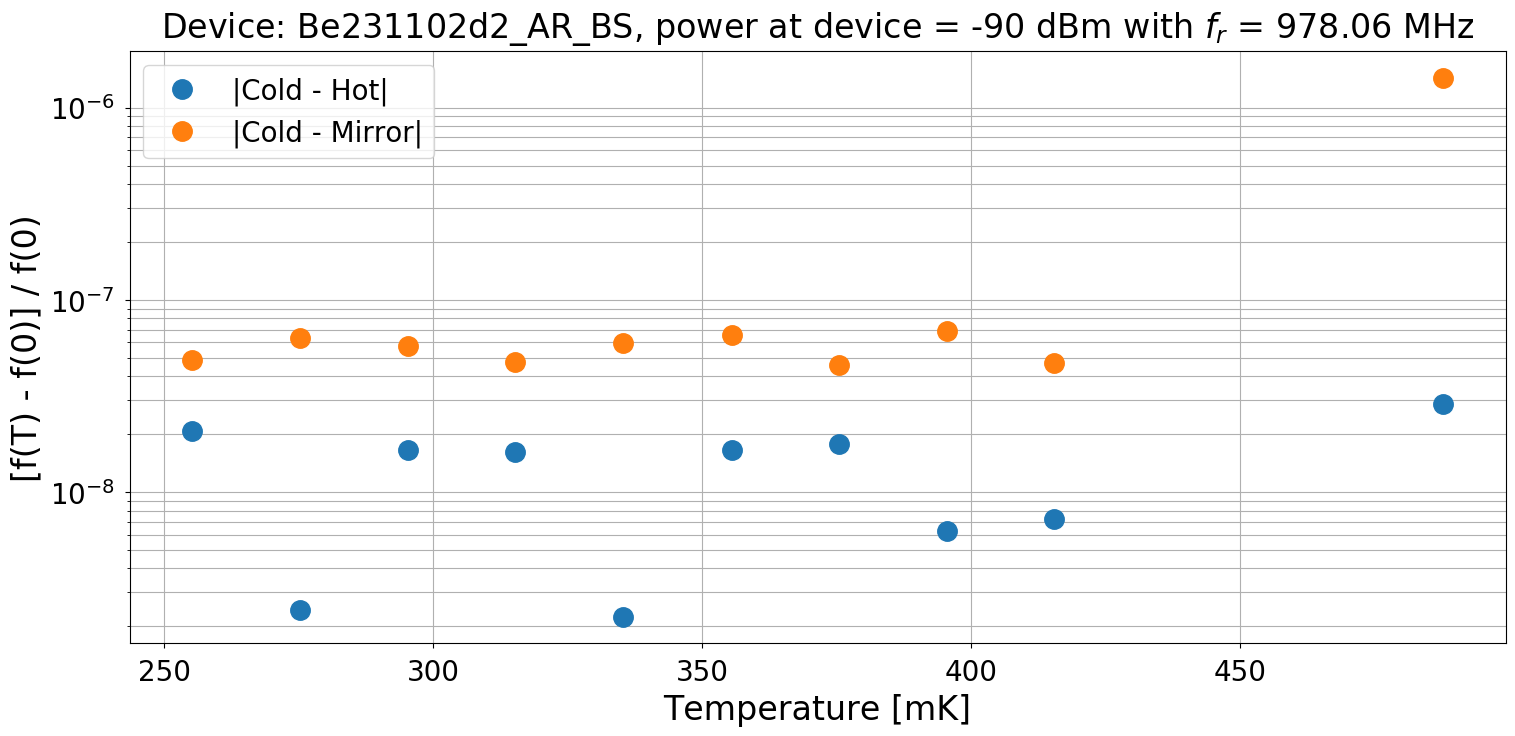}\hspace{3pt} \\
\end{tabular}
\end{center}
\vspace{-9pt}
\caption{\textbf{RF TLS loss tangent and direct absorption limits.} (Top)~Fractional frequency shift vs.\ $\Tb$ for three observed Nb LC resonators on each of the two devices under study here.  The low-power TLS RF loss tangent is given by the normalization of the fit.  The loss tangents, 2 and $3.7\,\times\,10^{-5}$, are excellent compared to most amorphous dielectrics.  The difference between the two is discussed in the text. (Bottom left)~Fractional frequency shift vs.\ $\Tb$ for the second device's 978~MHz resonator under three optical loads.  (Nb LC resonator data under optical load were not taken for the first device.) (Bottom right)~Absolute difference in fractional frequency shift vs.\ $\Tb$ between cold and hot and between cold and mirror optical loads for the same resonator.  It is clear the difference is limited by systematics because the cold--mirror difference is larger in magnitude than the cold--hot difference and, in the plot at right, the ordering from bottom to top is hot, cold, mirror when it should be cold, mirror, hot if the differences were due to true changes in substrate temperature.}
\label{fig:tls_loss}
\end{figure}

Every device includes lumped-element Nb LC resonators, identical in design to those used in our prior demonstrations of a-Si:H low-power TLS loss tangent as low as $\deltls^0 \approx 7\,\times\,10^{-6}$ near 1~GHz\cite{defrance_asihloss}, so that we may measure independently this loss tangent \textit{for the deposited material in the KIDs and microstripline on each device}.  There are two sets of resonators, one incorporating the 800~nm a-Si:H in the KID capacitors and the other the 1070~nm a-Si:H in the microstripline.  (See \S\ref{sec:ms_coupling} and \S\ref{sec:fab} for an explanation of the distinction between the two.)  As in that prior work, we infer $\deltls^0$ by measuring $\delta\fr/\fr$ vs.\ $\Tb$ in the 250--450~mK range, well below the temperatures at which quasiparticles cause any shift in $\fr$ in Nb, and fitting for $\deltls^0$  as the normalization for the known TLS loss dependence on temperature (e.g., \cite{gaonoise2, gao_thesis, jonas_arcmp}).  We show in Figure~\ref{fig:tls_loss} loss tangents of 2 and $3.7\,\times\,10^{-5}$ for the 800~nm KID a-Si:H for the two devices studied here.  (The resonators for the 1070~nm microstripline a-Si:H do not seem to be present.)  These results are 70\% and 25\% poorer than expected (\S\ref{sec:fab}), respectively, and worth tracking in future devices, but they do not significantly impact expectations for device optical efficiency or KID TLS noise.

We are in the midst of explicitly measuring the trans-mm wave loss of our a-Si:H using the dedicated test devices.  We may set a conservative upper limit using the data in hand by assuming the lowest efficiency resonators in each band for the second device (Figure~\ref{fig:opt_eff} (bottom right)) are part of a loss-test pair for that band and the highest efficiency resonators have twice the efficiency of the loss-test pair's reference detector.  The loss-test devices were designed to match the $1/e$ length for a loss tangent of $\approx$$10^{-3}$ and yield 10\% loss for a loss tangent of $\approx$$10^{-4}$.  The inferred efficiency ratio is approximately $0.6\approx \ln 0.5$ in all of B3, B4, and B5, so the loss tangent is at most $5\,\times\,10^{-4}$.  From this, we estimate a lower limit on the transmission for the three-scale antenna of 0.93--0.94, approximately independent of frequency.  This value is certainly good enough that it is subdominant to many other optical losses (\S\ref{sec:opt_eff}).  It is likely the constraints on $\delta$ will be improved by our ongoing measurements using the paired KIDs.  If $\delta \approx 10^{-4}$ is obtained, comparable to the best achieved with a-Si:H\cite{sron_asih_2021} and a-SiC:H\cite{sron_asich_2022}, then the transmission would be approximately 0.985.

\subsubsection{Direct Absorption Limits} 
\label{sec:nblc_direct}

A key feature of the device design is the many measures taken to limit direct absorption of trans-mm light by the KIDs (\S\ref{sec:design:kid}).  We are able to place limits on this absorption by including in our optical efficiency analysis the dark KIDs.  Their data are treated in the same way as those of the other KIDs, yielding an effective $\Popt$ and $d\Popt/d\Tl$ that can be compared to those of the optically sensitive KIDs.  ($\etaopt$ can be calculated too by assuming $\Delta \nu = 420 - 96 = 304$~GHz, the bandwidth between the metal mesh filter cutoff and $2\Delta_{Al}/h$, but $\Popt$ and $d\Popt/d\Tl$ are more important for determining what fraction of the signal observed by a given detector arrives via the antenna vs.\ being directly coupled.)  

Because it is physically allowed, to improve the fit quality, and to ensure we interpret any apparent response of the dark KIDs to varying optical load correctly, the model allows for the substrate to heat up due to broadband absorption in the silicon wafer itself.  We use a standard conductance power law, $P = g (\Ts^n - \Tb^n)$, where $\Ts$ is the substrate temperature and we fit for $g$ and $n$.  The measurement of $\delta\fr/\fr$ under dark, cold, and hot loads as a function of $\Tb$ breaks the degeneracy between optical efficiency and substrate heating.  We find that the data prefer substrate heating, but the modeling is problematic.  The preferred value is inconsistent among KIDs, even among the dark detectors, with values of up to $\Ts - \Tb \approx 15$~mK at low $\Tb$ for optical detectors but values of 60--70~mK for dark KIDs under the same conditions.  We also find that the value depends far more on $\Tb$ than on $\Tl$, which is unphysical.    

We can obtain a more useful constraint on $\Ts$ using the a-Si:H RF loss tangent diagnostic resonators discussed in \S\ref{sec:nblc_loss}.  The dependence of $\delta\fr/\fr$ on $\Tb$ due to TLS can be used as a differential thermometer by seeing how much $\fr$ changes between cold, mirror, and hot loads.  Figure~\ref{fig:tls_loss} shows those data.  While the desired difference is clearly limited by systematics, we can infer a conservative upper limit of about $10^{-7}$ on $(\fr^{hot} - \fr^{cold})/\fr$, and we observed $d(\delta \fr/\fr)/d\Tb \approx 5\,\times\,10^{-8}$/mK from the same data.  We may therefore conclude the substrate changes in temperature by no more than about 2~mK between exposure to cold and hot loads.  Thus, it is clear that the fitted deviation of $\Ts$ from $\Tb$ for the KIDs is indeed unphysical.  We suspect the model's preference for substrate heating reflects non-idealities in our Mattis-Bardeen modeling of $\delta\fr/\fr$ vs.\ $\Tb$ for dark data.

Fortunately, the fitted $\Popt(\Tl)$ is not significantly affected by whether substrate heating is included or not, so we use these data to set an upper limit $\Popt^{dark}/\Popt^{opt} \lesssim 1$\%.

\subsubsection{AlMn}
\label{sec:almn}

\begin{figure}[t!]
\begin{center}
\begin{tabular}{rr}
\includegraphics*[height=1.74in]{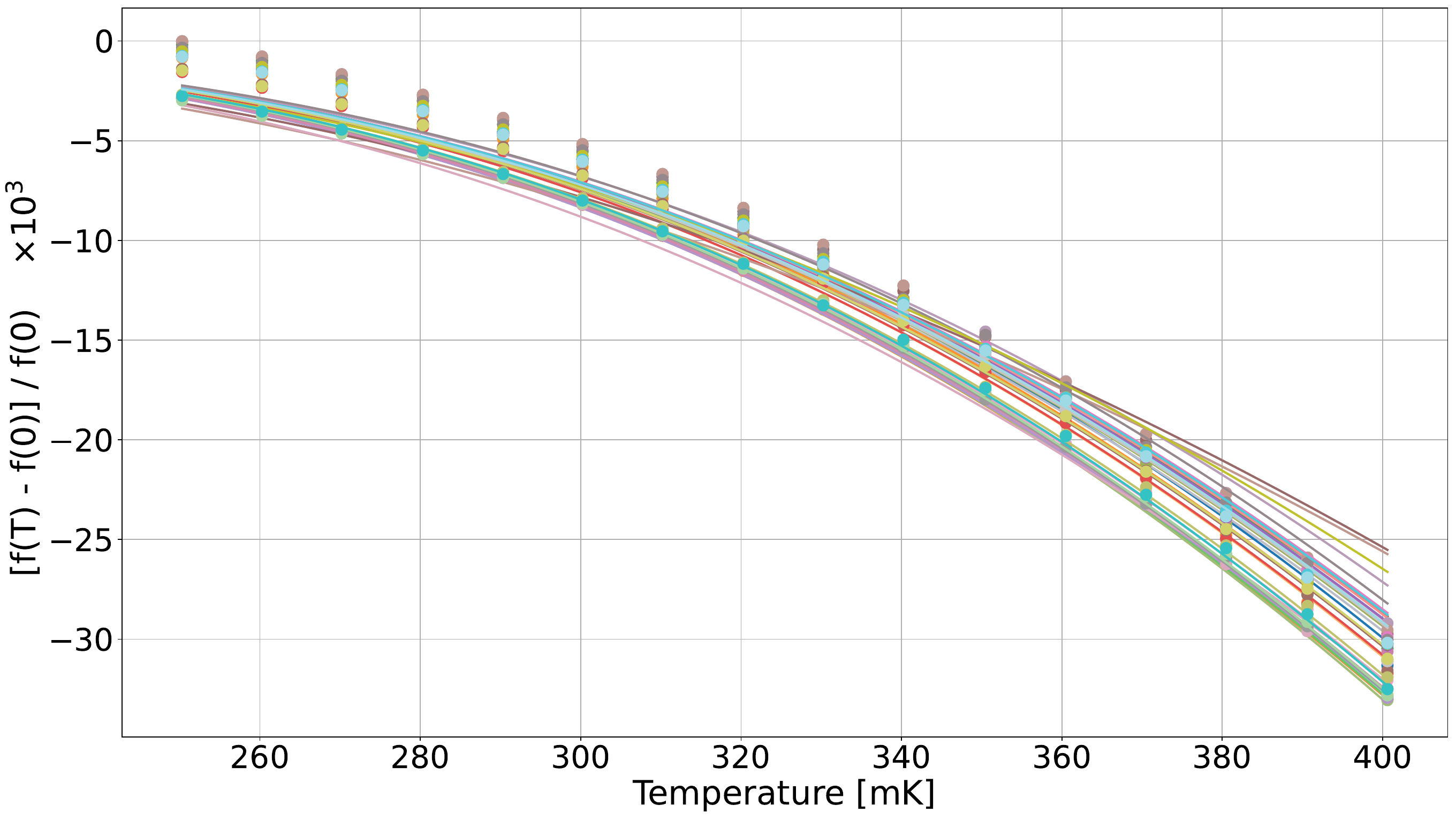} & 
\includegraphics*[height=1.74in]{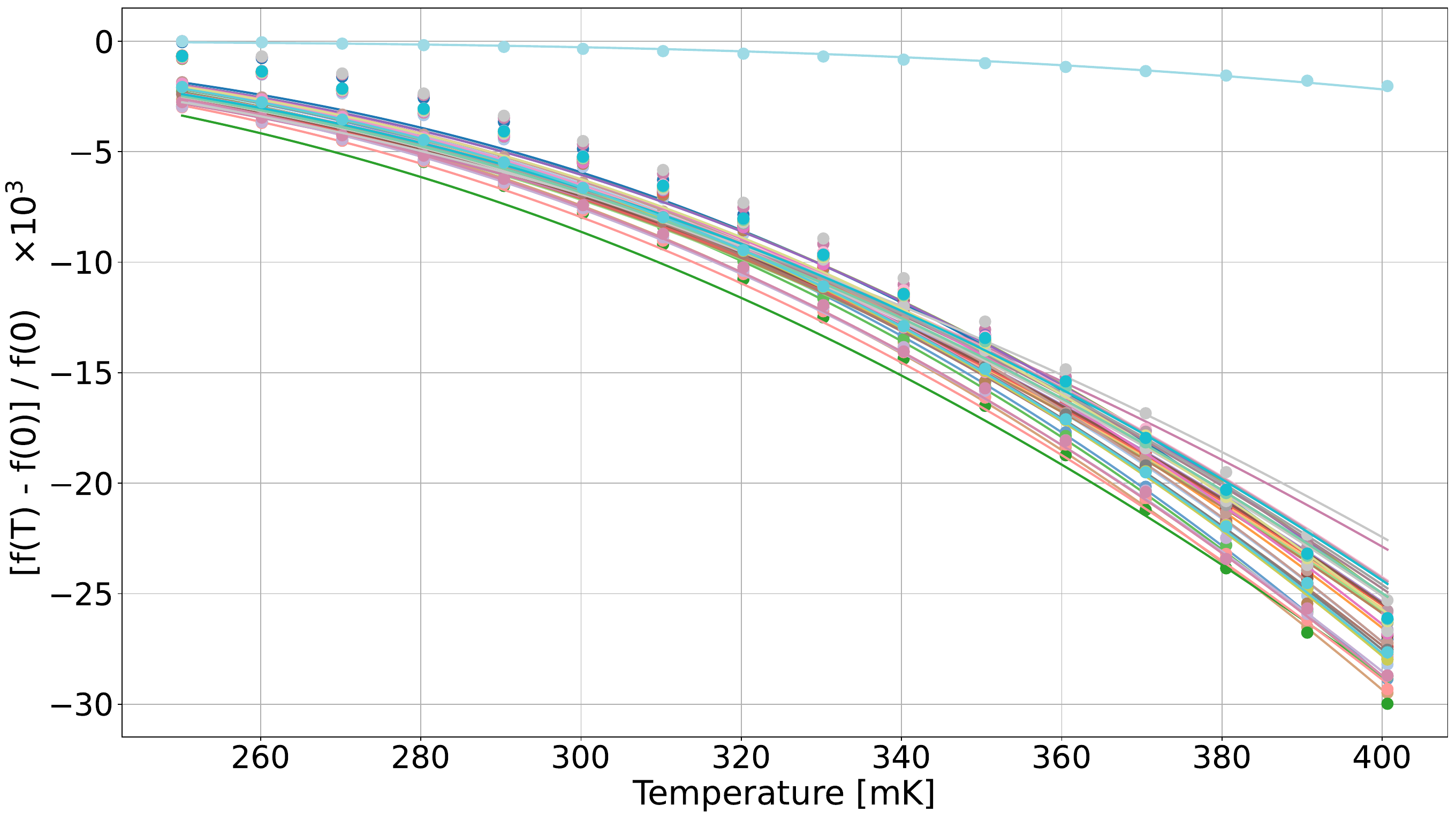} \\
\includegraphics*[height=1.74in]{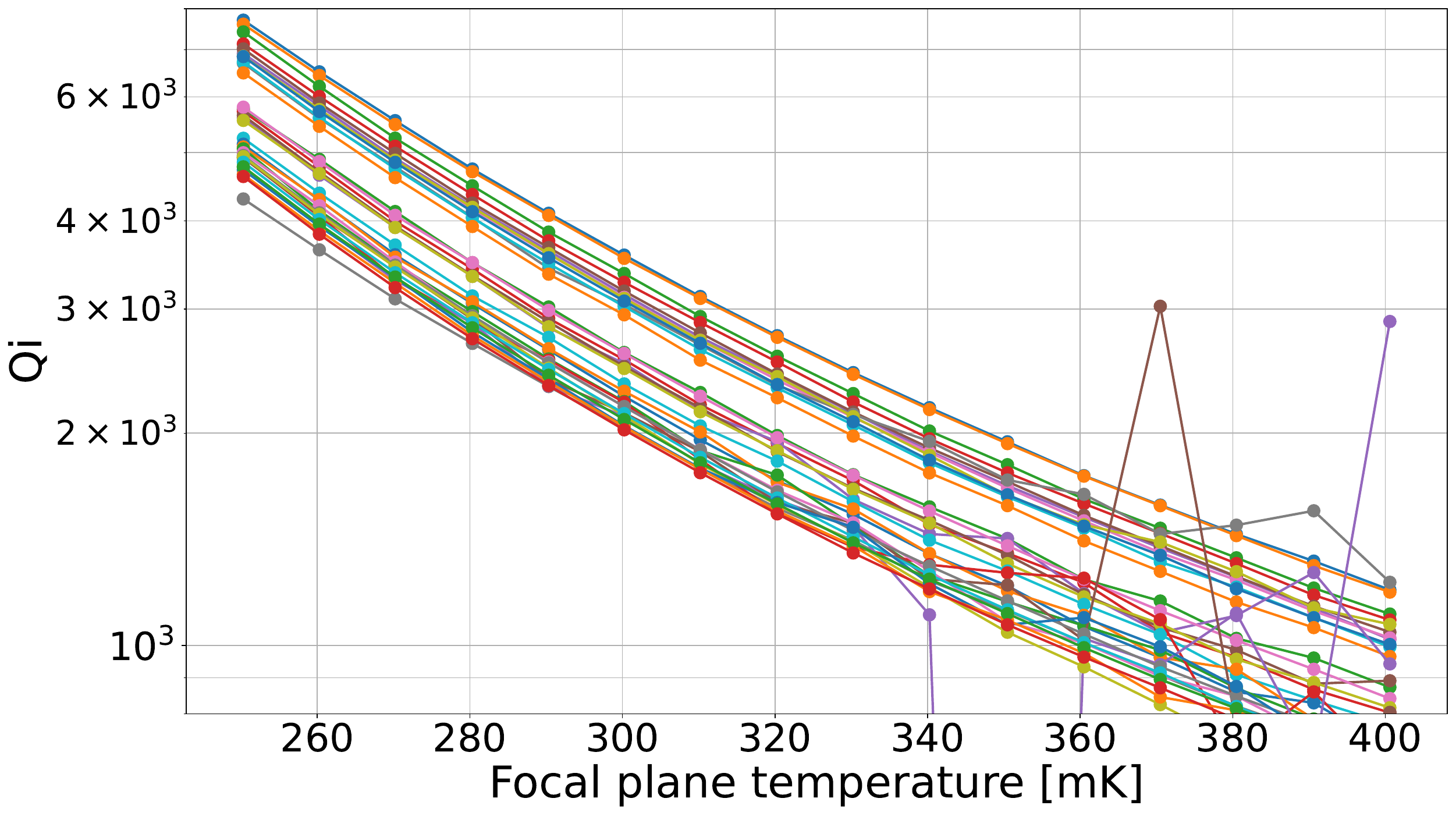} & 
\includegraphics*[height=1.74in]{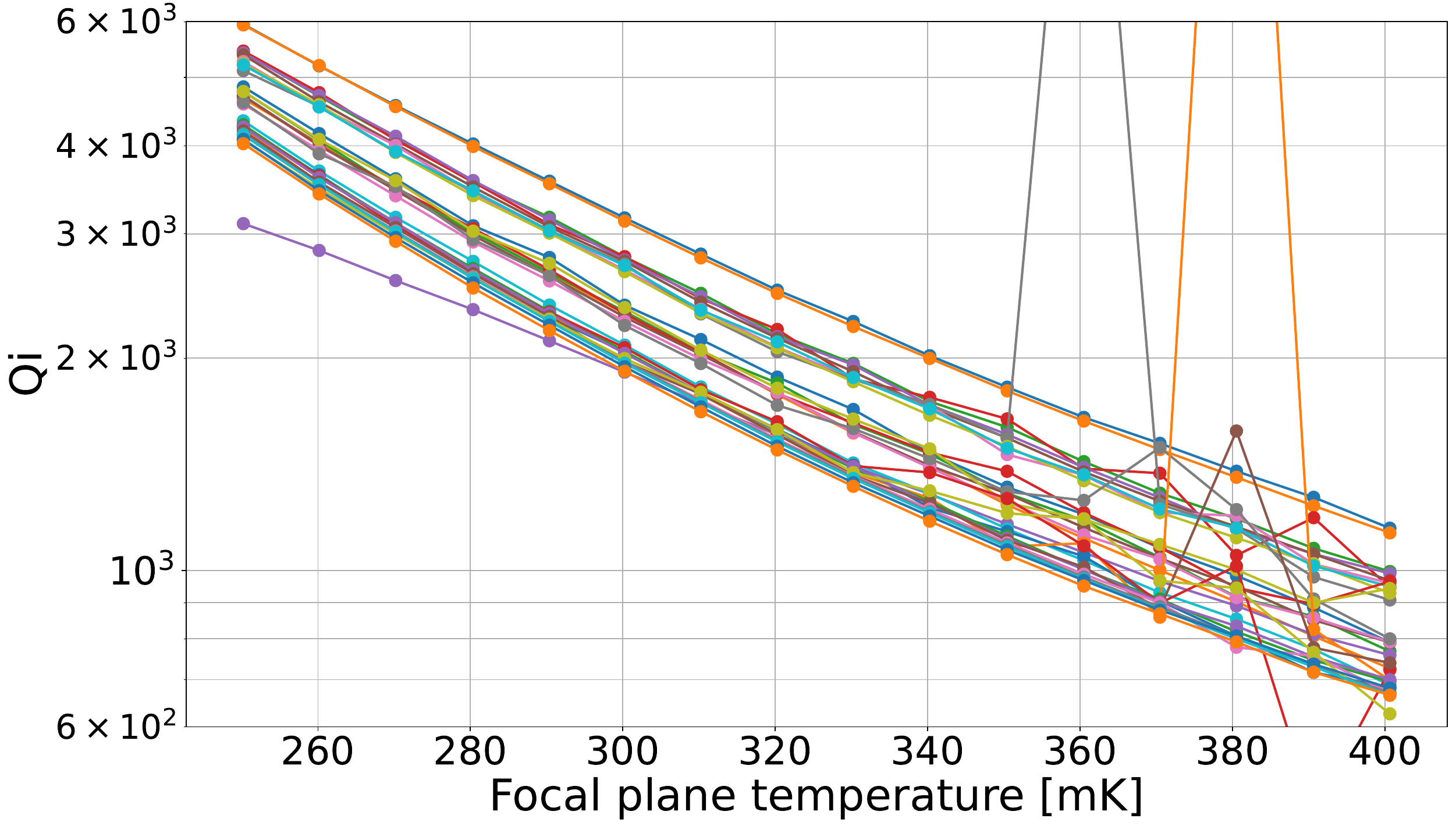} \\
\includegraphics*[height=1.75in]{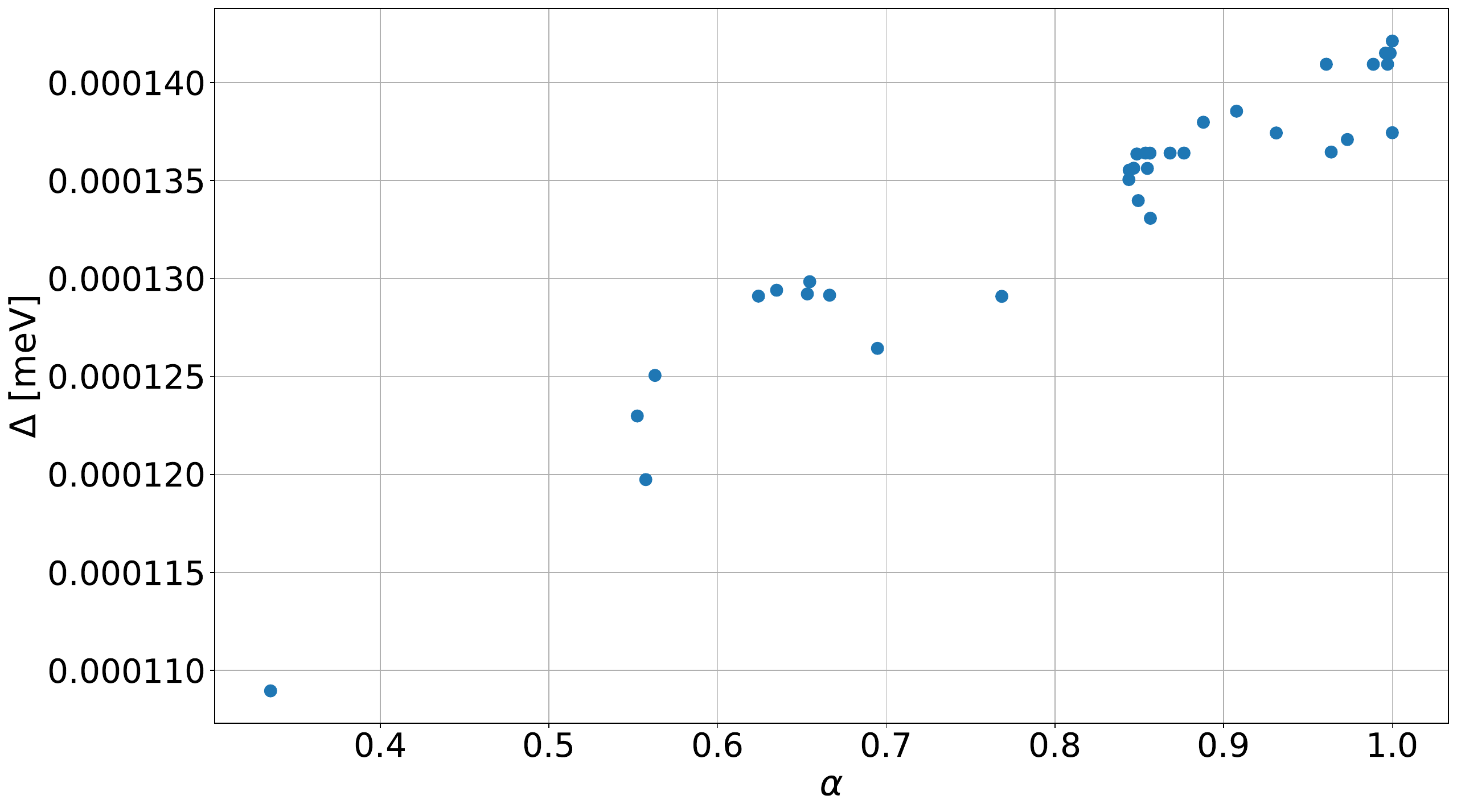} & 
\includegraphics*[height=1.75in]{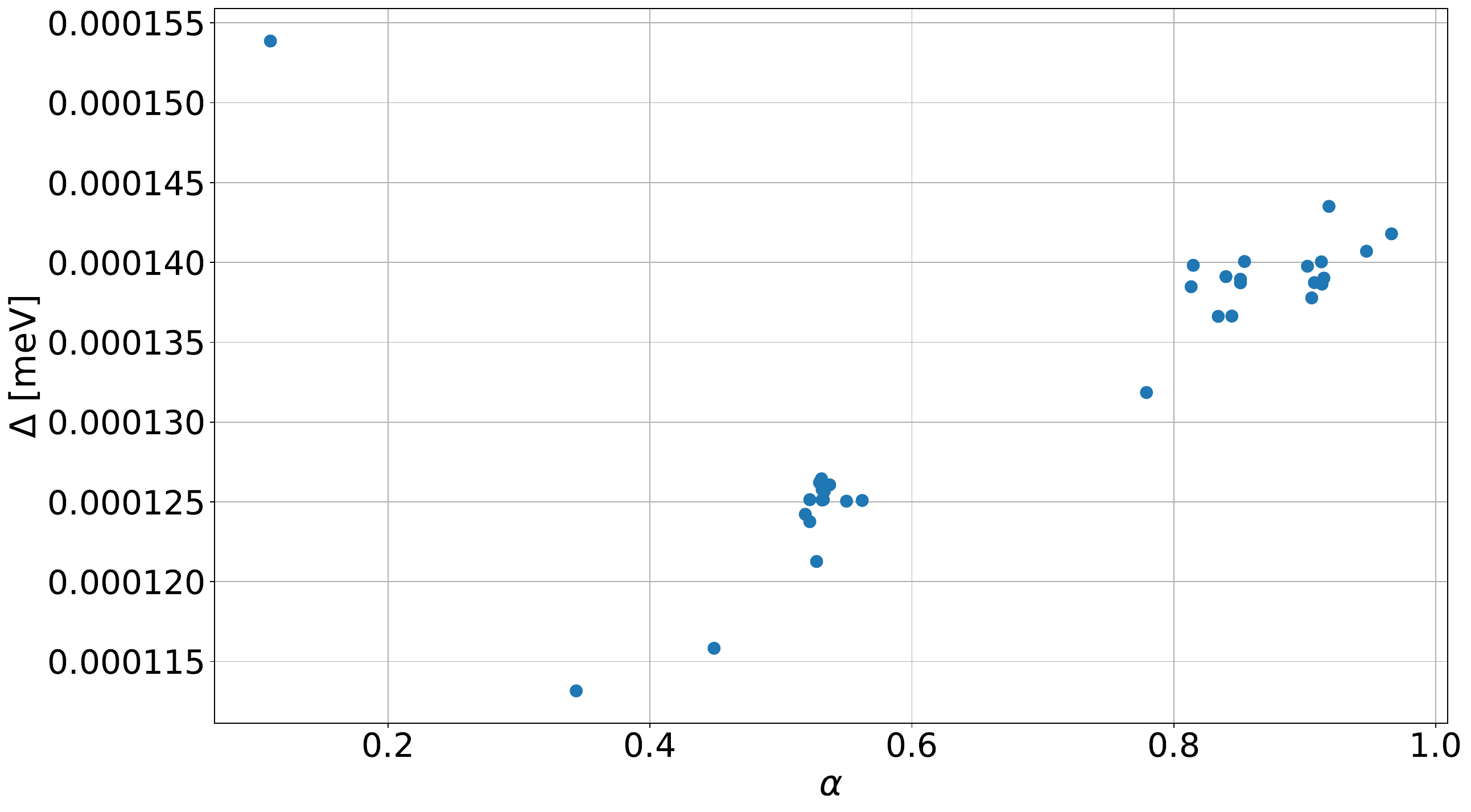} \\
\end{tabular}
\end{center}
\caption{\textbf{AlMn KIDs} (Top and middle)~$\delta\fr/\fr$ and $\Qi$ vs.\ $\Tb$ plots for two devices, from different wafers, replacing Al with AlMn.  The data do not saturate at low $\Tb$ due to the lower temperature limit of the apparatus being used.  Low $\Qr$ at high $\Tb$ makes it difficult to fit the resonance loops, resulting in some unreliable $\Qi$ values at high $\Tb$.  (Bottom)~Plot of $\alpha$ vs.\ $\Delta$.  A clear degeneracy is present due to the limited $\Tb$ range.  As discussed in the text, we can use the reduction of $\fr$ between Al and AlMn to constrain $\alpha_{AlMn}$ and thus narrow the range for $\Delta_{AlMn}$ to 0.125--0.135~meV ($2\,\Delta_{AlMn}/h = 60$--65~GHz).}
\label{fig:almn}
\end{figure}

We have recently demonstrated devices with the Al KID material replaced by AlMn and no other changes.  The yield on these first arrays appears to be around 60\%, a good starting point. Figure~\ref{fig:almn} shows $\delta\fr/\fr$ and $\Qi$ vs.\ $\Tb$ over a limited range of $\Tb$.  The low $\Qi$ reflects the high value of $\Tb/T_c$ and may make multiplexing and achieving fundamental-noise-limited performance challenging.  That said, the large B1 pixels imply very few B1 resonators are needed for the three-scale architecture (see \S\ref{sec:instr_params}), and $\etapb$ will be close to unity, so the low $\Qi$ may be acceptable. 

Because the data are not highly constraining, fits of $\delta\fr/\fr$ vs.\ $\Tb$ to Mattis-Bardeen theory yield a significant $\alpha$--$\Delta$ degeneracy, seen in Figure~\ref{fig:almn}.  
We may narrow the range for $\Delta_{AlMn}$ by constraining $\alpha_{AlMn}$ using $\alpha_{Al}$ and the observed $\fr$ ranges for Al and AlMn and the common resonator geometry: 
if $f_g$ is the resonant frequency assuming only the geometrical inductance $L_g$ and $\fr$ is the resonant frequency including the kinetic inductance $L_k$, then
\begin{align}
\fr = f_g \frac{\sqrt{L_g}}{\sqrt{L_g+L_k}} = f_g\sqrt{1 - \alpha}
\qquad \Longrightarrow \qquad \alpha_{AlMn} = 1 - (1 - \alpha_{Al}) \left(\frac{\fr^{AlMn}}{\fr^{Al}}\right)^2
\end{align}
Figure~\ref{fig:kid_params} shows the $\fr$ range for Al is 210-420~MHz (both devices) and the $\alpha_{Al}$ range is 0.25--0.40 (over both devices).  The observed $\fr$ ranges for the two AlMn devices are 136--265~MHz and 158--315~MHz.  Combining these ranges yields $\alpha_{AlMn} = 0.58$--0.75, which narrows $\Delta_{AlMn}$ to 0.125--0.135~meV, or $2\,\Delta_{AlMn}/h = 60$--65~GHz.
We will use index-test KIDs, which incorporate no BPFs, to explicitly measure $2\,\Delta_{AlMn}/h$.  

To improve $\Qi$ at the current $\Tb$, we could tune the AlMn $T_c$ to obtain higher $\Delta_{AlMn} \approx 0.165$~meV ($2\,\Delta_{AlMn}/h \approx 80$~GHz).  If $\Qi$ remains inconveniently low, it may be possible to deposit AlMn only for the B1 resonators, suffering low $\Qi$ for only $64/(2\times 64 + 2\times 256 + 2\times 1024) = 2.3$\% of the KIDs, by using coarse liftoff masks for Al and AlMn depositions: i.e., prepare a liftoff mask to only permit Al deposition in windows over the non-B1 resonators, deposit Al, remove the liftoff mask, prepare a second liftoff mask to only permit AlMn deposition in windows over the B1 resonators, deposit AlMn, and then proceed with the rest of the fabrication as before.  There is no need for the Al and AlMn to be in galvanic contact with each other, so this two-step deposition process would be acceptable.  

\subsubsection{Noise and Sensitivity} 

\paragraph{Generation-Recombination-Dominated Detector Noise}

\begin{figure}[t!]
\begin{center}
{\small \begin{tabular}{cc}
\hspace{0.5in}Dark, $\Tb = 310$~mK & 
\hspace{0.5in}$\Tl = 77~K + \Tx \approx 180$~K, $\Tb = 241$~mK\\
\includegraphics*[width=3.25in,viewport=0 300 600 590]{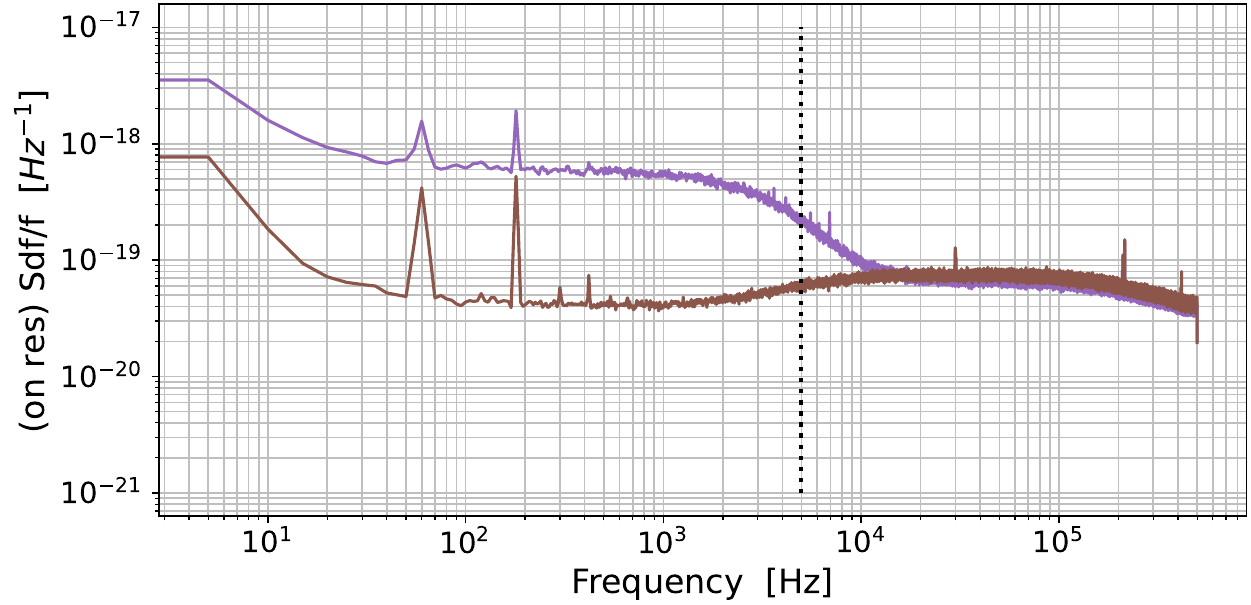}
& 
\includegraphics*[width=3.25in,viewport=0 300 600 590]{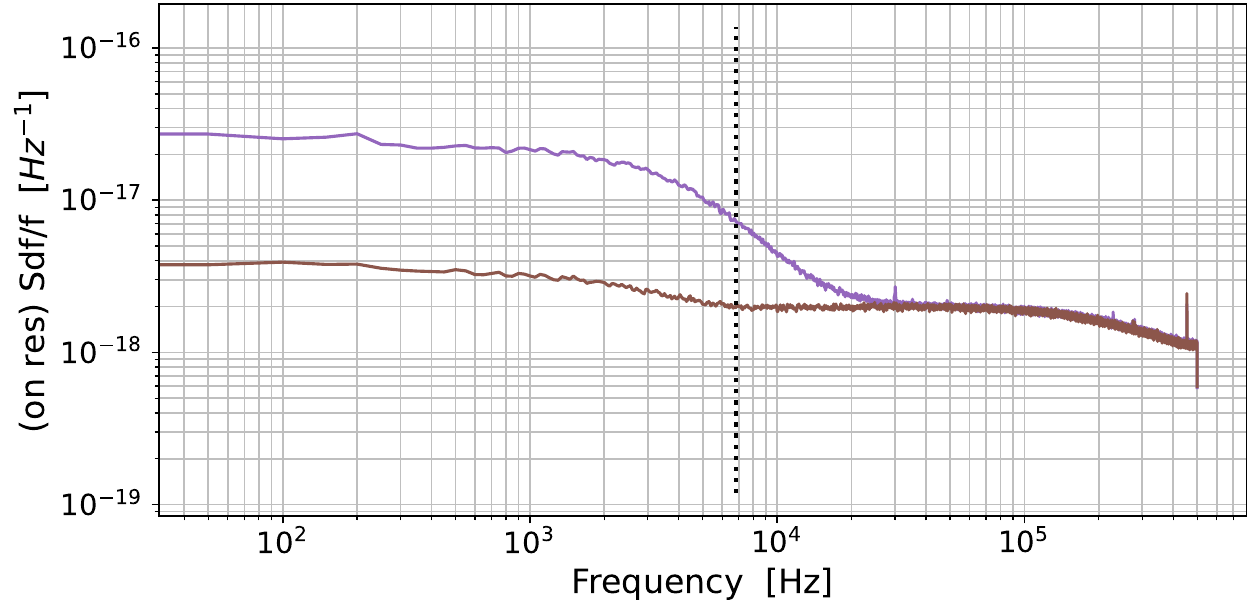} \\ 
\hspace{0.5in}Dark, varying $\Tb$ &  \\
\includegraphics*[width=3in,viewport=0 0 600 596]{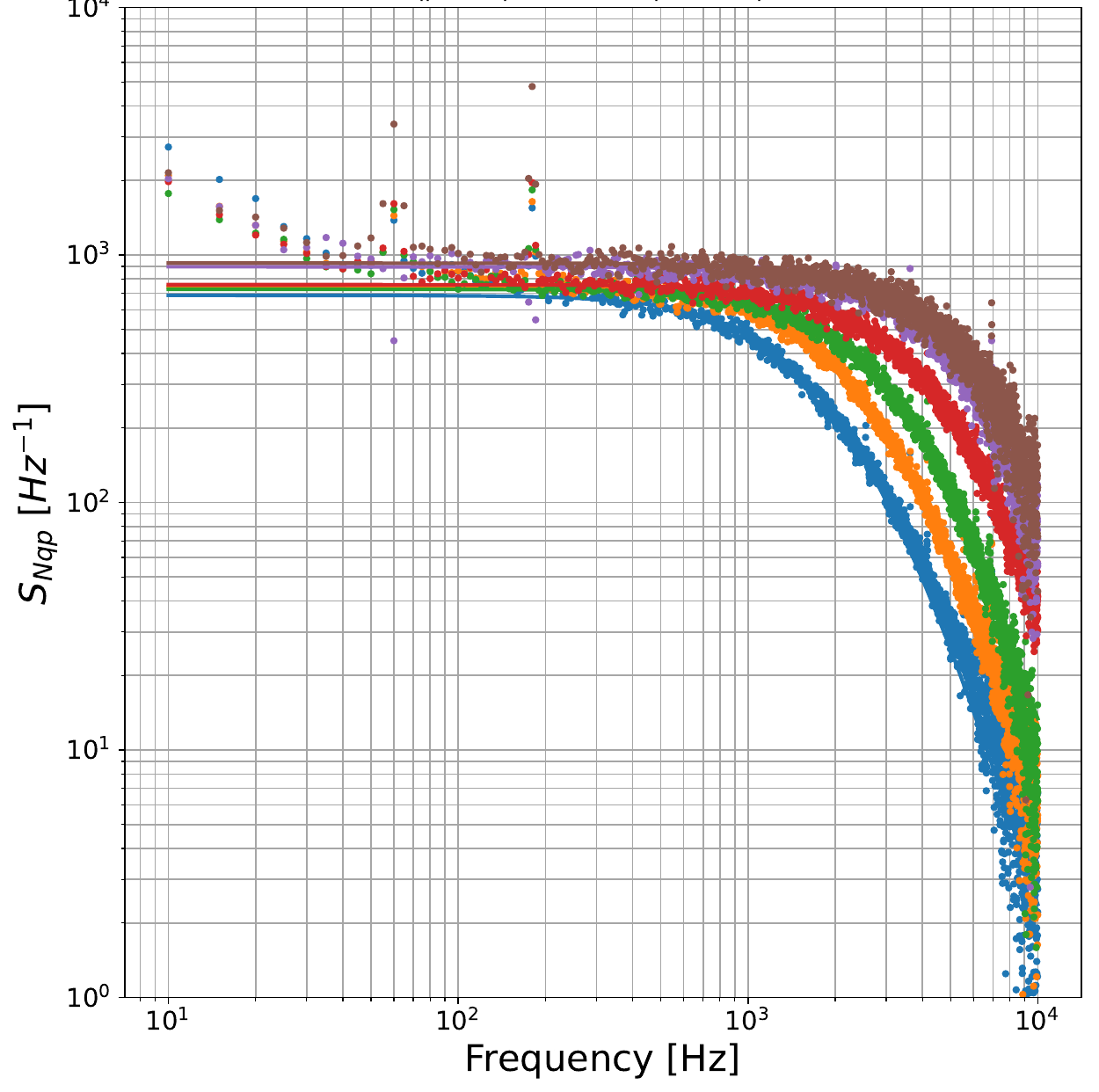}
& 
\includegraphics*[height=2.85in,viewport=30 0 58 288]{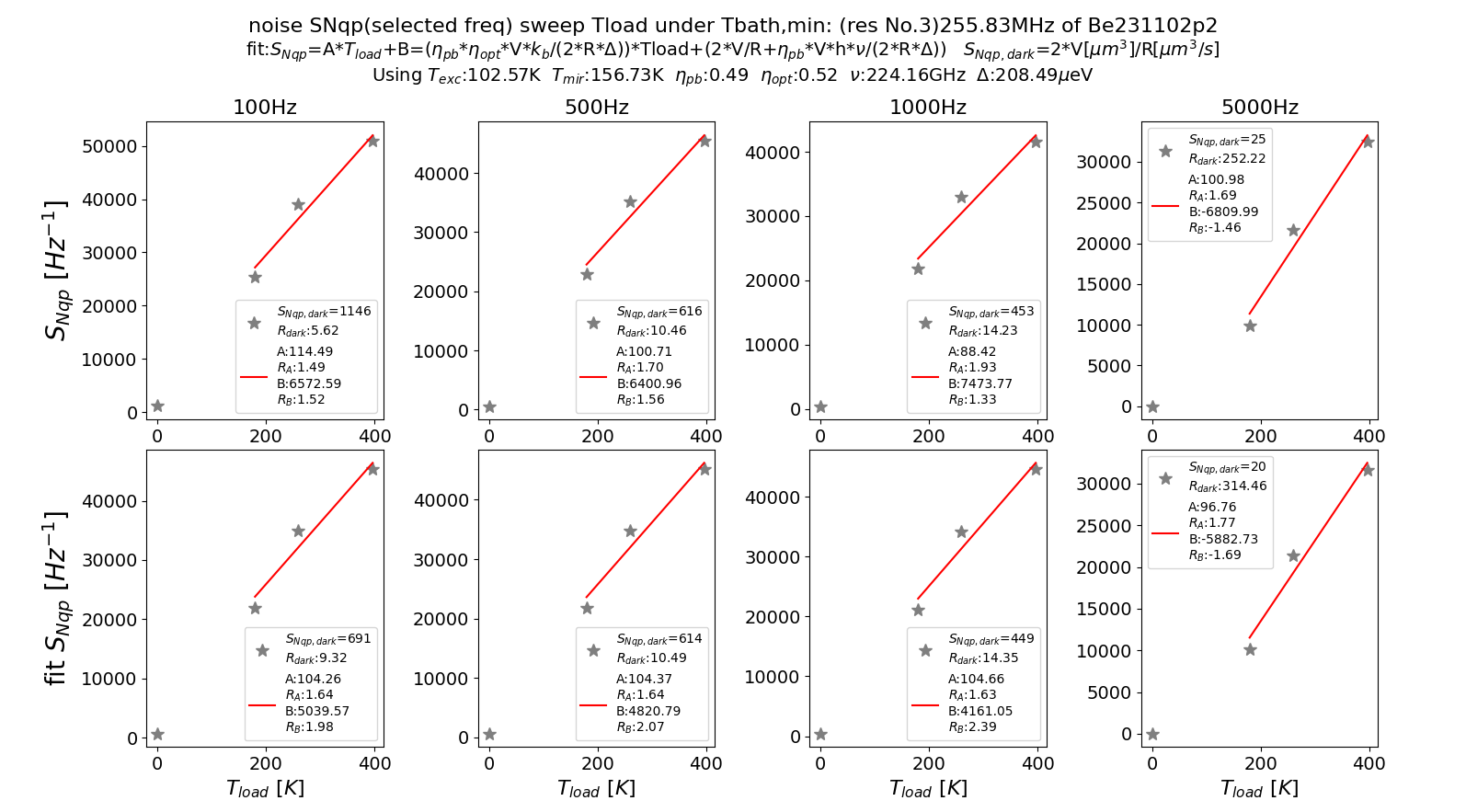}
\includegraphics*[height=2.85in,viewport=310 0 570 288]{figures/Be231102p2_SNqp_Tloadplot_addTexc7_3points_res3.png}
\end{tabular}
}
\end{center}
\caption{\textbf{Noise contributions}  (Top)~Fractional frequency noise power spectral density for one resonator under dark conditions at $\Tb = 310$~mK (left) and under a blackbody load $\Tl = 77~\text{K} + \Tx \approx  180$~K at $\Tb = 241$~mK (right), both at a feedline power of -81~dBm.  The noise in the frequency and dissipation directions are shown (both converted from voltage noise using the same $d(\delta f/f)/dV$, but which is different for the dark and optical data).  The dashed vertical lines indicate the resonator bandwidth ($\fr/2\Qr$).  The noise above 10~kHz is dominated by amplifier noise, while the noise at low frequency is likely dominated by electronics $1/f$ noise.  See text for discussion.  (Bottom left)~Quasiparticle number noise power spectral density $\SNqp$ (difference of on- and off resonance PSDs to subtract electronics white noise at high frequency) for the same resonator as a function of $\Tb$ from 250 to 350~mK in steps of 20~mK in the sequence blue-orange-green-red-purple-brown.  The flat noise level in $\SNqp$ is approximately independent of temperature as would be expected for quasiparticle lifetime $\tqp$ limited by recombination, $\SNqp^{GR} = 4\,\tqp\,\Nqp \approx 2\,V/R$, while the rolloff frequency moves up with temperature as $\tqp^{-1} \approx 2 R\,\Nqp/V$ increases. (Bottom right)~$\SNqp$ as a function of Rayleigh-Jeans optical load temperature.  The optical load includes both the intended load applied outside the dewar and the excess loading from the dewar $\Tx = 103$~K inferred as described in the text.  The behavior is consistent with Equation~\ref{eqn:SNqp_Tload}, with $A$ and $B$ being the slope and offset, indicating our detectors are GR or photon-noise dominated over this full range of optical loads.  (The dark point at $\Tl = 0$ is not included in the fit because it does not include the shot noise term.)}
\label{fig:noise}
\end{figure}

Figure~\ref{fig:noise} shows an example of the noise power spectral densities $\Sdff$ and $\SdQinv$ measured in a dark run at $\Tb = 310$~mK, a temperature at which the quasiparticle density is similar to what we expect under optical load at a telescope on the sky for the most demanding bands (B1--B3), $\Tl \approx 30$--40~K (Table~\ref{tbl:parameters}).  We focus on the region above 100~Hz, where electronics $1/f$ noise is negligible.  The noise in the frequency direction is far in excess of that in the dissipation direction but is flat from 100~Hz until it begins to roll off just above 1~kHz.  The noise in the dissipation direction is approximately flat from 100~Hz to 100~kHz.  (The slight dip seen below a few kHz is likely due to slight miscalibration of the frequency and dissipation directions.  Correcting this miscalibration would have little effect on the frequency-direction noise given the logarithmic vertical scale.)  The resonator ring-down bandwidth is indicated.  We interpret the approximately white dissipation-direction noise as amplifier noise given that it maintains approximately the same level above the resonator bandwidth.  We interpret the higher flat noise in the frequency direction as generation-recombination (GR) noise, rolling off with the quasiparticle lifetime ($\f3db \approx 3.3$~kHz, $\tqp \approx 50$~\mustxtnosp) and the resonator bandwidth ($\f3db \approx 5$~kHz).   The plot thus shows that, under  dark conditions at $\Tb = 310$~mK, the device responsivity is large enough for GR noise to dominate over amplifier noise by about a factor of about 3, the square root of the ratio of the flat noise spectral densities in the frequency and dissipation directions.  

The above conclusion that the noise seen in the frequency direction is GR noise is reinforced by the observation that, when plotted in $\SNqp$ for various temperatures under dark conditions (Figure~\ref{fig:noise}), the flat level is approximately independent of temperature.  We would expect this behavior in the recombination-dominated limit, where $\SNqp^{GR} = 4\,\tqp\,\Nqp \approx 2\,V/R$ with $V$ the inductor volume and $R$ the recombination constant, while the rolloff frequency, given by $\f3db = (2\,\pi\,\tqp)^{-1}$, moves up with temperature as $\tqp$ decreases.  With $V \approx 3500$~
\mumcutxt for this B3 detector and the flat noise level, we infer $R \approx 6.5$--9.5~\mumcutxtnosp/sec, not too different from the canonical value for Al, $R = 10$~\mumcutxtnosp/sec.

\paragraph{Photon-Noise-Dominated Performance Under Optical Load}
\label{sec:noise_opt}

We have also measured the noise under optical load.  We used 77~K and room-temperature (295~K) blackbody loads outside the dewar window as for the optical efficiency measurements, and we also put a reflective cover in front of the dewar window (``mirror''), which we calibrated to yield an effective outside-the-dewar optical load $\Tl \approx 150$~K.  The fits of the dark, cold, and hot load data also allow us to calibrate the excess loading from the dewar to be $\Tx \approx 150$~K in B3, which adds to the loads applied outside the dewar window.  We show $\Sdff$ under cold load ($\Tl \approx 77~\mathrm{K} + \Tx \approx 180$~K) in Figure~\ref{fig:noise}.   While the amplifier noise level in $\Sdff$ increases substantially due to the decrease in $\partial V/\partial(\delta \fr/\fr) = 2\,\Qr^2/\Qc$ under increased optical loading, the frequency direction noise increases by about the same factor.  We interpret this increase as due to the addition of photon noise, which we corroborate below via its dependence on $\Tl$.

To more clearly establish that the increase in the flat noise level observed under optical load is indeed due to photon noise, it is particularly useful to plot $\SNqp$ as a function of $\Tl$ because, in the recombination-dominated regime, it satisfies
\begin{align}
\SNqp^{tot} & = \SNqp^{GR} + S_{N_{qp}}^{shot} + S_{N_{qp}}^{Bose} 
= \frac{V}{R} \left[ 2 + \etapb 
\left( \frac{h\,\nu}{2\,\Delta} + \frac{\etaopt\,k_B\,\Tl}{2\,\Delta} \right) \right]
\label{eqn:SNqp_Tload}
\end{align}
That is, we expect a simple linear dependence of $\SNqp$ on $\Tl$, with the intercept providing the sum of the GR and photon shot noise terms. (Beyond this simple behavior, another benefit of the $\SNqp(\Tl)$ analysis is that it does not depend on converting $\SNqp$ to noise-equivalent power at the input to the KID, $\textrm{NEP}_{opt} = (\Delta/\etapb \tqp)\sqrt{\SNqp}$, which relies on theoretical calculations for $\etapb$ and may suffer systematic uncertainties if $\tqp$ cannot be measured well, such as when it is too similar to the resonator ring-down time constant (as is seen in Figure~\ref{fig:noise}).)  Figure~\ref{fig:noise} shows the expected linear dependence, demonstrating that the additional noise observed under optical load is indeed photon noise.  

We can determine from the $\SNqp$ vs.\ $\Tl$ plot, by empirical extrapolation rather than calculation, the expected $\SNqp$ under the expected optical load on sky at a telescope and the relative contributions of GR, shot, and Bose noise.  For the B3 detector shown, we expect $\Tl = 40$~K on sky, yielding an expected total noise of $\SNqp \approx 9000$/Hz.  Given $\SNqp \approx 750$/Hz measured under dark conditions, it would appear the photon noise is well in excess of the GR noise.  However, a detailed comparison reveals the dark and optical data do not yield a consistent recombination constant.  Using $\etapb = 0.49$ as discussed earlier, estimated values of $\nu = 225$~GHz (B3) and $2\,\Delta_{Al} = 96$~GHz, $V = 3500$~\mumcutxtnosp, and $R = 8$~\mumcutxtnosp/sec, we find the shot noise term should be 60\% of the GR noise term, or about 450/Hz.  The fit yields an offset, corresponding to the sum of the GR and shot noise terms, of $\approx$4800/Hz, about four times larger than the expected $750 + 450 = 1200$/Hz.  Similarly, using $\etaopt = 0.5$ (\S\ref{sec:opt_eff}), we find $\etaopt\,k_B/2\,\Delta_{Al} \approx 0.11$ and thus the slope should be 23/(K~Hz) while the fit yields a slope of 104/(K~Hz), 4.5 times larger than expected.  We can only reconcile the dark and optical data by positing that somehow $R$ is reduced when trans-mm-generated quasiparticles are present, a hypothesis we will return to momentarily.  Assuming $R = 1.8$--2~\mumcutxtnosp/sec is more appropriate for optical data, then we infer $\SNqp^{GR} \approx 3000$--3400/Hz, $\SNqp^{shot} \approx 1800$--2000/Hz, and $\SNqp^{Bose}(\Tl = 40~\mathrm{K}) \approx 3500$--4000/Hz.  The result is that the GR noise is about 60\% of the photon noise and about 1/3 of the total noise --- our detectors will be quite close to photon-noise-limited at expected optical loads.

As to how $R$ can decrease under optical load, one hypothesis would be that: 1)~the phonons arising from pair recombination of trans-mm-generated quasiparticles have a distinctly different spectrum than the Wien tail of thermal phonons that thermally generate quasiparticles; and, (2) this different spectrum may be more easily trapped by acoustic mismatch in the Al KID film  than thermal phonons.  It is known that the recombination constant and quasiparticle lifetime measured in a purely thermal environment for thin films already reflect some level of phonon trapping --- lifetimes calculated from BCS theory~\cite{kaplan1976} are much faster than those observed --- so we hypothesize that our data can be explained by an enhancement of this effect.  The effect may have be enhanced by the 800~nm a-Si:H layer between the KID film and the susbtrate. With more exhaustive data over all the frequency bands, we can test for a systematic dependence of $R$ on $\nu$, which would impact the trans-mm-generated phonon spectrum and which varies by a factor of 2.5 over B2--B5.

While we have already demonstrated performance limited by GR+photon noise, it is useful to convert to noise-equivalent power, NEP$_{opt}$ (\wrthztxtnosp), because it is the standard performance metric for detectors of this type.  Under the expected on-sky optical loads from 
Table~\ref{tbl:parameters}, the photon-noise-limited sensitivity, $\NEP^\gamma_{opt}$, varies from 6.4 to $16\,\times\,10^{-17}$~\wrthztxt for B2--B6.  (B1 requires AlMn, which will have different $\SNqp$.)  If we map from the 
expected $\tqp$ under optical load ($\approx 30$~\mustxtnosp) to the $\Tb$ with matching $\tqp$, then $\Tb = 310$~mK and we see that $\NEP^{GR}_{opt} \le \NEP^\gamma_{opt}$.  The margin is not as great as the one from 
the empirical extrapolation above because it fails to account for the decrease in $R$ observed under optical load.  Both $\SNqp$ and $\tqp$ are inversely proportional to $R$, so $\NEP^{GR}_{opt} \propto \sqrt{R}$, decreasing as $R$ decreases.  The factor of 4 reduction in $R$ between dark and optical data implies a factor of $\sqrt{4} = 2$ in $\NEP^{GR}_{opt}$, implying $\NEP^{GR}_{opt} \le \NEP^\gamma_{opt}/2$ and ensuring photon noise is dominant.

\begin{figure}[t!]
\begin{center}
\includegraphics*[width=3in,viewport=0 0 620 596]{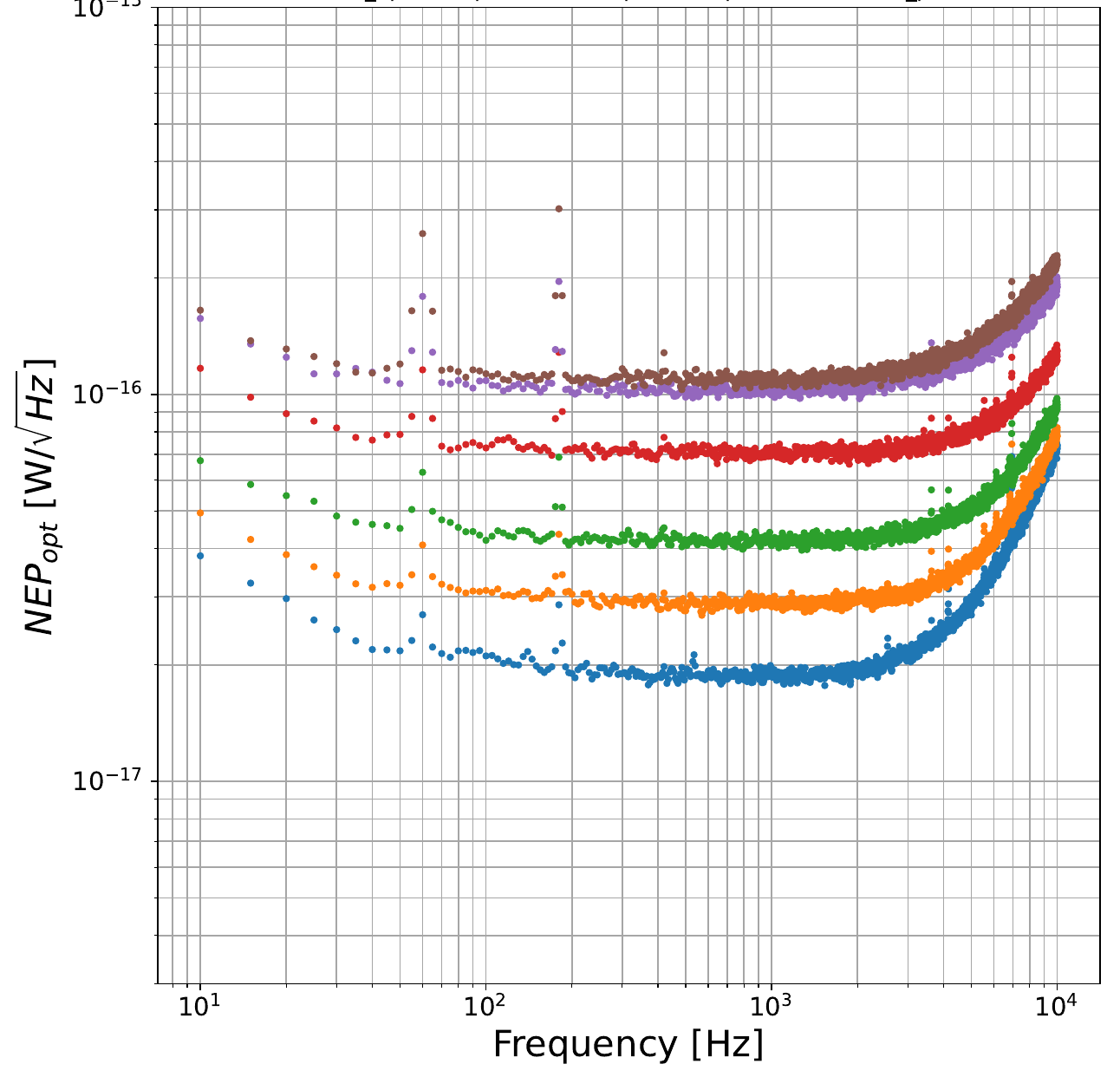}
\end{center}
\vspace{-9pt}
\caption{\textbf{Sensitivity} $NEP^{GR}_{opt} = (\Delta/\etapb\tqp)\sqrt{\SNqp}$ using $\SNqp$ from dark data (Figure~\ref{fig:noise} (bottom left)) as a function of $\Tb$ assuming $\etapb = 0.5$, same legend as Figure~\ref{fig:noise} (bottom left).  See discussion in text for comparison to expected photon noise; it is not trivial given the apparent reduction of $R$ under optical load.}
\label{fig:sensitivity}
\end{figure}

\paragraph{TLS Noise}

%
%
%
%
Electronics $1/f$ noise currently makes it difficult to measure the noise below 100~Hz, but we may estimate the expected TLS noise level using existing measurements\cite{defrance_asihnoise}, which yield $\STLSdff = 0.8$, 0.4, and $0.25\,\times\,10^{-18}$/Hz at 10, 100, and 1000~Hz, an electric field of 130~V/m, at $\Tb = 250$~mK, and for a capacitor area $A_C = 0.3$~\mmsqtxt (single top plate).  (We do not correct the TLS noise for its $T^{-1.7}$ temperature dependence\cite{kumar_noise} because the capacitor dielectric would be at $\Tb = 250$~mK even if the quasiparticle density under load is comparable to what is seen thermally at $\Tb = 330$~mK.)  Comparing to the dark data in Figure~\ref{fig:noise} (upper left), we scale to the expected B3 KID capacitor area $A_C = 0.55$~\mmsqtxt and the electric field $E = 1430$~V/m at the applied feedline power of --81~dBm to obtain $\STLSdff = 3.9$, 2.0, and $1.2\,\times\,10^{-20}$/Hz at 10, 100, and 1000~Hz, below the amplifier noise and well below the GR noise.  Comparing to the cold load data in Figure~\ref{fig:noise} (upper right), the same feedline power now yields $E = 990$~V/m because of the reduced $\Qi$ under load.  The TLS noise scales to $\STLSdff =5.6$, 2.8, and $1.8\,\times\,10^{-20}$/Hz at 10, 100, and 1000~Hz, slightly higher but now even further below the amplifier noise and photon noise because the latter have increased substantially in $\Sdff$ units while the TLS noise has not.   Extrapolating to lower frequencies more relevant for astronomical observations, assuming a conservative $\Sdff \propto 1/f$ scaling, yields $\STLSdff = 3.9$--$5.6\,\times\,10^{-19}$/Hz at 1~Hz and 3.9--$5.6\,\times\,10^{-18}$/Hz at 0.1~Hz.  Since the amplifier and fundamental (GR+photon) noise under sky loading will likely be somewhere between the dark and cold load noise $\Sdff$ PSDs from Figure~\ref{fig:noise}, TLS noise will likely be subdominant to amplifier noise and certainly well below fundamental noise.  Actual measurements of TLS noise under optical load close to expectations will likely motivate reducing the capacitor area to reduce focal plane dead area.

\subsection{Supporting Subsystems}
\label{sec:supporting}

NEW-MUSIC builds on the heritage of the MUSIC instrument~\cite{music_spie2012_golwala, music_spie2014_sayers}, and we will reuse much of MUSIC for NEW-MUSIC.  We review the extant MUSIC sub-systems and describe the necessary modifications.

\paragraph*{Cryostat}   
\label{sec:cryostat} 
The MUSIC cryostat~\cite{music_spie2010_hollister} consists of a 1.5~m tall, 0.6~m diameter dewar with two internal stages cooled by a Cryomech PT-415 pulse-tube cooler, providing 40~W at 50~K and 1.35~W at 4~K.  A Chase Cryogenics $^4$He/$^3$He/$^3$He closed-cycle sorption cooler provides 3~\muwtxt at 240~mK and 100~\muwtxt at 350~mK, sufficient to accommodate conductive and radiative heat loads while providing well in excess of 24 hrs hold time after a 6~hr cycle.  It incorporates a two-layer A4K shield at 4~K to provide magnetic shielding against Earth's field, which varies with time as the telescope moves.  The cryostat originally included eight 2-8~GHz HEMT amplifiers, which will be replaced as needed with SiGe amplifiers that provide excellent performance below 2~GHz.  The cryostat is already equipped with the necessary RF coaxes as well as DC wiring for thermometry.

\paragraph*{Optical Train} 
\label{sec:optics}
 
\begin{figure}[t!]
\begin{center}
\includegraphics*[width=6.5in,viewport=145 735 611 951,page=2]{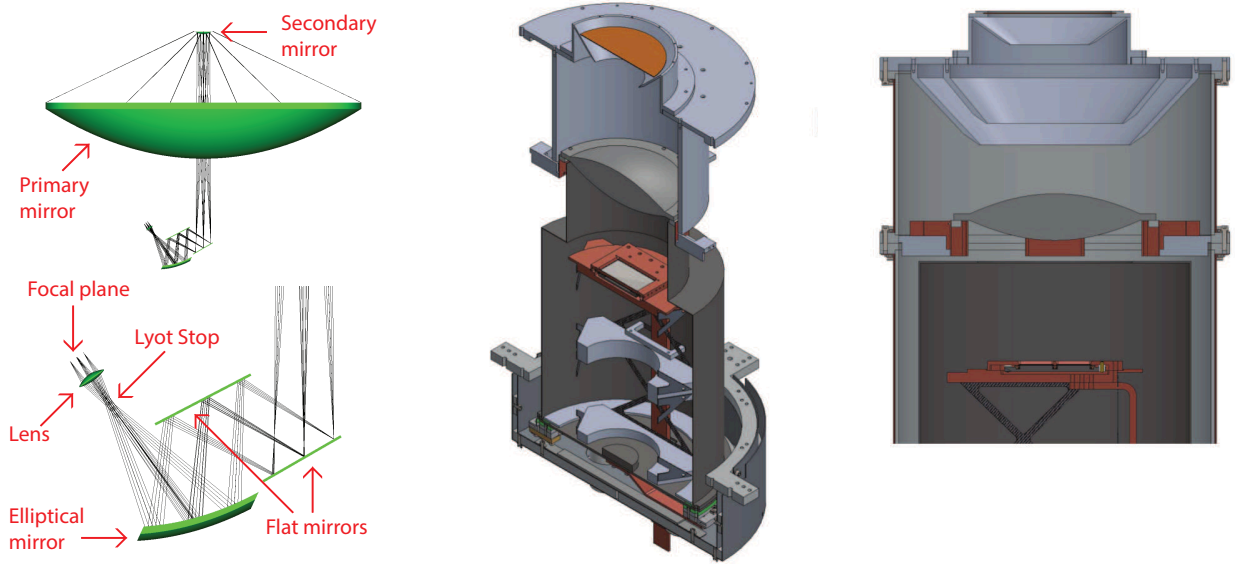}
\end{center}
\vspace{-6pt}
\caption{\textbf{MUSIC optical design.} (Top left) Full optical model including primary and secondary mirrors of Leighton Telescope.  (Bottom left) Close-up on relay optics include two folding flats, one 300~K ellipsoid, a cold lens, and the focal surface.  The Cassegrain focus is visible just above the first fold mirror.  (Center) CAD model of internal optics at 4~K (including Lyot stop and lens) and focal plane location for $f/2.19$ configuration.  Some baffles are removed.  (Right) CAD model showing all 4~K baffles for $f/2.19$ configuration.  The Lyot stop is at the bottom of the top conical baffle.}
\label{fig:music_optics}
\end{figure}

The MUSIC optical train consists of one powered mirror and two flat mirrors at 300~K followed by a cold Lyot stop and cold lens at 4~K; see Figure~\ref{fig:music_optics}.  The original cold optics reimaged the $f/12.6$ Cassegrain focus of the CSO to $f/3.45$ with a 15\arcmintxt diameter (135~mm) field-of-view and a flat focal surface using a single-layer Porex-coated HDPE lens.   A second version of the cold optics instead reimaged to $f/2.19$, making the field-of-view 85~mm in diameter.  We will revise the 4~K lens slightly to provide a focal ratio of $f/1.72$ to match our antenna pixels, making the 14\arcmintxt field-of-view roughly 60~mm across.  We will also AR-coat the lens with two layers of Porex to provide broader bandwidth.  Ongoing development work on broadband AR-structured metamaterial silicon gradient-index lenses~\cite{defrance_grin_2024} may be implemented later to enhance optical efficiency.  At the focal plane, NEW-MUSIC will use a proven three-layer silicon AR structure~\cite{defrance_ltd18_2020, defrance_grin_2024}, or an in-development four-layer AR.


In terms of windows and filtering, the existing MUSIC design is quite similar to the one used for the work presented here, with the differences being that MUSIC used a HDPE rather than UHMWPE window, metal-mesh 14~THz and 3.75~THz low-pass blackbody filters  instead of Zotefoam sheets to limit 300~K radiation incident on 50~K, and PTFE instead of Nylon at 4~K.  As noted earlier, we plan to revisit this stack to increase its efficiency without degrading the optical load on 4~K or the sub-Kelvin cooler.

The $f/2.19$ optical train incorporated significant baffling near the Lyot stop, which ensures that, in a time-reversed sense, rays emitted by the focal plane have many opportunities to intersect cold, absorbing surfaces before exiting the cryostat.  See Figure~\ref{fig:music_optics}.  The absorber used was ``steelcast''~\cite{wollack_steelcast2008}.  Without these baffles, rays that reflect off the Lyot stop can reflect off the focal plane and exit through the optical path.   In spite of this baffling, MUSIC suffered a significant ($\sim$50\%) loss of beam to wide angles~\cite{siegel_thesis}, suggesting the baffling was not completely effective in absorbing wide-angle beam.  We will incorporate more sophisticated absorber materials that have been developed in recent years~\cite{wollack_absorber_2016,xu_simons_metamaterial_2021,inoue_absorber_2023,pisano_metamaterial_absorber_2023}, and we plan an extensive baffling test campaign.  

\paragraph*{KID Readout and Data Acquisition} 
\label{sec:readout} 

Our initial deployment plan (for a quarter-scale focal plane; see \S\ref{sec:conclusion}) is to reuse the MUSIC readout but without the the up/down-converting IF system.  Each MUSIC readout module\cite{mchugh_readout} consists of three boards: a ROACH-1 board~\cite{roach_casper} providing a Virtex5 FPGA, a PowerPC for a simple linux operating system, and an ethernet transceiver; an ADC/DAC board with two pairs of TI DAC5618 1000 MHz 16-bit DACs and TI ADS5463 500-MHz 12-bit ADCs, which we operate at 491.52 MHz; and, an ``IF'' (intermediate frequency) board that provides 300~K amplification, a variable attenuator, IQ mixers, baseband amplification, and anti-alias filtering.  The board set accepts signals from a Rb frequency standard locked to GPS to provide a stable frequency reference (10~MHz) for generation of the CPU and ADC/DAC clocks and for absolute time synchronization (1~Hz).  Each module has firmware~\cite{music_spie2010_duan, music_readout_duan} that handles 192 readout tones across 450~MHz of usable bandwidth, using a $2^{16}$-sample FFT to output data at 7.5~kHz for each tone.  On-board filtering and decimation reduce the data rate to 100~Hz, matching the telescope pointing stream, for transport off the board.  The entire planned MUSIC focal plane required 16 boards: 144 KIDs per board with 48 monitor tones for removal of low-frequency gain and phase noise.  The boards were housed in two fan-cooled electronics crates that mount on the telescope near the cryostat.  The vented air is ducted away from the telescope.  DC power supplies sat in the CSO dome, away from the telescope. 

The 100-400 MHz readout band motivates replacing the MUSIC IQ mixer (``IF'') board with a much simpler one that eliminates the IQ mixers and uses diplexers at the Nyquist frequency to separately populate the first and second Nyquist zones of different ADC/DAC pairs.  It would still include additional gain, variable attenuation, and anti-alias filtering.  The new boards will be compatible with the MUSIC FPGA command interface to obviate firmware changes. 

RF system-on-chip (RFSoC) implementations for KID readout are making fast progress~\cite{sinclair_rfsoc_spie2020, sinclair_rfsoc_spie2022, sinclair_rfsoc_isstt2023, sinclair_rfsoc_spie2024}, so we anticipate switching to such a system for a full NEW-MUSIC focal plane.

\paragraph*{Data Acquisition} 

MUSIC used an industrial server running Linux from internal SSDs with a hard-drive RAID for data.  CASPER~\cite{roach_casper} provided python tools for communicating with the ROACHs, which we used wholesale.  We wrote our own Matlab-based data acquisition software to accept the streams of data from the 16 ROACHs, combine it with pointing and slow monitoring data, and package it into the HDF5 format.  Initially, we plan to port this existing DAQ to modern computing hardware and a Python platform and use SSDs for storage as well as the operating system.  When we switch to a RFSoC system, we may adapt DAQ software built for such systems and only replace the low-level DAQ-KID readout drivers.

\paragraph*{Data Reduction} An IDL-based pipeline was used for reduction of MUSIC data for extensive instrument characterization~\cite{music_spie2012_golwala, music_spie2014_sayers, siegel_music_noise_2015, siegel_thesis} and  science publications~\cite{music_rxj1347, music_hatlas}.  This pipeline was an extension of the pipeline developed for Bolocam, which was the mm-wave facility camera on the CSO for over a decade and resulted in tens of publications.  We will either port this code base to Python or adopt data reduction packages developed over the past decade for KID-based instruments (e.g., \texttt{citlali}~\cite{toltec_citlali_2022} for TolTEC).

\subsection{Sensitivity and Mapping Speed}
\label{sec:sensitivity}

Given uncertainties on the recombination constant $R$, we take a semi-empirical approach to calculating the expected sensitivity and mapping speed of NEW-MUSIC, making heavy use of  Equation~\ref{eqn:SNqp_Tload} and the measurements in \S\ref{sec:noise_opt}.  We calculate $\SNqp^{GR}$ from $V = 3500$~\mumcutxt and the empirical noise-based value of $R = 1.9$~\mumcutxtnosp/sec from \S\ref{sec:noise_opt}.  We calculate $\SNqp^{shot}$ using the same $V$ and $R$ as for $\SNqp^{GR}$ along with $\Delta_{AlMn} = 165$~\muevtxt for B1 (\S\ref{sec:almn}) and $\Delta_{Al} = 200$~\muevtxt (Figure~\ref{fig:kid_params}) for the other bands along with $\etapb$ from\cite{guruswamy_etapb_2014} (listed in Table~\ref{tbl:sensitivity}; we take $\etapb = 1$ for B1 because $h\,\nu \approx 2\,\Delta_{AlMn}$).  For $\SNqp^{Bose}$, we start with the measurement of the $\SNqp^{Bose}(\Tl)$ slope in Equation~\ref{eqn:SNqp_Tload} shown in Figure~\ref{fig:noise}.  Because of the $\etaopt$ factor in that term, we correct it by the ratio between the optical efficiency expected for NEW-MUSIC (which accounts for the Lyot stop) and the optical efficiency we measure here (the average of the two devices' ``max'' values of $\etaopt^{design}$, with the use of the maximum motivated as in \S\ref{sec:opt_eff}).  We then multiply by $\Tl$ given in Table~\ref{tbl:parameters}.  We convert these three $\SNqp$ values to noise-equivalent power (NEPs) using another recombination-limit relation:
\begin{align}
\frac{d\Popt}{d\Nqp} = \sqrt{\frac{\Delta \,\Popt}{\etapb}\,\frac{R}{V}}
\end{align}
This approach is mathematically equivalent to calculating $\Nqp$ from $\Popt$, $R$, $V$, and $\Delta$ using the GR equation, calculating $\tqp$ from $\Nqp$, $R$, and $V$, and using $d\Popt/d\Nqp = \Delta/\etapb \tqp$, but it obviates the intermediate result for $\tqp$ and makes it clear how the derivative depends on input parameters.  We calculate $\Popt$ for the expected $\Tl$ and assuming $\Delta \nu$ from Table~\ref{tbl:parameters} and $\etaopt$ for the instrument as given in Table~\ref{tbl:sensitivity}.  Consistent with expectations from \S\ref{sec:noise_opt}, Table~\ref{tbl:sensitivity} shows this semi-empirical model indicates NEW-MUSIC will be quite close to photon-noise-limited.  

Table~\ref{tbl:sensitivity} also shows calculations of noise-equivalent flux density (NEFD) and mapping speed ($N_{pix} \Omega_{beam}/\text{NEFD}^2$).  We calculate NEFD from NEP$_{tot}$ using the 10.4~m diameter of the Leighton Telescope along with a degradation factor to account for primary illumination, including the effect of the Lyot stop.  We then calculate mapping speed from the NEFD, $N_{pix}$, and $\Omega_{beam} = \text{FWHM}^2\,\pi/(4 \ln 2)$ ($N_{pix}$ and FWHM from Table~\ref{tbl:parameters}).

\begin{table}
\begin{center}
    \begin{tabular}{r|cccccc}
                         &   B1 &   B2 &   B3 &   B4 &   B5 &    B6 \\ \hline\hline
$\etaopt$ w/Lyot stop    & 0.26 & 0.40 & 0.34 & 0.38 & 0.26 &  0.29 \\
$\etapb$                 &    1 & 0.64 & 0.49 & 0.45 & 0.41 &  0.41 \\ 
$\Popt$ w/Lyot stop [pW] &  4.1 &  8.7 &  8.4 &  9.8 &   12 &    18 \\ 
$\Delta$ [$\mu$eV]       &  165 &  200 &  200 &  200 &  200 &   200 \\ \hline
$\SNqp^{GR}$ [/Hz]       & \multicolumn{6}{c}{3700}                \\
$\SNqp^{shot}$ [/Hz]     & 2100 & 1800 & 2200 & 2500 & 2700 &  3200 \\
$\SNqp^{Bose}$ [/Hz]     & 3000 & 3100 & 2800 & 3800 & 8500 & 12600 \\
$d\Popt/d\Nqp$ [pW]      & 0.48 & 0.97 & 1.09 & 1.23 & 1.45 &  1.74 \\ \hline
NEP$_{GR}$ [aW/$\sqrt{\text{Hz}}$]      &   29 &   59 &   67 &   75 &   88 &  106 \\
NEP$_{photon}$ [aW/$\sqrt{\text{Hz}}$]  &   35 &   68 &   77 &   98 &  153 &  219 \\
NEP$_{tot}$ [aW/$\sqrt{\text{Hz}}$ ]    &   45 &   90 &  102 &  123 &  176 &  244 \\
NEP$_{tot}$/NEP$_{photon}$              & 1.31 & 1.32 & 1.32 & 1.26 & 1.15 & 1.11 \\ \hline
NEFD [mJy$\sqrt{\text{sec}}$ ]  &   13 &   16 &   20 &   22 &   69 &  115 \\ 
Mapping Speed [arcmin$^2$/mJy/hr]       & 2500 &  810 &  730 &  500 &  100 &   26\\ \hline
    \end{tabular}
    \end{center}
\caption{\textbf{NEW-MUSIC expected sensitivity and mapping speed under best-case conditions.} The sensitivity is estimated using the semi-empirical approach discussed in the text using the ``best-case conditions'' optical loads provided in Table~\ref{tbl:parameters}.  Amplifier and TLS noise are not listed in the noise budget because Figure~\ref{fig:noise} shows they are highly subdominant.  The ratio NEP$_{tot}$/NEP$_{photon}$ shows that the degradation relative to fully photon-noise-limited performance will be small.  The mapping speeds provided are for the full pixel-count focal plane, while the NEPs and NEFDs are independent of the FPU size.  We assume AlMn KIDs for B1 and Al KIDs for the other bands (see \S\ref{sec:almn}). }
\label{tbl:sensitivity}
\end{table}

\section{Conclusion and Future Plans}
\label{sec:conclusion}

We have motivated, described, and provided significant technology validation for the Next-generation, Extended Wavelength, MUlti-band Sub/mm Inductance Camera, NEW-MUSIC, a six-band, polarization-sensitive, trans-mm camera for the 10.4~m Leighton Chajnantor Telescope.  NEW-MUSIC will provide SEDs from 80 to 420~GHz for: a variety of time-domain sources to probe new frontiers in energy, density, time, and magnetic field: the Sunyaev-Zeldovich effects in galaxies and galaxy clusters to study accretion, feedback, and dust content in their hot gaseous haloes; and to provide new insights into stellar and planetary nurseries via dust thermal emission and polarization. 

Hierarchical summing of our slot-dipole, phased-array antennas works as expected and that microstripline loss is acceptable.  We have demonstrated reasonable spectral bandpasses and competitive optical efficiency in four of NEW-MUSIC's six bands, including validation of NEW-MUSIC's groundbreaking microstripline-coupled, parallel-plate-capacitor, lumped-element KIDs.  The detectors' generation-recombination noise dominates over amplifier and two-level-system noise above 100~Hz, and scaling predictions  indicate this performance should continue to hold down to 0.1~Hz.  The NEW-MUSIC detectors are demonstrably photon-background-limited.  Direct absorption is at the $\lesssim 1$\% level.  We have fabricated and observed reasonable first-try yield for AlMn KIDs necessary for B1. 
 
In the near term, we expect to provide the following key remaining demonstrations: beams for the three-scale antenna; similar sensitivity down to the 0.1--1~Hz audio frequencies necessary for astronomical scanning observations; AlMn KIDs with photon-background-limited sensitivity for B1; improved optical efficiency and spectral bandpasses; explicit measurements of loss and wave-speed for our a-Si:H microstripline; and, resonator yield and  collision statistics.  These results will enable the final design of NEW-MUSIC.  Funding-permitting, NEW-MUSIC will deploy with a quarter-scale focal plane on LCT in 2027, with a focus on time-domain sources and object-oriented science.  On-sky validation will motivate construction of the full focal-plane and readout system to enable wide-area surveys.

\acknowledgments 
 
This work has been supported by the JPL Research and Technology Development Fund, the National Aeronautics and Space Administration under awards 80NSSC18K0385 and 80NSSC22K1556, the Department of Energy Office of High-Energy Physics Advanced Detector Research program under award DE-SC0018126, and the Wilf Foundation. The research was carried out in part at the Jet Propulsion Laboratory, California Institute of Technology, under a contract with the National Aeronautics and Space Administration (80NM0018D0004).  The authors acknowledge the work of numerous former students and collaborators in the development of the technologies presented here and thank Liam Connor, Anna Ho, Mansi Kasliwal, Shri Kulkarni, Sterl Phinney, and Vikram Ravi for development of the time-domain astronomy science targets.

\bibliography{report} 
\bibliographystyle{spiebib} 

\end{document}